\documentclass[usenatbib]{aa}
\usepackage[varg]{txfonts}
\usepackage[english,english]{babel}
\usepackage{amsmath}
\usepackage{amssymb,amsfonts,textcomp}
\usepackage{array}
\usepackage{supertabular}
\usepackage{hhline}
\usepackage{hyperref}
\usepackage[usenames]{color}
\hypersetup{dvips, colorlinks=true, linkcolor=black, citecolor=blue, filecolor=blue, urlcolor=blue}
\usepackage{graphicx}
\usepackage{xspace}
\bibpunct{(}{)}{;}{a}{}{,}

\def\gtrsim{\lower.5ex\hbox{$\; \buildrel > \frac \sim \;$}}

\newcommand{\msun}{\mbox{$M_\odot$}}

\newcommand{\hagn}{\mbox{{\sc \small Horizon-AGN}}}

\newcommand{\nh}{\mbox{{\sc \small NewHorizon}}}
\newcommand{\newh}{\mbox{{\sc \small NewHorizon}}}

\newcommand{\sigmagas}{\mbox{$\Sigma_{\rm gas}$}\xspace}
\newcommand{\sigmasfr}{\mbox{$\Sigma_{\rm SFR}$}\xspace}

\begin{document}

\title{Introducing the NewHorizon simulation: Galaxy properties with resolved internal dynamics across cosmic time}

\titlerunning{The NewHorizon simulation}
\authorrunning{Y. Dubois et al.}

\author{Yohan Dubois\inst{1} \and Ricarda~Beckmann\inst{1} \and Fr\'ed\'eric~Bournaud\inst{2,3} \and Hoseung~Choi\inst{4} \and Julien~Devriendt\inst{5} \and Ryan~Jackson\inst{6}  \and Sugata~Kaviraj\inst{6}  \and Taysun~Kimm\inst{4} \and Katarina~Kraljic\inst{7,8} \and Clotilde~Laigle\inst{1}  \and Garreth~Martin\inst{9,10} \and Min-Jung~Park\inst{4} \and S\'ebastien~Peirani\inst{11,1}  \and Christophe~Pichon\inst{1,12,13} \and Marta Volonteri\inst{1} \and Sukyoung K.~Yi\inst{4}}
\institute{Institut d'Astrophysique de Paris, UMR 7095, CNRS, UPMC Univ. Paris VI, 98 bis boulevard Arago, 75014 Paris, France \\\email{dubois@iap.fr} 
\and AIM, CEA, CNRS, Université Paris-Saclay, Université Paris Diderot, Sorbonne Paris Cité, 91191 Gif-sur-Yvette, France
\and IRFU, CEA, Université Paris-Saclay, 91191 Gif-sur-Yvette, France
\and Department of Astronomy and Yonsei University Observatory, Yonsei University, Seoul 03722, Republic of Korea
\\\email{yi@yonsei.ac.kr}
\and Department of Physics, University of Oxford, Keble Road, Oxford OX1 3RH, United Kingdom
\and Centre for Astrophysics Research, University of Hertfordshire, College Lane, Hatfield, Herts AL10 9AB, United Kingdom
\and Aix Marseille Université, CNRS, CNES, UMR 7326, Laboratoire d’Astrophysique de Marseille, Marseille, France
\and Institute for Astronomy, University of Edinburgh, Royal Observatory, Blackford Hill, Edinburgh, EH9 3HJ, United Kingdom
\and Steward Observatory, University of Arizona, 933 N. Cherry Ave, Tucson, AZ 85719, USA
\and Korea Astronomy and Space Science Institute, 776 Daedeokdae-ro, Yuseong-gu, Daejeon 34055, Republic of Korea
\and Université Côte d’Azur, Observatoire de la Côte d’Azur, CNRS, Laboratoire Lagrange, Nice, France
\and IPHT, DRF-INP, UMR 3680, CEA, Orme des Merisiers Bat 774, 91191 Gif-sur-Yvette, France
\and Korea Institute of Advanced Studies (KIAS) 85 Hoegiro, Dongdaemun-gu, Seoul, 02455, Republic of Korea
} 

\date{Received / Accepted }

\abstract{Hydrodynamical cosmological simulations are increasing their level of realism by considering more physical processes and having greater resolution or larger statistics.
However, usually either the statistical power of such simulations or the resolution reached within galaxies are sacrificed.
Here, we introduce the \nh\ project in which  we simulate at high resolution a zoom-in region of $\sim(16 \,\rm Mpc)^3$ that is larger than a standard zoom-in region around a single halo and is embedded in a larger box.
A resolution of up to $34\, \rm pc$, which is typical of individual zoom-in, up-to-date resimulated halos, is reached within galaxies; this allows the simulation to capture the multi-phase nature of the interstellar medium and the clumpy nature of the star formation process in galaxies. 
In this introductory paper, we present several key fundamental properties of galaxies and their black holes, including the galaxy mass function, cosmic star formation rate, galactic metallicities, the Kennicutt-Schmidt relation, the stellar-to-halo mass relation, galaxy sizes, stellar kinematics and morphology, gas content within galaxies and its kinematics, and the black hole mass and spin properties over time.
The various scaling relations are broadly reproduced by \nh\ with some differences with the standard observables.
Owing to its exquisite spatial resolution, \nh\ captures the inefficient process of star formation in galaxies, which evolve over time from being more turbulent, gas rich, and star bursting at high redshift. 
These high-redshift galaxies are also more compact, and they are more elliptical and clumpier until the level of internal gas turbulence decays enough to allow for the formation of discs.
The \nh\ simulation gives access to a broad range of galaxy formation and evolution physics at low-to-intermediate stellar masses, which is a regime that will become accessible in the near future through surveys such as the LSST.
}


\maketitle

\section{Introduction}

The origin of the various physical properties of galaxies, such as their mass content, size, kinematics, or morphology, emerges from the complex multi-scale and highly non-linear nature of the problem.
It involves a strong connection between the small-scale star formation embedded in large molecular complexes and the gas that is accreted from the intergalactic medium and ejected into large-scale galactic outflows.
To draw a theoretical understanding of the process of galaxy formation and evolution, it is necessary to connect cosmological structure formation---which leads to gas accretion into galaxies, that is the fuel of star formation---to the relevant small-scale processes that lead to the formation of the stars.
Therefore, cosmological simulations are now a key tool in this theoretical understanding by allowing us to track the anisotropic non-linear cosmic accretion~\cite[which spectacularly results in filamentary gas accretion; e.g. ][]{keresetal05,dekel&birnboim06,ocvirketal08} in a self-consistent fashion.

Important challenges exist in the field of galaxy formation that need to be addressed, such as the global inefficiency of the star formation process on galactic scales~\cite[e.g.][]{mosteretal13,behroozietal13}, the morphological diversity of galaxies across the whole mass range~\citep[e.g.][]{conselice06,martinetal20}, and the important evolution of the nature of galaxies over time; galaxies are more gas rich~\citep[e.g.][]{daddietal10a} and turbulent~\citep[e.g.][]{kassinetal07}, clumpy and irregular~\citep[e.g.][]{genzeletal11}, and star forming~\citep[e.g.][]{elbazetal07} at early time than they are in the local Universe. 

High-redshift galaxies substantially differ in nature from low-redshift galaxies because cosmic accretion is more efficiently funnelled to the centre of dark matter (DM) halos owing to higher large-scale densities~\citep{dekeletal09}, bringing gas into galaxies with lower angular momentum, higher surface densities, and, hence, more efficient star formation.
However, for this high-redshift Universe that is naturally more efficient at feeding intergalactic gas into structures, a significant amount of galactic-scale feedback has to regulate the gas budget. 
On the low-mass end, it is generally accepted that stellar feedback as a whole, and more likely feedback from supernovae (SNe), is able to efficiently drive large-scale galactic winds~\citep[e.g.][]{dekel&silk86, springel&hernquist03, dubois&teyssier08winds, dallavecchia&schaye08}, although the exact strength of that feedback, and hence, how much gas is driven in and out of galaxies is still largely debated and relies on several important physical assumptions~\citep[e.g.][]{hopkinsetal12, agertzetal13, kimmetal15, rosdahletal17, dashyan&dubois20}.
On the high-mass end, because of deeper potential wells, stellar feedback remains largely inefficient and gas regulation relies on the activity of central supermassive black holes~\citep[e.g.][]{silk&rees98,dimatteoetal05,crotonetal06,duboisetal10,duboisetal12,kavirajetal17,beckmannetal17}.

Low-mass and low surface-brightness regimes are becoming important frontiers for the study of galaxy evolution \citep[e.g.][]{Martin2019} as surveys such as the LSST will allow us to observe very faint structures such as tidal streams and, for the first time, thousands of dwarfs at cosmological distances (mostly at $z<0.5$). Complementary high-resolution cosmological simulations and deep observational datasets will enable us to start addressing the considerable tension between theory and observations in the dwarf regime \citep[e.g.][]{boylankolchinetal11,pontzen&governato12,naab&ostriker17,silk17,kavirajetal19,jacksonetal21a} as well as in the high-mass regime, where faint tidal features encode information that can aid in understanding the role of galaxy mergers and interactions in the formation, evolution, and survival of discs \citep{Jackson2020,Park2019} and spheroids \citep[][]{toomre&toomre72,bournaudetal07,naabetal09,Kaviraj14,duboisetal16,martinetal18}.

Owing to their modelling of the most relevant aspects of feedback, SNe
and supermassive black holes, which occur at the two mass ends
of galaxy evolution, respectively, and thanks to their large statistics, large-scale hydrodynamical cosmological simulations with box sizes of $\sim 50-300\,\rm Mpc$  have made a significant step towards a more complete understanding of the various mechanisms  (accretion, ejection, and mergers) involved in the formation and evolution of galaxies; these large-scale simulations include
Horizon-AGN~\citep{duboisetal14}, Illustris~\citep{vogelsbergeretal14}, EAGLE~\citep{schayeetal15},
IllustrisTNG~\citep{pillepichetal18}, SIMBA~\citep{daveetal19}, Extreme-Horizon~\citep{chabanieretal20},
and Horizon Run 5~\citep{leeetal21}.
However, as a result of their low spatial resolution in galaxies (typically of the order of 1 kpc), and therefore owing to their intrinsic inability to capture the multi-phase nature of the interstellar medium (ISM), their sub-grid models for star formation or the coupling of feedback to the gas has had to rely on cruder effective approaches than what a higher-resolution simulation might allow.
A couple of simulations with an intermediate volume and a better mass and spatial resolution stand out; these are  the TNG50 simulation~\citep{pillepichetal19} from the IllustrisTNG suite and the Romulus25 simulation~\citep{tremmeletal17}, which offer sub-kiloparsec resolution of 100 and 250 pc, respectively.

An important aspect of the evolution of galaxies is that rather than occurring in a homogeneous medium of diffuse interstellar gas, star formation proceeds within clustered molecular complexes; these range from pc to 100 pc in size and have properties that vary from one galaxy to another~\citep[e.g.][]{hughesetal13,sunetal18}.
This has several important consequences. A clumpier star formation affects the stellar distribution via a more efficient migration of stars; it can be locally efficient while globally inefficient, and it can also enhance the effect of stellar feedback by driving more concentrated input of energy.
Therefore, the necessity of capturing this minimal small-scale clustering of gas in galaxies has constrained numerical simulations to either rely on isolated set-ups (i.e. an isolated disc of gas and stars or isolated spherical collapsing halos; see e.g.~\citealp{dobbsetal11,bournaudetal14,semenovetal18}) or on zoomed-in cosmological simulations with a handful of objects~\citep[e.g.][]{ceverinoetal10,hopkinsetal14,hopkinsetal17,duboisetal15snbh,nunezetal20,agertzetal20}; this is because of the strong requisite on spatial resolution, that is typically below the 100 pc scale.
Since star formation occurs in molecular clouds that are gravitationally bound or marginally bound with respect to turbulence, a consistent theory of a gravo-turbulence-driven star formation efficiency can be built considering that this shapes the probability density function (PDF) of the gas density within the cloud~\citep[see e.g.][and references therein]{federrath&klessen12}. Such a theory can only be used in  simulations in which the largest-scale modes of the interstellar medium turbulence are captured~\citep{hopkinsetal14,kimmetal17,nunezetal20}.
Similarly, less ad hoc models for SN feedback can be used to accurately reproduce the distinct physical phases of the blown-out SN bubbles~\citep[the so-called Sedov and snowplough phases; e.g.][]{kimm&cen14}, depending on the exact location of these explosions in the multi-phase ISM.

Our approach in this new numerical hydrodynamical cosmological simulation called \nh, which we introduce in this work\footnote{See~\cite{Park2019,parketal21,volonterietal20,martinetal21,jacksonetal21a} and \cite{jacksonetal21b} for early results on the origin of discs and spheroids, the thickness of discs, the mergers of black holes,  the role of interactions in the evolution of dwarf galaxies, the DM deficient galaxies, and low-surface brightness dwarf galaxies, respectively.}, is to provide a complementary tool between these two standard techniques, that is between the few well-resolved objects versus a large ensemble of poorly resolved galaxies. The
\newh\ tool is designed to capture the basic features of the multi-scale, clumpy, ISM with a spatial resolution of the order of $34\,\rm pc$ in a large enough high-resolution, zoomed-in volume of $(16\,\rm Mpc)^3$. This is larger than a standard zoomed-in halo, has a standard cosmological mean density, and  is embedded in the initial lower-resolution $(142\,\rm Mpc)^3$ volume of the Horizon-AGN simulation~\citep{duboisetal12};  at $z=0.25$, the mass density in that zoom-in region is 1.2
times that of the cosmic background density.
Although still limited in terms of statistics over the entire range of galaxy masses (in particular galaxies in clusters are not captured), this volume offers sufficient enough statistics -- in an average density region -- to meaningfully study the evolution of galaxy properties at a resolution sufficient to apply more realistic models of star formation and feedback.

This paper introduces the \nh\ simulation with its underlying physical model and reviews the main fundamental properties of the simulated galaxies, including their mass budget, star formation rate (SFR), morphology, kinematics, and the mass and spin properties of the hosted black holes in galaxies.

The paper is organised as follows. 
Section~\ref{section:simulation} presents the numerical technique, resolution, and physical models implemented in \nh. 
Section~\ref{section:results} presents the various results of the properties of the galaxies in the simulation and their evolution over time.
Finally, we conclude in Section~\ref{section:conclusions}.

\section{The NewHorizon simulation: Prescription}
\label{section:simulation}

We describe the \nh\, simulation employed in this work\footnote{\href{http://new.horizon-simulation.org}{http://new.horizon-simulation.org}}, which is a sub-volume extracted from its parent \hagn\, simulation~\citep{duboisetal14}\footnote{\href{http://horizon-simulation.org}{http://horizon-simulation.org}}, and the procedure we use to identify halos and galaxies.
A number of physical sub-grid models have been substantially modified compared to the physics implemented in \hagn\, \citep[see e.g.][]{volonterietal16,kavirajetal17}, in particular regarding the models for star formation, feedback from SNe and from active galactic nuclei (AGN).
A comparison with simulated galaxies in \hagn\ within the same sub-volume will be the topic of a dedicated paper. 
Nonetheless, we describe the corresponding differences with \hagn\  at the end of each of the subsections of the sub-grid model.

\subsection{Initial conditions and resolution}
\label{section:ics}

The \nh\, simulation is a zoom-in simulation from the $142\,\rm Mpc$ size \hagn\ simulation~\citep{duboisetal14}.
The \hagn\ simulation initial conditions had $1024^3$ DM  particles, a $1024^3$ minimum grid resolution, and a $\Lambda$CDM cosmology.
The total matter density is $\Omega_{\rm m}=0.272$, dark energy density $\Omega_\Lambda=0.728$, amplitude of the matter power spectrum $\sigma_8=0.81$, baryon density $\Omega_{\rm b}=0.045$, Hubble constant $H_0=70.4 \, \rm km\,s^{-1}\,Mpc^{-1}$, and $n_s=0.967$ is compatible with the WMAP-7 data~\citep{komatsuetal11}. 
Within this large-scale box, we define an initial spherical patch of $10 \,\rm Mpc$ radius, which is large enough to sample multiple halos at a $4096^3$ effective resolution, that is with a DM mass resolution of $M_{\rm DM,hr}=1.2\times 10^6 \,\rm M_\odot$.
The high-resolution initial patch is embedded in buffered regions with decreasing mass resolution of $10^7\,\rm M_\odot$, $8\times 10^7\,\rm M_\odot$, $6\times 10^8\,\rm M_\odot$ for spheres of $10.6\,\rm Mpc$, $11.7\,\rm Mpc$, and $13.9\,\rm Mpc$ radius, respectively, and a resolution of $5\times 10^9\,\rm M_\odot$ in the rest of the simulated volume.
In order to follow the Lagrangian evolution of the initial patch, we fill this initial sub-volume with a passive colour variable with values of 1 inside and zero outside, and we only allow for refinement when this passive colour is above a value of $0.01$.
Within this coloured region, refinement is allowed in a quasi-Lagrangian manner down to a resolution of $\Delta x=34 \,\rm$ pc at $z=0$: refinement is triggered if the total mass in a cell becomes greater than eight times the initial mass resolution. The minimum cell size is kept roughly constant by adding an extra level of refinement every time the expansion factor is doubled (i.e. at $a_{\rm exp}=0.1,0.2,0.4$ and $0.8$);  the minimum cell size is thus between $\Delta x =27$ and 54 pc. We also added a super-Lagrangian refinement criterion to enforce the refinement of the mesh if a cell has a size shorter than one Jeans' length wherever the gas number density is larger than $5\, \rm H\, cm^{-3}$.

The \nh\, simulation is run with the adaptive mesh refinement {\sc ramses} code~\citep{teyssier02}.
Gas is evolved with a second-order Godunov scheme and the approximate Harten-Lax-Van Leer-Contact~\citep[HLLC][]{toro} Riemann solver with linear interpolation of the cell-centred quantities at cell interfaces using a minmod total variation diminishing scheme.
Time steps are sub-cycled on a level-by-level basis, that is each level of refinement has a time step that is twice as small as the coarser level of refinement, following a Courant-Friedrichs-Lewy condition with a Courant number of 0.8. 
The simulation was run down to $z=0.25$ using a total amount of 65 single core central processing unit (CPU) million hours.
The simulation contained typically 0.5-1 billion of leaf cells in total. With 30-100 millions of leaf cells per level of refinement in the zoom-in region from level 12 to level 22, the region had a total of $3.3\times 10^8$ star particles formed and completed $4.7\times 10^6$ fine time steps (the number of time steps of the maximum level of refinement), thus, corresponding to an average fine time step of size $\Delta t\simeq2.3\,\rm kyr$), by $z=0.25$. 

\begin{figure*}
\centering \includegraphics[width=\textwidth]{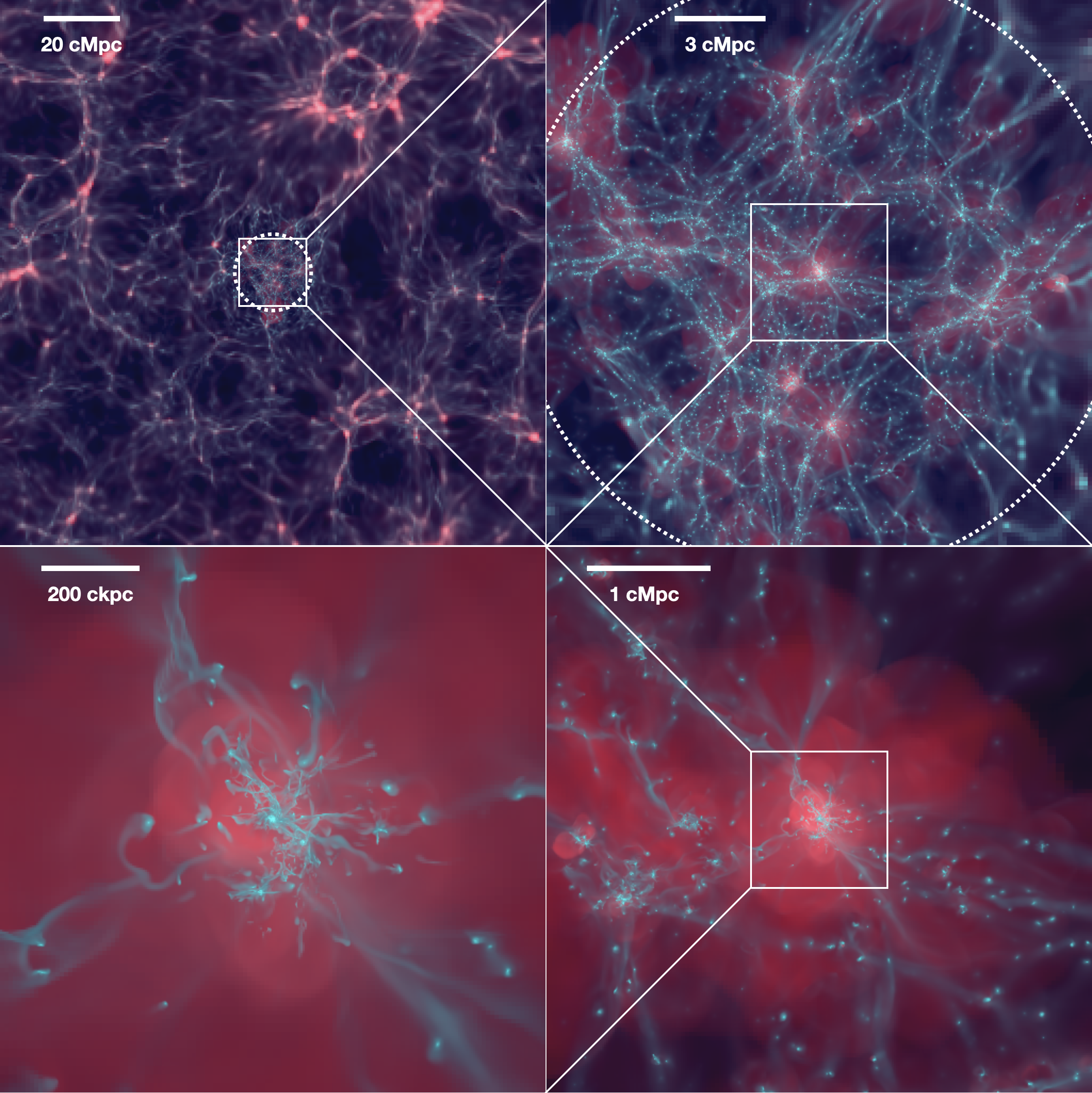}
\caption{Sequential zoom  (clockwise from top left) over the projected density (silver blue colours) and projected temperature (red) of the \nh\, simulation at redshift $z=2$. The dashed white circles encompass the initial high-resolution volume. Each panel is a zoomed-in version of the previous panel (identified by the white square in the previous panel) with the panel sizes of 142, 18, 4.4, and 1.1 comoving Mpc width, respectively. The two top panels encompass the zoom-in region, with its network of filaments. The two bottom panels illustrate how narrow filaments break up and mix once they connect to one of the most massive galaxies of that zoom-in region. }
\label{fig:keke}
\end{figure*}

\begin{figure}
\centering \includegraphics[width=0.5\textwidth]{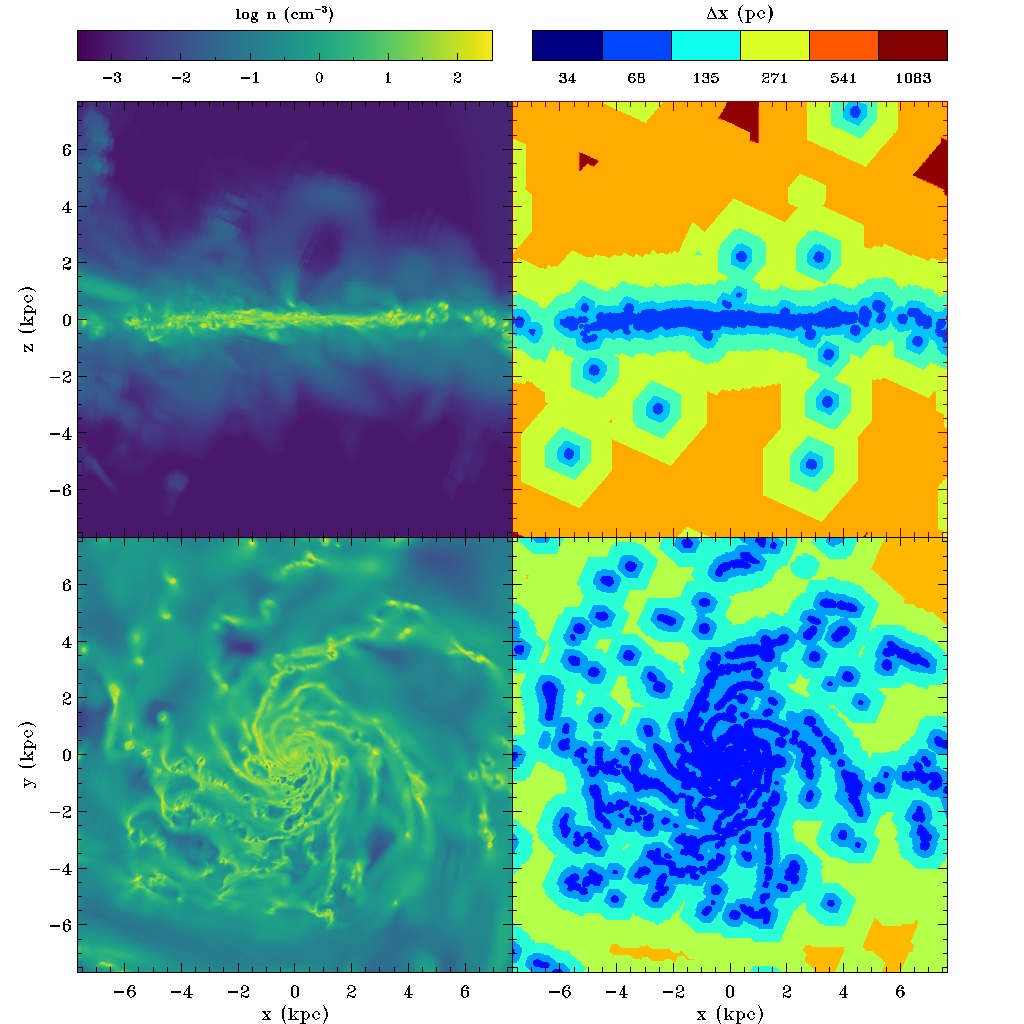}
\caption{Illustration of the structure of the gas density (left panels) and the corresponding spatial resolution (right panels) in a massive galaxy of $M_{\rm s}=6\times 10^{10}\,\rm M_\odot$ at $z=1$ seen edge-on (top panels) or face-on (bottom panels).
}
\label{fig:illustrate_nhres}
\end{figure}

Fig.~\ref{fig:keke} shows a projection of the high-resolution region.
Fig.~\ref{fig:illustrate_nhres} illustrates the typical structure of the gas density achieved in one of the massive galaxies at $z=1$ and the corresponding gas resolution. 
The diffuse ISM (0.1-1 cm$^{-3}$) is resolved with a $\sim$100 pc resolution or such, while the densest clouds reach the maximum level of refinement corresponding to 34 pc and the immediate galactic corona is resolved with cells of size 500 pc.
In terms of mass and spatial resolution, \nh\ is comparable to TNG50~\citep{pillepichetal19} ($4.5\times 10^5\,\rm M_\odot$ DM mass resolution and a spatial resolution in galaxies of 100 pc) or zoomed-in cosmological simulations (such as for the most massive galaxies of the FIRE-2 runs;~\citealp{hopkinsetal17}).

\subsection{Radiative cooling and heating}

We adopt the equilibrium chemistry model for primordial species (H and He) assuming collisional ionisation equilibrium in the presence of a homogeneous UV background. The primordial gas is allowed to cool down to $\approx 10^4\, \rm K$ through collisional ionisation, excitation, recombination, Bremsstrahlung, and Compton cooling. Metal-enriched gas can cool further down to $0.1\, \rm K$ using rates tabulated by \cite{sutherland&dopita93} above $\approx 10^4\,\rm K$ and those from~\cite{dalgarno&mccray72} below $\approx 10^4\,\rm K$.
The heating of the gas from a uniform UV background takes place after redshift $z_{\rm reion} = 10$, following~\cite{haardt&madau96}. Motivated by the radiation-hydrodynamic simulation results that the UV background is self-shielded in optically thick regions ($n_{\rm H}\ga 0.01 \,\rm H\, cm^{-3}$)  \citep{rosdahl&blaizot12}, we assume that UV photo-heating rates are reduced by a factor $\exp\left(-n_{\rm H}/n_{\rm shield}\right)$, where $n_{\rm shield}= 0.01 \,\rm H\, cm^{-3}$.

Compared to \hagn, the gas can now cool below $10^4\,\rm K$.

\subsection{Star formation}

Star formation occurs in regions with hydrogen gas number density above $n_0=10\, \rm H\, cm^{-3}$  (the stellar mass resolution is $n_0 m_{\rm p} \Delta x^3=1.3\times 10^4 \,\rm M_\odot$) following a Schmidt law: $\dot \rho_\star= \epsilon_\star {\rho_{\rm g} / t_{\rm ff}}$, where $\dot \rho_\star$ is the SFR mass density, $\rho_{\rm g}$ the gas mass density, $t_{\rm ff}=\sqrt{3\pi/(32G\rho_{\rm g})}$ the local free-fall time of the gas, $G$ the gravitational constant, and $\epsilon_\star$ ia varying star formation efficiency~\citep{kimmetal17,trebitschetal17,trebitschetal20}.

The current theory of star formation provides a framework for working out the efficiency of the star formation where  the gas density PDF is well approximated by a log-normal PDF~\citep{krumholz&mckee05,padoan&nordlund11, hennebelle&chabrier11,federrath&klessen12}.
This PDF is related to the star-forming cloud properties through the cloud turbulent Mach number $\mathcal{M}=u_{\rm rms}/c_{\rm s}$, where $u_{\rm rms}$ is the root mean square velocity, $c_{\rm s}$ the sound speed. The virial parameter $\alpha_{\rm vir}=2E_{\rm kin}/E_{\rm grav}$ and the efficiency is fully determined by integrating how much mass passes above a given density threshold using the multi-free fall approach of~\cite{hennebelle&chabrier11} as follows:
\begin{equation}
  \label{eq:sfr_ff}
  \epsilon_{\star} = \frac{\epsilon}{2\phi_{\rm t}} \exp\left(\frac{3}{8}\sigma_s^2\right)\left[1 + \mathrm{erf}\left(\frac{\sigma_s^2 - s_{\rm crit}}{\sqrt{2\sigma_s^2}}\right)\right],
\end{equation}
where $s=\ln(\rho/\rho_0)$ is the logarithmic density contrast of the PDF with mean $\rho_0$ and variance $\sigma_{\rm s}^2=\ln (1+b^2\mathcal{M}^2)$. In this expression
$b=0.4$ conveys the fractional amount of solenoidal to compressional modes of the turbulence. 
The critical density contrast $s_{\rm crit}$ is determined by~\cite{padoan&nordlund11}
as follows:\begin{equation}
  \label{eq:s_crit}
  s_{\rm crit} = \ln \left (\frac{0.067}{\theta^2} \alpha_{\rm vir}\mathcal{M}^2\right).
\end{equation}
In the \nh\ simulation, the turbulent Mach number is given by the local three-dimensional instantaneous velocity dispersion $\sigma_{\rm g}$ (obtained by computing $\sigma_{\rm g}^2={\rm sum} (\nabla \otimes u {\rm d}x)^2$), and the virial parameter also takes the thermal pressure support $\alpha_{\rm vir,0}=5(\sigma_{\rm g}^2+c_{\rm s}^2)/(\pi\rho_{\rm g}G\Delta x^2)$ into account.
In this case, $\phi_{\rm t}^{-1}=0.57$ and $\theta=0.33$ are empirical parameters of the model determined by the best-fit values between the theory and the numerical experiments~\citep{federrath&klessen12}.
The different values of $\phi_t^{-1}$ and $\theta$ we use compared to those given in ~\cite{federrath&klessen12} arise from the difference between the definition of $\alpha_{\rm vir}$ (measured over time, which are the values given in~\citealp{federrath&klessen12}) and $\alpha_{\rm vir,0}$ (the homogeneous cloud initial conditions). As our measurements of the virial parameter are meant to correspond to the initial cloud value $\alpha_{\rm vir,0}$, that is to the virial parameter of a spherical gas cloud with the same mass, radius, and thermo-turbulent velocity dispersion~\citep{bertoldi&mckee92,krumholz&mckee05} of the gas cell, we use the best-fit values from~\cite{federrath&klessen12} corresponding to this definition of the virial parameter (Fedderath, private communication).
We ignore the role of the magnetic field in this model despite the effect it has on the critical density and variance of the density PDF due to its large pressure with respect to the thermal pressure in the cold neutral medium~\citep[e.g.][]{heiles&troland05,crutcher12}. In Eq. (1)
$\epsilon=0.5$ is a proto-stellar feedback parameter that controls the actual amount of gas above $s_{\rm crit}$ that is able to form stars~\citep[typical estimates of $\epsilon$ are around $0.3-0.5$; see][]{matzner&mckee00,alvesetal07,andreetal10}. 

Such a star formation law shows a significantly different behaviour on galactic scales with respect to simulations with constant (usually low) efficiencies since the efficiency can now vary by orders of magnitude. For instance, for gravitationally bound ($\alpha_{\rm vir}<1$) and highly turbulent regions ($\mathcal{M}>1$), the efficiency can go well above 1, while regions that are marginally bound have an efficiency that quickly drops to very low values. 
Star formation efficiency, in conjunction with stellar feedback, plays a key role in shaping galaxy properties~\citep[e.g.][]{agertzetal11,nunezetal20}, and such potentially higher and more bursty star formation participates in driving stronger outflows and self-regulation of galaxy properties.
We note that our gravo-turbulent model of SF is somewhat reminiscent of those adopted in~\cite{hopkinsetal14,hopkinsetal17} or~\cite{semenovetal18}. In those models,  $\alpha_{\rm vir}$ is used as a criterion to trigger star formation (gas needs to be sufficiently bound), but star formation proceeds with a constant efficiency in contrast to our model.

Compared to \hagn, star formation in \nh\, occurs at above a hundred times larger gas density, and a varying gravo-turbulent-based star formation efficiency is used instead of assuming a constant 2 per cent efficiency.

\subsection{Feedback from massive stars}

We include feedback from Type II SNe assuming that each explosion initially releases the kinetic energy of $10^{51}\,{\rm erg}$. Because the minimum mass of a star particle is $10^4\,\rm M_\odot$, each particle is assumed to represent a simple stellar population with a Chabrier initial mass function (IMF) \citep{chabrier05} where the lower (upper) mass cut-off is taken as $M_{\rm low}=0.1$ ($M_{\rm upp}=150$) $\rm M_\odot$, respectively. We further assume that the minimum mass that explodes is $6 \,\rm M_\odot$ in order to include electron-capture SNe \citep[][see also \citealt{crain15}]{chiosi92}. The corresponding specific frequency of SN explosion is $0.015 \, \rm M_\odot^{-1}$. We increase this number by a factor of 2 ($0.03 \, \rm M_\odot^{-1}$) because multiple clustered SN explosions can increase the total radial momentum, with respect to the total momentum predicted by the accumulation of individual SNe~\citep{thorntonetal98}, by decreasing the ambient density into which subsequent SNe explode  \citep[][Na et al. {\sl in prep.}]{kimetal17,gentryetal19}.
Supernovae are assumed to explode instantaneously when a star particle becomes older than $5\,\rm Myr$. The mass loss fraction of a stellar particle from the explosions is 31\% and has a metal yield (mass ratio of the newly formed metals over the total ejecta) of 0.05.

We employ the mechanical SN feedback scheme \citep{kimm&cen14,kimmetal15}, which ensures the transfer of a correct amount of radial momentum to the surroundings.
Specifically, the model examines whether the blast wave is in the Sedov-Taylor energy-conserving or momentum-conserving phase \citep{chevalier74,cioffietal88,blondinetal98} by calculating the mass swept up by SN.
If the SN explosion is still in the energy-conserving phase, the assumed specific energy is injected into the gas since hydrodynamics naturally capture the expansion of the SN and imparts the correct amount of radial momentum. 
However, if the cooling length in the neighbouring regions is under-resolved owing to finite resolution, radiative cooling takes place rapidly, thereby suppressing the expansion of the SN bubble. This leads to an under-estimation of the radial momentum, hence weaker feedback. In order to avoid this artificial cooling, the mechanical feedback model directly imparts the radial momentum expected during the momentum-conserving phase if the mass of the neighbouring cell exceeds some critical value. This is done by first measuring the local ratio of the swept-up gas mass over the ejecta mass and examining whether the ratio is greater than the critical ratio corresponding to the energy-to-momentum phase transition. That is to say 
$70 \,E_{\rm 51}^{-2/17}\,n_{\rm 1}^{-4/17}\,Z'^{-0.28}$, where $E_{\rm 51}$ is the total energy released in units of $10^{51}\,\rm erg$, $n_1$ is the hydrogen number density in units of $\rm cm^{-3}$, and $Z'={\rm max}[Z/Z_\odot,0.01]$ is the metallicity,  normalised to the solar value ($Z_\odot=0.02$). The final momentum  in the snowplough phase per SN explosion is taken from \citet{thorntonetal98} as
\begin{equation}
q_{\rm SN}=3\times10^5 \, \rm{km\,s^{-1} M_\odot} \, E_{51}^{16/17} n_{\rm 1}^{-2/17} Z'^{-0.14}\, .
\end{equation}
We further assume that the UV radiation from the young OB stars over-pressurises the ambient medium near to young stars and increases the total momentum per SN to 
\begin{equation}
q_{\rm SN+PH}=5\times10^5 \, \rm{km\,s^{-1} M_\odot} \, E_{51}^{16/17} n_{\rm 1}^{-2/17} Z'^{-0.14} ,
\end{equation}
following~\cite{geenetal15}.

It is worth noting that the specific energy used for SN II explosion in this study is larger than previously assumed. A~\cite{chabrier03} IMF with a low- to high-mass cut-off of $M_{\rm low}=0.1$ and $M_{\rm upp}=100 \, \rm M_\odot$ and an intermediate-to-massive star transition mass at $M_{\rm IM}=8\, \rm M_\odot$ gives $e_{\rm SN}=1.1\times 10^{49}\,\rm erg\,M_\odot^{-1}$. However, $e_{\rm SN}$ can be increased up to $3.6\times 10^{49}\,\rm erg\, M_\odot^{-1}$ if a non-negligible fraction ($f_{\rm HN}=0.5$) of hypernovae~\citep[with $E_{\rm HN}\simeq 10^{52}\,\rm erg$ for stars more massive than $20\,\rm M_\odot$; e.g.][]{iwamotoetal98,nomotoetal06} is taken into account. This is necessary to reproduce the abundance of heavy elements, such as zinc~\citep{kobayashietal06}, or if a lower transition mass $M_{\rm IM}=6\, \rm M_\odot$ and a shallower (Salpeter) slope of $-2.1$ at the high-mass end~\cite[reflecting that early star formation should lead to a top-heavier IMF; e.g.][]{treuetal10,cappellarietal12,martin-navarroetal15} are assumed.
Furthermore, various sources of stellar feedback that would contribute to the overall formation of large-scale outflows including type Ia SNe, stellar winds, shock-accelerated cosmic rays~\citep[e.g.][]{uhligetal12,salem&bryan14,dashyan&dubois20}, multi-scattering of infrared photons with dust~\citep[e.g.][]{hopkinsetal11,roskaretal14,rosdahl&teyssier15}, or Lyman-$\alpha$ resonant line scattering~\citep{kimmetal18,smithetal17} are neglected.
In addition runaway OB stars~\citep{ceverino&klypin09,kimm&cen14,anderssonetal20} or the unresolved porosity of the medium~\citep{iffrig&hennebelle15} are also ignored. In this regard, the \nh\ simulation is unlikely to overestimate the effects of stellar feedback, as described in Section 3.

Unlike \hagn, feedback from stars in \nh\, only includes Type II SNe and ignores stellar winds and Type Ia SNe. In addition, \nh\ adopts a mechanical scheme for SNe instead of a kinetic solution \citep{dubois&teyssier08winds}. The assumed IMF is also changed from the Salpeter IMF to a Chabrier type, and thus the mass loss, energy, and yield are all increased.

\subsection{MBHs and AGN}

We now briefly describe the models corresponding to massive black hole (MBH) formation and their AGN feedback. 

\subsubsection{Formation, growth, and dynamics of MBH}

In \nh, MBHs are assumed to form in cells that have gas and stellar densities above the threshold for star formation, a stellar velocity dispersion larger than $20\,\rm km\,s^{-1}$, and that are located at a distance of at least 50 comoving kpc from any pre-existing MBH.

Once formed, the mass of MBHs grows at a rate $\dot M_{\rm MBH}=(1-\epsilon_{\rm r})\dot M_{\rm Bondi}$, where $\epsilon_r$ is the spin-dependent radiative efficiency (see Eq: \ref{eq:epsilonr_spin}) and $\dot M_{\rm Bondi}$ is the Bondi-Hoyle-Lyttleton rate, that is
\begin{equation}
\frac{dM_{\rm Bondi}}{dt}=4\pi \bar \rho \frac{(GM_{\rm MBH})^2}{(\bar u^2+\bar{c_{\rm s}}^2)^{3/2}}\, ,
\end{equation}
where $\bar u$ is the average MBH-to-gas relative velocity, $\bar{c_{\rm s}}$ the average gas sound speed, and $\bar \rho$ the average gas density. All average quantities are computed within  $4 \Delta x$ of the MBH, using mass weighting and a kernel weighting as specified in~\cite{duboisetal12}. We do not employ a boost factor in the formulation of the accretion rate, as is commonly done in cosmological simulations, because we have sufficient spatial resolution to model part of the multi-phase structure of the ISM of galaxies directly.

The Bondi-Hoyle-Lyttleton accretion rate is capped at the Eddington luminosity rate for the appropriate $\epsilon_{\rm r}$
\begin{equation}
\frac{dM_{\rm Edd}}{dt}=\frac{4\pi G M_{\rm MBH} m_{\rm p}}{\epsilon_{\rm r}\sigma_{T}c}\, ,
\end{equation}
where $\sigma_{T}$ is the Thompson cross-section, $m_p$ the proton mass, and $c$ the speed of light.

To avoid spurious motions of MBHs around high-density gas regions as a result of finite force resolution effects, we include an explicit drag force of the gas onto the MBH, following \cite{Ostriker1999}. This drag force term includes a boost factor with the functional form $\alpha=(n/n_0)^2$ when $n>n_0$, and $\alpha=1$ otherwise. The use of a sub-grid drag force model is justified by our larger-than-Bondi-radius spatial resolution~\citep{beckmannetal18}. We also enforce maximum  refinement within a region of radius $4\Delta x$ around the MBH, which improves the accuracy of MBH motions~\citep{lupietal15}. 

The MBHs are allowed to merge when they get closer than $4\Delta x$ ($\sim 150$~pc) and when the relative velocity of the pair is smaller than the escape velocity of the binary. A detailed analysis of MBH mergers in \nh\ is presented in~\cite{volonterietal20}.

\subsubsection{Spin evolution of MBH}

\begin{figure}
\centering \includegraphics[width=0.45\textwidth]{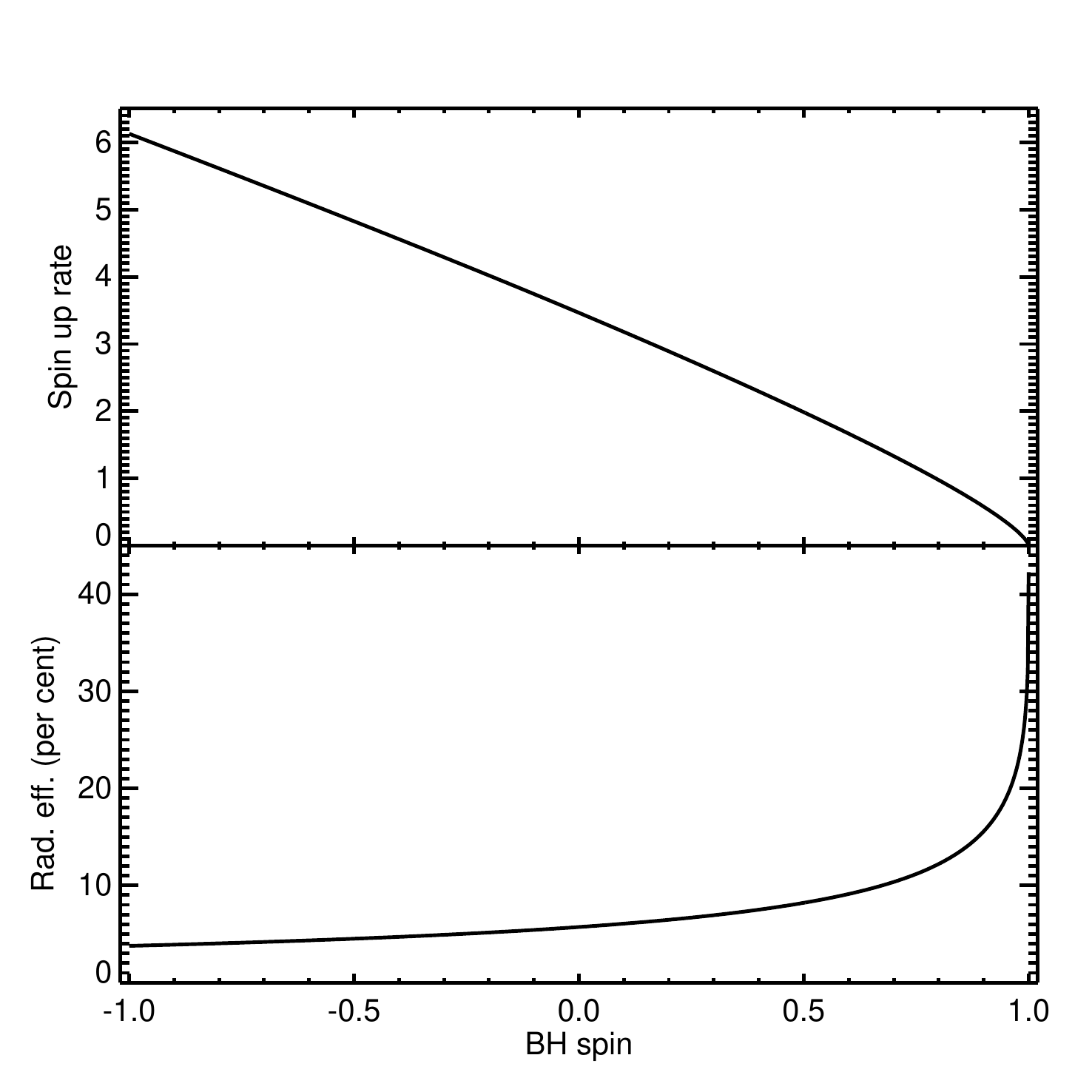}
\caption{Spin-up rate (top panel) and radiative efficiency $\epsilon_{\rm rad}/f_{\rm att}$ (bottom panel) as a function of the MBH spin for the thin disc solution~\citep{shakura&sunyaev73} applied to the quasar mode. At negative values of the MBH spin, the gas accreted from the thin accretion disc decreases the MBH spin, while for a positive MBH spin, the gas increases. For the thin disc solution, the radiative efficiency is an increasing function of the MBH spin with a sharp increase (by 4) between a MBH spin of 0.7 and 0.998.}
\label{fig:thindisc}
\end{figure}

\begin{figure}
\centering \includegraphics[width=0.45\textwidth]{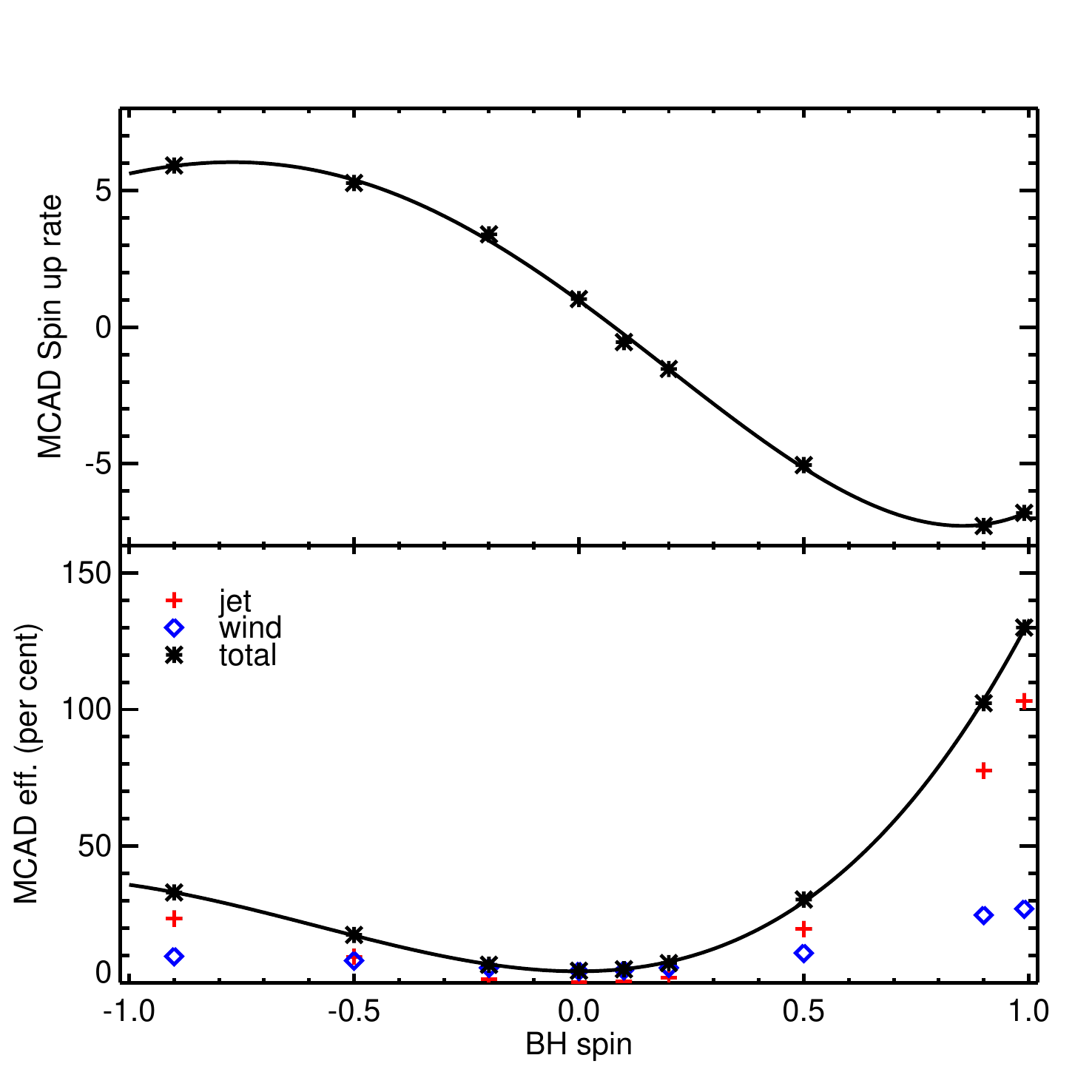}
\caption{Spin-up rate (top panel), jet (red plus signs), wind (blue diamonds), and total (black) efficiencies (bottom panel) as a function of the MBH spin for the MCAD solution applied to the radio mode. The symbols represent the results from the simulations of~\citet{mckinneyetal12} and the solid lines indicate the interpolated functions used in the \nh\, simulation. As opposed to the thin disc solution (Fig.~\ref{fig:thindisc}), gas accreted from the thick accretion disc always decreases the MBH spin. The MCAD feedback efficiency is an increasing function of the absolute value of the BH spin with a minimum efficiency for non-rotating MBHs.}
\label{fig:madjet}
\end{figure}

The evolution of the spin parameter $a$ is followed on-the-fly in the simulation, taking the effects of gas accretion and MBH-MBH mergers into account.
The model of MBH spin evolution is introduced in~\cite{duboisetal14bhspin}, and technical details of the model are detailed in that paper. The only change is that  we now use a different MBH spin evolution model at low accretion rates: $\chi=\dot M_{\rm MBH}/\dot M_{\rm Edd}<\chi_{\rm trans}$, where $\chi_{\rm trans}=0.01$.
At high accretion rates ($\chi\ge \chi_{\rm trans}$), a thin accretion disc solution is assumed~\citep{shakura&sunyaev73}, as in \cite{duboisetal14bhspin}. The angular momentum direction of the accreted gas is used to decide whether the accreted gas feeds an aligned or misaligned Lense-Thirring disc precessing with the spin of the MBH~\citep{kingetal05}, thereby spinning the MBH up or down for  co-rotating and counter-rotating systems, respectively (see top panel of Fig.~\ref{fig:thindisc}).
At low accretion rates ($\chi< 0.01$), we assume that jets are powered by energy extraction from MBH rotation \citep{blandford&znajek77} and that the MBH spin magnitude can only decrease. The change in the spin magnitude $da/dM$ follows the results from~\cite{mckinneyetal12}, where we fitted a fourth-order polynomial to their sampled values; from their table 7, $s_H$ for A$a$N100 runs, where $a$ is the value of the MBH spin. The functional form of the spin evolution as a function of MBH spin  at low accretion rates is represented in the top panel of Fig.~\ref{fig:madjet}, where the dimensionless spin-up parameter $s\equiv d(a/M)/dt$ is shown, where if $s$ and $a$ have opposite signs the black hole spins down. 

In addition, MBH spins change in magnitude and direction during MBH-MBH coalescences, with the spin of the remnant depending on the spins of the two merging MBHs and the orbital angular momentum of the binary, following analytical expressions from~\cite{rezzollaetal08}.

The evolution of the spin parameter is a key component of the AGN feedback model because it controls the radiative efficiency of the accretion disc and the jet efficiency.
Therefore, the Eddington mass accretion rate, used to cap the total accretion rate, and the AGN feedback efficiency in the jet and thermal modes vary with  spin values.
The spin-dependent radiative efficiency (see bottom panel of Fig.~\ref{fig:thindisc}) is defined as 
\begin{equation}
    \epsilon_{\rm r}=f_{\rm att}\left(1-e_{\rm isco} \right)=f_{\rm att}\left(1-\sqrt{1-2/(3r_{\rm isco})}\right)
\label{eq:epsilonr_spin}
,\end{equation}
where $e_{\rm isco}$ is the energy per unit rest mass energy of the innermost stable circular orbit (ISCO), $r_{\rm isco}=R_{\rm isco}/R_{\rm g}$ is the radius of the ISCO in reduced units, and $R_{\rm g}$ is half the Schwarzschild radius of the MBH. The parameter $R_{\rm isco}$ depends on spin $a$.
For the radio mode, the radiative efficiency used in the effective growth of the MBH is attenuated by a factor $f_{\rm att}=\min(\chi/\chi_{\rm trans},1)$  following~\cite{benson&babul09}. 
The MBH seeds are initialised with a zero spin value and a maximum value of the BH spin at $a_{\rm max}=0.998$~\citep[due to the emitted photons by the accretion disc captured by the MBH;][]{thorne74} is imposed.

\subsubsection{Radio and quasar modes of AGN feedback}

Active galactic nuclei feedback is modelled in two different ways depending on the Eddington rate~\citep{duboisetal12}: below $\chi< \chi_{\rm trans}$ the MBH powers jets (a.k.a. radio mode) continuously releasing mass, momentum, and total energy into the gas~\citep{duboisetal10}, while above $\chi\ge \chi_{\rm trans}$ the MBH releases only thermal energy back into the gas (a.k.a. quasar mode,~\citealp{teyssieretal11}).
The AGN releases a power that is a fraction of the rest-mass accreted luminosity onto the MBH, $L_{\rm AGN, R,Q}=\eta_{\rm R,Q}\dot M_{\rm MBH}c^2$, where the subscripts R and Q stand for the radio jet mode and quasar heating mode, respectively.

For the jet mode of AGN feedback, the efficiency $\eta_{\rm R}$ is not a free parameter. This value scales with the MBH spin, following the results from magnetically chocked accretion discs (MCAD) of~\cite{mckinneyetal12}, where we fitted a fourth-order polynomial to the sampled values of jet plus wind efficiencies of this work (from their table 5, $\eta_j$ plus $\eta_{w,0}$ for runs A$a$N100). This fit is shown in the bottom panel of Fig.~\ref{fig:madjet}.
When active in our simulation, the bipolar AGN jet deposits mass, momentum, and total energy within a cylinder of size $\Delta x$ in radius and semi-height, centred on the MBH, whose axis is (anti)aligned with the MBH spin axis (zero opening angle). Jets are launched with a speed of  $10^4\,\rm km\, s^{-1}$, whose exact value has little impact on MBH growth or galaxy mass content~\citep{duboisetal12}.

The quasar mode of AGN feedback deposits internal energy into its surrounding within a sphere of radius $\Delta x$, within which the specific energy is uniformly deposited (uniform temperature increase). Because only a fraction of the AGN-driven wind is expected to thermalise and only some of the multiwavelength radiation emitted from the accretion disc couples to the gas on ISM scales~\citep{bierietal16}, we scale the  feedback efficiency in quasar mode by a coupling factor of $\eta_c=0.15$, which is calibrated on the local $M_{\rm MBH}$-$M_{\rm s}$ in lower resolution ($\sim$kpc) simulations~\citep{duboisetal12}. The effective feedback efficiency in quasar mode is therefore $\eta_{\rm Q}=\epsilon_r\eta_c$.

Compared to \hagn, \nh\, now includes MBH spin evolution, which affects several compartments of MBH mass growth and feedback. The MBH accretion is changed owing to the spin-dependent radiative efficiency, thereby changing the maximum Eddington accretion rate. The AGN feedback is also changed by the spin-dependent radiative efficiency in the quasar mode. For the radio mode, the jet closely follows the spin-dependent mechanical efficiency of the MCAD model instead of a constant efficiency of 1, and the jet direction is now along the BH spin axis instead of along the accreted gas angular momentum.

\subsection{Identification of halos and galaxies}
\label{section:catalogs}

Halos are identified with the AdaptaHOP halo finder~\citep{aubertetal04}.
The density field used in AdaptaHOP is smoothed over 20 particles. The minimum number of particles in a halo is 100 DM particles. We only consider halos with an average overdensity with respect to the critical density $\rho_{\rm c}$, which is larger than $\delta_{\rm t}=80$ and which overcomes the Poissonian noise filtering density threshold at $(1+5/\sqrt{N})\delta_{\rm t}\rho_{\rm c}$ \citep[where $N$ is the number of particles in the (sub)structure; see][for details]{aubertetal04}.
For a substructure, it is only kept if the maximum density is 2.5 times its mean density.
The centre of the halo  is recursively determined by seeking the centre of mass in a shrinking sphere,  while decreasing its radius by 10 per cent recurrently down  to a minimum radius of 0.5 kpc~\citep{poweretal03}.
The maximum DM density in that radius is defined as the centre of the halo.
The shrinking sphere approach is used since strong feedback processes can significantly flatten the central DM density and smaller, but denser, substructures can be misidentified as being the centre of the main halo.

We run the same identification technique, using either AdaptaHOP or HOP, on stars  to identify the galaxies in the simulation, except that we only consider galaxies with more than 50 star particles and a value of $\delta_{\rm t}$ twice as large.
The AdaptaHOP tool separates substructures that include  in situ star-forming clumps as well as satellites already connected to a galaxy, while HOP keeps all substructures connected to the main structure (i.e. it does not detect substructures).
Appendix~\ref{appendix:size_hop_vs_ahop} shows examples of how using HOP or AdaptaHOP affect the segmentation of galaxies.
Both tools can be employed depending on context, as indicated in the corresponding text.
For the centring of the galaxies at the low-mass end, particular attention has to be taken, since  these galaxies tend to be extremely turbulent structures where bulges cannot be easily identified.

Since the \nh\, simulation is a zoom simulation embedded in a larger cosmological volume filled with lower DM resolution particles, we also need to remove halos of the zoom regions polluted with low-resolution DM particles.
To that end, we only consider halos as well as the embedded galaxies and MBHs encompassed in their virial radius, which are found devoid of low-resolution DM particles up to some threshold (see Appendix~\ref{appendix:halopurity} for the halo mass function for different purity levels).
With 100 \% purity,  there are, respectively, 626, 245, 53, and 5 main galaxies (which are not substructures in the sense of AdaptaHOP) at $z=2$ with stellar mass above $10^7$, $10^8$, $10^9$, and $10^{10}\,\rm M_\odot$; 403, 191, 70, and 12 at $z=1$; and 276, 145, 58, and 16 at $z=0.25$.
For comparison, considering a contamination lower than 1 per cent in number of DM, the number of galaxies typically doubles at $z=0.25$ (see Table~\ref{table:1} for detailed numbers). 
We note that the most massive unpolluted halo obtained at $z=0.25$ has a DM virial mass of $8\times 10^{12}\,\rm M_\odot$.

\begin{table}
\caption{Number of galaxies for different stellar mass thresholds. Purity is indicated as a threshold in the percentage of high resolution DM particles of the host halo (in number of DM particles). This work employs the 100$\%$ purity sample by default except when indicated.}
\label{table:1}
\centering
\begin{tabular}{c c c c c c}
\hline\hline
Purity & Redshift & $M_{\rm s}>$ &$M_{\rm s}>$&$M_{\rm s}>$&$M_{\rm s}>$ \\
($\%$) &          & $10^7\,\rm M_\odot$ &$10^8\,\rm M_\odot$&$10^9\,\rm M_\odot$&$10^{10}\,\rm M_\odot$ \\
\hline
100 & 4 & 688 & 148 & 12 & 0 \\
99.9 & 4 & 697 & 152 & 12 & 0 \\
99 & 4 & 722 & 157 & 12 & 0 \\
\hline
100 & 2 & 626 & 245 & 53 & 5 \\
99.9 & 2 & 884 & 342 & 75 & 5 \\
99 & 2 & 931 & 364 & 84 & 7 \\
\hline
100 & 1 & 403 & 191 & 70 & 12 \\
99.9 & 1 & 649 & 310 & 112 & 18 \\
99 & 1 & 732 & 362 & 132 & 23 \\
\hline
100 & 0.25 & 276 & 145 & 58 & 16 \\
99.9 & 0.25 & 443 & 238 & 99 & 28 \\
99 & 0.25 & 531 & 285 & 121 & 32 \\
\hline
\end{tabular}
\end{table}

\section{Cosmic evolution of baryons }
\label{section:results}

In this section we present several standard properties of the simulated galaxies including their stellar and gas mass content, SFR, morphological and structural properties, and  kinematics. We also present their hosted MBHs and compare these to observational relations down to the lowest redshift reached out by the simulation ($z=0.25$).

\begin{figure*}
\centering \includegraphics[width=\textwidth]{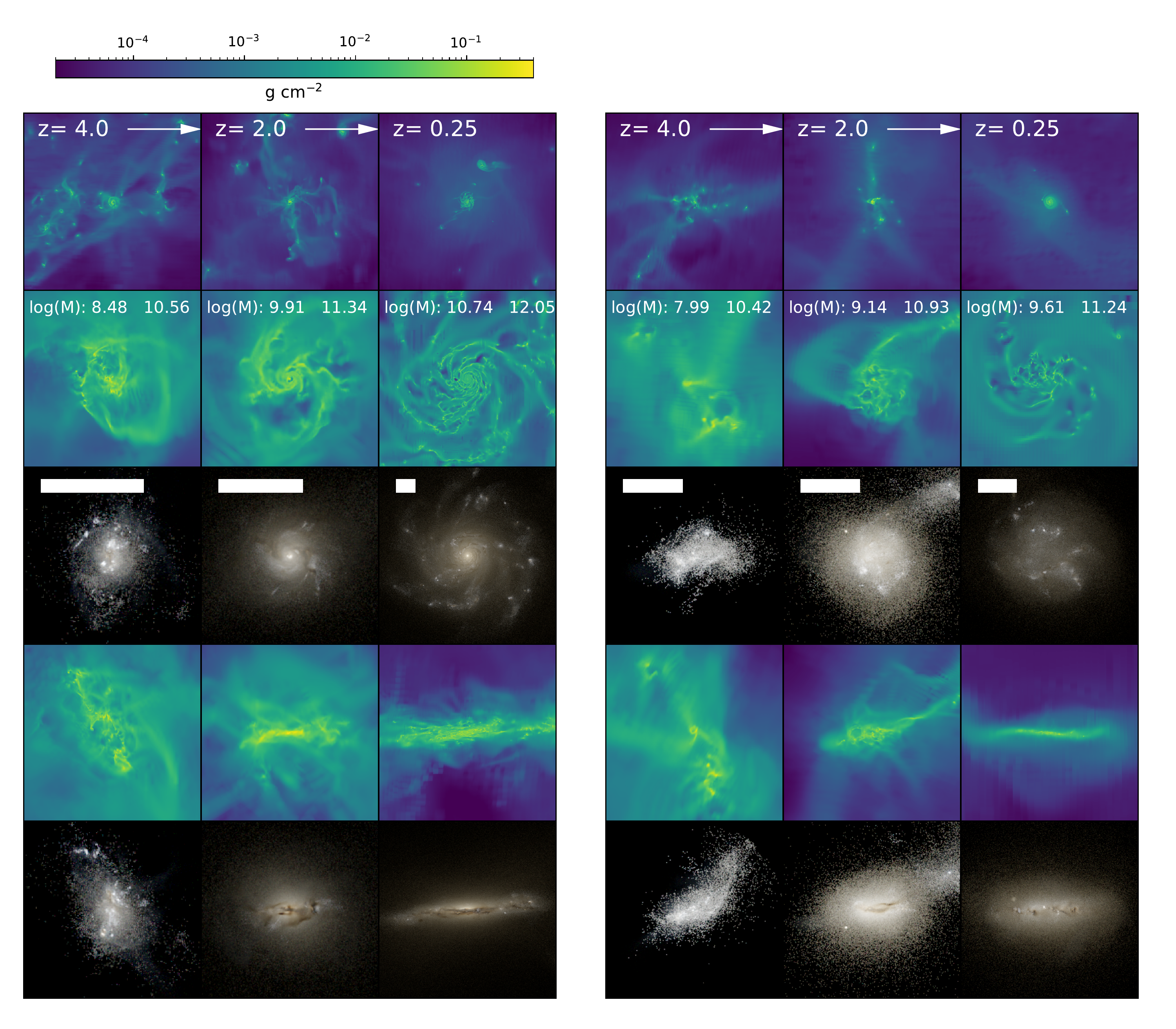}
\caption{Projection of the gas density and mock observations. Panels on the left are images of a massive galaxy at different epochs and the panels on the right are for a less massive galaxy. The first row shows gas density projections a of $1.4 \,\rm Mpc$ at different epochs, while the second and fourth rows are zoomed-in gas density projections with face-on and edge-on views of the galaxy, respectively. Third and fifth rows are SKIRT mock observations in face-on and edge-on direction. Stellar mass and halo mass of each galaxy at each epoch are given in the second row in log scale. For a galaxy at a given epoch, the second to fifth panels are on the same scale and the white bar in third row indicates $5 \,\rm kpc$. Gas density maps share the same colour scheme as given in the colour bar.}
\label{fig:galim_off}
\end{figure*}

\subsection{Synthetic galaxy morphology}

In order to qualitatively illustrate the variety of galaxy properties simulated in \nh, we show in Fig.~\ref{fig:galim_off} a couple of galaxies at $z=4$, $z=2$ and $z=0.25$ with their gas density and stellar emission.
The 15 panels on the left show the images of a massive galaxy (stellar mass $M_{\rm s} = 3.0 \times 10^{8} \,\rm M_\odot$ at $z=4$, $8.2 \times 10^{9} \,\rm M_\odot$ at $z=2$ and $5.5 \times 10^{10} \,\rm M_\odot$ at $z=0.25$) and the 15 panels on the right represent a less massive galaxy ($9.7 \times 10^{7} \,\rm M_\odot$ at $z=4$, $1.4 \times 10^{9}  \,\rm M_\odot$ at $z=2$ and $4.0 \times 10^{9}  \,\rm M_\odot$ at $z=0.25$). 
While the first, second, and fourth rows show their gas density maps, the third and fifth rows show the mock images; the second and third rows are shown with a face-on view (with respect to the stellar angular momentum of the galaxy) and the fourth and fifth rows an edge-on view. The mock images are in SDSS g-r-i bands and are generated via the SKIRT9 code \citep{Camps2020A&C....3100381C}, which computes radiative transfer effects based on the properties and positions of stars and the dusty gas assuming a dust fraction $f_{\rm dust} = 0.4$ following \citet{Saftly2015}.
The high resolution of \nh \ (34 pc) reveals the detailed structure of the cosmologically simulated galaxies, and it is clearly evident that star formation (highlighted by the young blue region in the stellar maps) proceeds in clustered regions of dense gas.
The massive galaxy settles its disc around $z \approx 2.5$ and appears as a regular disc galaxy with well-defined spiral arms and a central bulge if witnessed at $z=0.25$. We used the visual inspection as well as $(V/\sigma)_{\rm gas}>3$ \citep{Kassin2012TheNow} for disc settling criteria; the calculation of the kinematics is detailed in Section~\ref{section:gaskin}.
The less massive galaxy, on the other hand, exhibits an extremely irregular morphology at $z=2$ with strong asymmetries in both gas and stars, and prominent off-centred (blue) star-forming clusters.
This low-mass galaxy, which only grows moderately  by $z=0.25$, develops a galactic-scale disc at $z \approx 1.0$ and maintains the marginally stable disc for the rest of the cosmic history. 

\subsection{Galaxy mass function}

We compare the $z=0.25$ mass function obtained from \nh\ with the mass function obtained from an equivalent volume in the HSC-SSP survey \citep{Aihara2019}. In order to do this we take 100 random pointings from the HSC-SSP deep layer (encompassing the SXDS, COSMOS, ELIAS, and DEEP-2 fields), where each pointing has an equivalent volume to the \nh\ box. The central redshift of each volume is varied by up to 0.02 around a central redshift of $z=0.25$ for each random pointing. Since the photometric redshift errors are typically larger than the 20~Mpc box length, it is likely that we do not capture the full variance in the mass function since cosmic variance would be underestimated along the radial axis.

To infer the stellar masses of the HSC-SSP sample, we use the spectral energy distribution (SED) fitting code {\sc{LePhare}} \citep{Arnouts1999,Ilbert2006,Arnouts2011} with the \cite{bruzual&charlot03} (BC03 here and after) templates to estimate galaxy stellar masses from the $g$, $r$, $i,$ and $z$ cModel magnitudes. We then use the luminosity function tool {\sc{alf}} \citep{Ilbert2005} to construct galaxy stellar mass functions for each pointing using the method of \citet{Sandage1979}. Galaxies are selected in the $r$ band with an apparent magnitude cut of 26. We first constrain the knee of the mass function ($M_{\rm s}$) by computing the mass function for each pointing in a larger redshift slice ($0.1<z<0.4$) before re-fitting the mass function with $M_{\rm s}$ fixed for the smaller volume. For the simulated sample we follow a similar procedure, first obtaining dust-attenuated $g$, $r$, $i,$ and $z$ magnitudes for galaxies identified with HOP using \textsc{Sunset} \citep[see][section 2.2.1]{martinetal21}. To approximate the selection effects present in real data, we select galaxies by their effective surface brightness, where the probability of selecting a galaxy is proportional to the surface brightness completeness of the HSC survey; this value is estimated by assuming that the true number of objects continues to rise exponentially as a function of effective surface brightness after the turnover in the number of galaxies observed by HSC. We again use {\sc{LePhare}} and {\sc{alf}} to construct the galaxy stellar mass function using the same 26 mag cut in the $r$ band. Because of the more limited volume of \nh, the number of galaxies that are considerably more massive than the knee of the mass function is too small to effectively constrain this value, thereby leading to unrealistic fits. We therefore fix $M_{\rm s}$ at a value of $10^{10.8}\,\rm M_{\odot}$, which is calculated from the full volume of the \hagn\ simulation. While varying $M_{\rm s}$ also necessarily affects the slope at the low-mass end, this is not significant enough to qualitatively alter our comparison to the observed mass functions within a reasonable range of values (e.g. $10^{10.6}$M$_{\odot}$ to $10^{11}$M$_{\odot}$).

The galaxy mass function, which is a volume-integrated quantity poses a conceptual challenge to a zoom-in simulation. 
Indeed, galaxies within halos polluted with low-resolution DM particles continue to form stars, and it is questionable whether or not their contribution to the overall cosmic star formation should be taken into account. In addition, we have to determine the actual corresponding volume of the zoom-in region, which can expand or contract over time.
For the volume entering the calculation of the galaxy mass function (and in other volume-integrated quantities measured in this work), we take the entire initial volume of the zoom-in region of the simulation, hence, $(16 \,\rm Mpc)^3$. We could alternatively use the sum of each individual leaf cell that passes a given threshold value of the passive scalar colour value (see Section~\ref{section:ics}). The corresponding initial volume can be reduced by 20-40 per cent for a threshold value of resp. 0.1- 0.9, depending on redshift. 
We decided to simplify the problem by taking the initial zoom-in volume, but we note that the presented volume-integrated quantities are only a lower limit and can be a few tens of per cent higher.

\begin{figure}
\centering \includegraphics[width=0.49\textwidth]{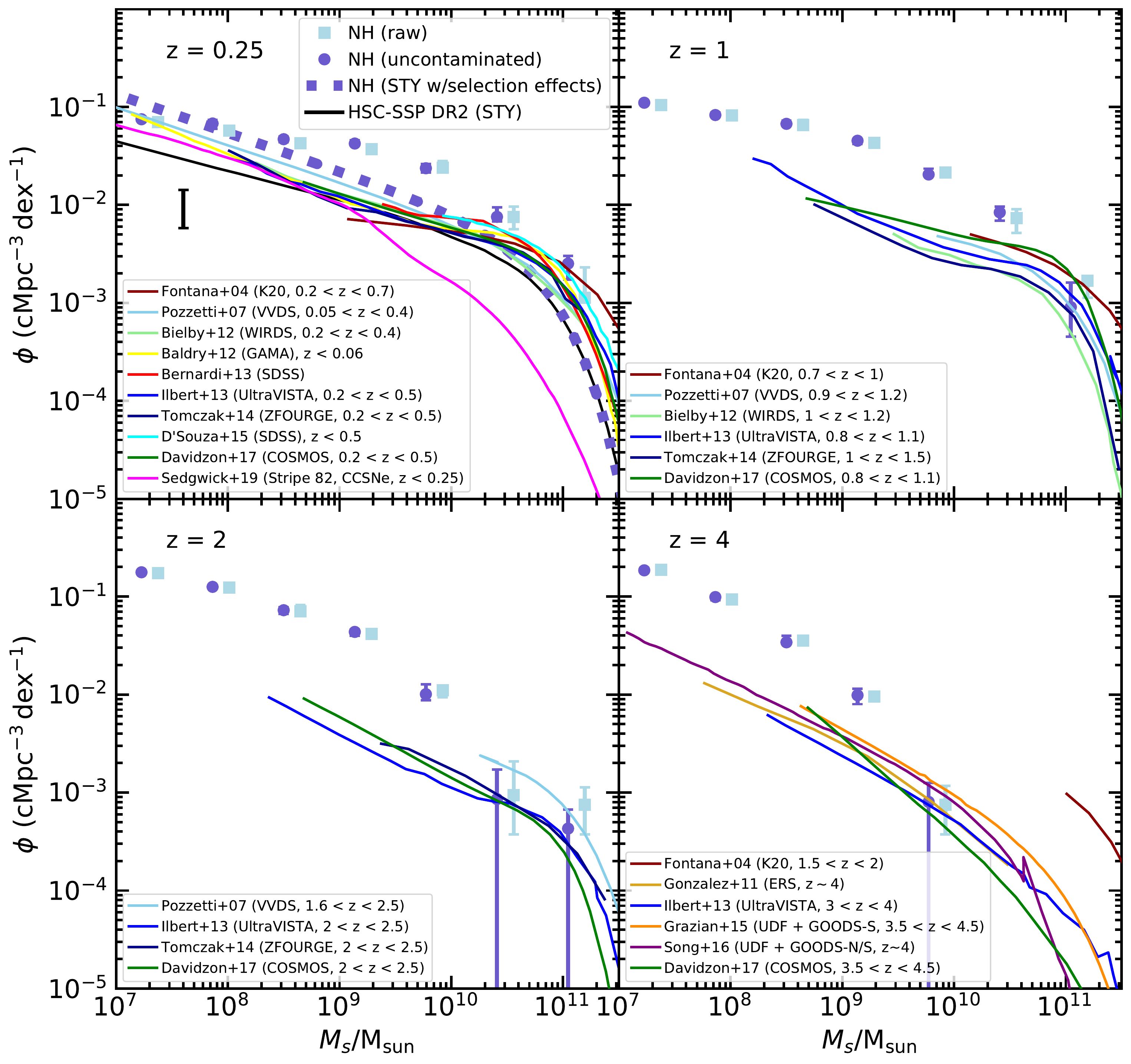}
\caption{Galaxy stellar mass function at $z=0.25$ in \nh\ and the HSC-SSP survey. The light blue squares and dark blue circles with error bars indicate the \nh\ stellar mass function for all galaxies (offset by 0.15 dex) and only uncontaminated galaxies (with a volume correction applied). The black line indicates the median galaxy stellar mass function from 100 random pointings from a volume of the HSC-SSP deep layer with the same volume as \nh\ and the black error bar indicates the 90 percentile scatter in the mass function at the low-mass end.  A comparison is made with additional mass functions from literature \citep[][]{Sedgwick2019b,Davidzon2017,tomczaketal16,Song2016,Grazian2015,DSouza2015,Fontana2014,Bernardi2013,Ilbert2013,Baldry2012,Bielby2012,Gonzalez2011,Pozzetti2007}, which are shown as thin coloured lines. The thick purple dotted line indicates the \nh\ mass function constructed using the \citet{Sandage1979} method (STY) and including selection effects.}
\label{fig:mass_fn}
\end{figure}

Figure \ref{fig:mass_fn} shows the galaxy stellar mass function from \nh, HSC-SSP, and from the literature \citep[][]{Sedgwick2019b,Davidzon2017,tomczaketal16,Song2016,Grazian2015,DSouza2015,Fontana2014,Bernardi2013,Ilbert2013,Baldry2012,Bielby2012,Gonzalez2011,Pozzetti2007}. We note that \citet{Sedgwick2019b} includes only star-forming galaxies.  The light blue squares with error bars represent the \nh\ stellar mass function with Poisson errors for all galaxies. The purple circles show the same, but include only galaxies whose halos are not contaminated by low-resolution particles from outside of the highest resolution zoom region -- a simple correction is made to account for the smaller effective volume by dividing the mass function by the fraction of uncontaminated galaxies. Additionally, the mass function (with selection effects) for \nh\ that is constructed using the \citet{Sandage1979} method (STY) is shown as a thick purple dotted line. The black line indicates the median galaxy stellar mass function from the 100 random pointings from the HSC-SSP deep layer. Various other mass functions from the literature are also indicated as thin coloured lines in each panel.

Once selection effects are included, the \nh\ mass function lies within the upper range of the observational mass functions shown. The discrepancy between the raw mass function likely emerges from incompleteness in the observed data at low surface brightness, meaning the observed mass function may be underestimated towards lower mass. The effect of selection effects and environment on the galaxy stellar mass function will be explored in more detail in Noakes-Kettel et al. ({\sl in prep.}).

The 90 per cent variance in the low-mass end of the HSC mass function is indicated by a black error bar. Over such a limited volume, the normalisation of the galaxy stellar mass function varies significantly. We note that our estimate of the variance may be an underestimate as redshift uncertainties are significantly larger than the selected volume. Additionally the location of the four HSC-SSP deep fields were chosen to enable certain science goals and not necessarily to sample representative volumes of the Universe as a whole.

\subsection{Surface brightness-to-galaxy mass relation}

\begin{figure}
\centering
\includegraphics[width=0.8\columnwidth]{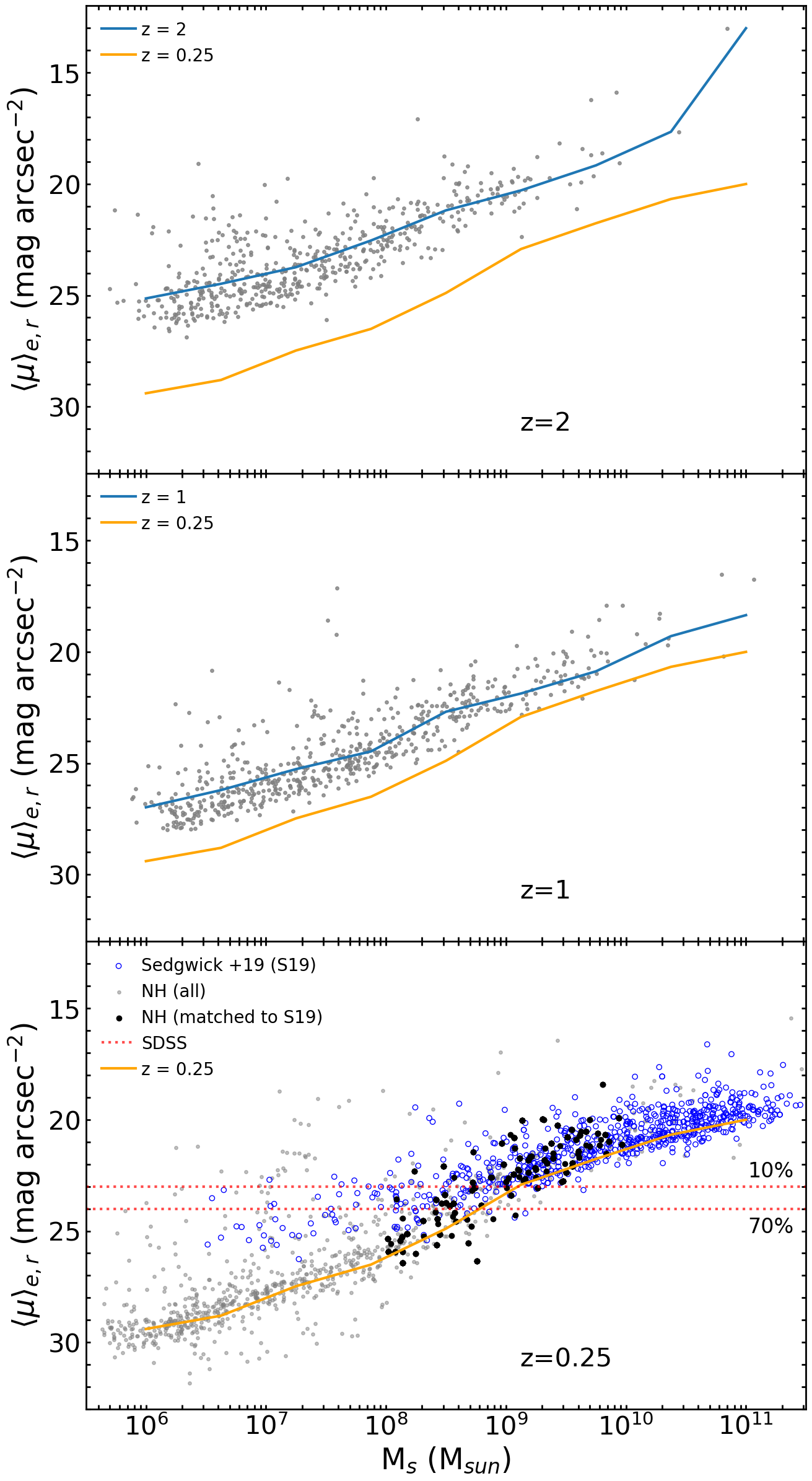}
\caption{Surface brightness vs. stellar mass in the NewHorizon simulation for 3 redshifts. The grey points indicate the entire galaxy population of NewHorizon in all 3 panels with the median lines for the redshift in question and z=0.25 shown in blue and orange, respectively. In the bottom panel the open blue points indicate galaxies from \citet{Sedgwick2019b} and black points are \nh\ galaxies. The predicted surface brightness vs. stellar mass plane in \nh\  corresponds well to that where the Sedgwick et al. galaxies are complete. The red dotted lines in the bottom panel indicate the 70$\%$ and 10$\%$ completeness limits from the SDSS \citep[see e.g. Table 1 in][]{Blanton2005}. The overwhelming majority of galaxies in the Universe lie below the surface brightness thresholds of surveys such as the SDSS; only those galaxies that depart strongly from the typical surface brightness vs. stellar mass relation are likely to be detectable in these datasets.}
\label{fig:locus_comparison}
\end{figure}

While many comparisons between theory and observation treat galaxies as one-dimensional points, the two-dimensional distribution of baryons (e.g. as summarised by the effective surface brightness and effective radius) are important points of comparison. Since high-resolution cosmological simulations now offer predictions of the distribution of baryons, it is worth comparing these predictions to observational data of galaxy surface brightnesses. 

Here, we obtain the dust attenuated surface brightness for each \nh\ galaxy using the intensity-weighted second order central moment of the stellar particle distribution \citep[e.g.][]{Bernstein2002} and we compare these values with \citet{Sedgwick2019b}.
 
For each star particle, we first obtain the full SED from a grid of dust attenuated BC03 simple stellar population models corresponding to the age and metallicity of the star particle. We redshift each BC03 template to match the overall redshift distribution of the observed sample to account for surface-brightness dimming. Since low-mass galaxies in this sample are biased towards lower redshift, we also account for this by drawing a redshift for each galaxy from the conditional probability distribution $P_{z}(M_{s})$ (i.e. the redshift probability distribution at a given stellar mass as found in \citealp{Sedgwick2019b}) so that they are redshifted and their flux and size are calculated according to a redshift whose probability is related to the stellar mass of the object. The SEDs are then convolved with the response curve for the SDSS $r$-band filter. We then weight by the particle mass to obtain the luminosity contribution of each star particle, and obtain the apparent $r$-band magnitude by converting the flux to a magnitude and adding the distance modulus and zero point.

The second moment ellipse is obtained by firstly constructing the covariance matrix of the intensity-weighted second order central moments (sometimes called the moment of inertia) for all the star particles, that is
\begin{equation}
    \mathrm{cov}[I(x,y)]=\begin{bmatrix}
        Ix^{2}&Ixy\\
        Ixy&Iy^{2}\\
    \end{bmatrix},
\end{equation}
where $I$ is the flux, and $x$ and $y$ are the projected positions from the barycentre in arcseconds of each star particle in the galaxy. The major ($\alpha=\sqrt{\lambda_{1}/\Sigma I}$) and minor ($\beta=\sqrt{\lambda_{2}/\Sigma I}$) axes of the ellipse are obtained from the covariance matrix, where $\lambda_{1}$ and $\lambda_{2}$ are its eigenvalues and $\Sigma I$ is the total flux. The scaling factor, $R$, scales the ellipse so that it contains half the total flux of the object. Finally, the mean surface brightness within the effective radius can be calculated from $\langle\mu\rangle_{e,r} = m - 2.5\mathrm{log_{10}}(2) + 2.5\mathrm{log_{10}}(A)$, where $A=R^{2} \alpha  \beta \pi$ and $m$ is the $r$-band apparent magnitude of the object. We repeat this process in multiple orientations ($xy$, $xz$ and $yz$), taking the mean surface brightness for each object. The method presented in this work is equivalent to the method employed in  \textsc{SExtractor} \citep[][]{bertin1996sextractor} to derive basic shape parameters, which \cite{Sedgwick2019b} use to derive their measurements.

In Fig.~\ref{fig:locus_comparison}, we show the evolution of the surface brightness $\langle\mu\rangle_{e,r}$ versus stellar mass plane in \nh. In the bottom panel we compare the predicted surface brightnesses to a recent work that uses the IAC Stripe 82 Legacy Survey project \citep{Sedgwick2019b,Sedgwick2019a}. This study is one of few that probes the surface brightnesses of galaxies down into the dwarf regime, which is only possible at low redshift using past and current surveys, which are typically very shallow. To probe galaxy surface brightness down to faint galaxies Sedgwick et al. introduce a novel technique with core-collapse SNe (CCSNe). Using custom settings in SExtractor \citep{bertin1996sextractor} they extract the host galaxies of these CCSNe, including those that are not detected in the IAC Stripe 82 Legacy survey. The resultant sample is free of incompleteness in surface brightness in the stellar mass range M$_{\rm s}$ > 10$^8$ M$_{\odot}$; a host is identified for all 707 CCSNe candidates at $z<0.2$. Given the high completeness of the sample at low surface brightness and the relative ease with which we can model the selection function and apply it to our simulated data, this dataset is an ideal choice to compare to the \nh\ data. More details on how the matching between the two datasets has been completed are available in \citealt{jacksonetal21a}.

Figure \ref{fig:locus_comparison} shows that the surface brightness versus stellar mass plane in \nh\ corresponds well to~\cite{Sedgwick2019b}, where the observational data is complete; we note that the simulation is not calibrated to reproduce galaxy surface brightness. The flattening seen in the observations is due to high levels of incompleteness at $M_{\rm s} < 10^8\,\rm M_\odot$. The prediction for the evolution of this plane to higher redshifts shows that \nh\ galaxies have increasing brightness at higher redshift for a fixed stellar mass (i.e. galaxies are more concentrated, see
Section~\ref{section:reff}) can be tested using data from future instruments such as the LSST.

\subsection{Star formation rates}

\begin{figure}
\centering \includegraphics[width=0.49\textwidth]{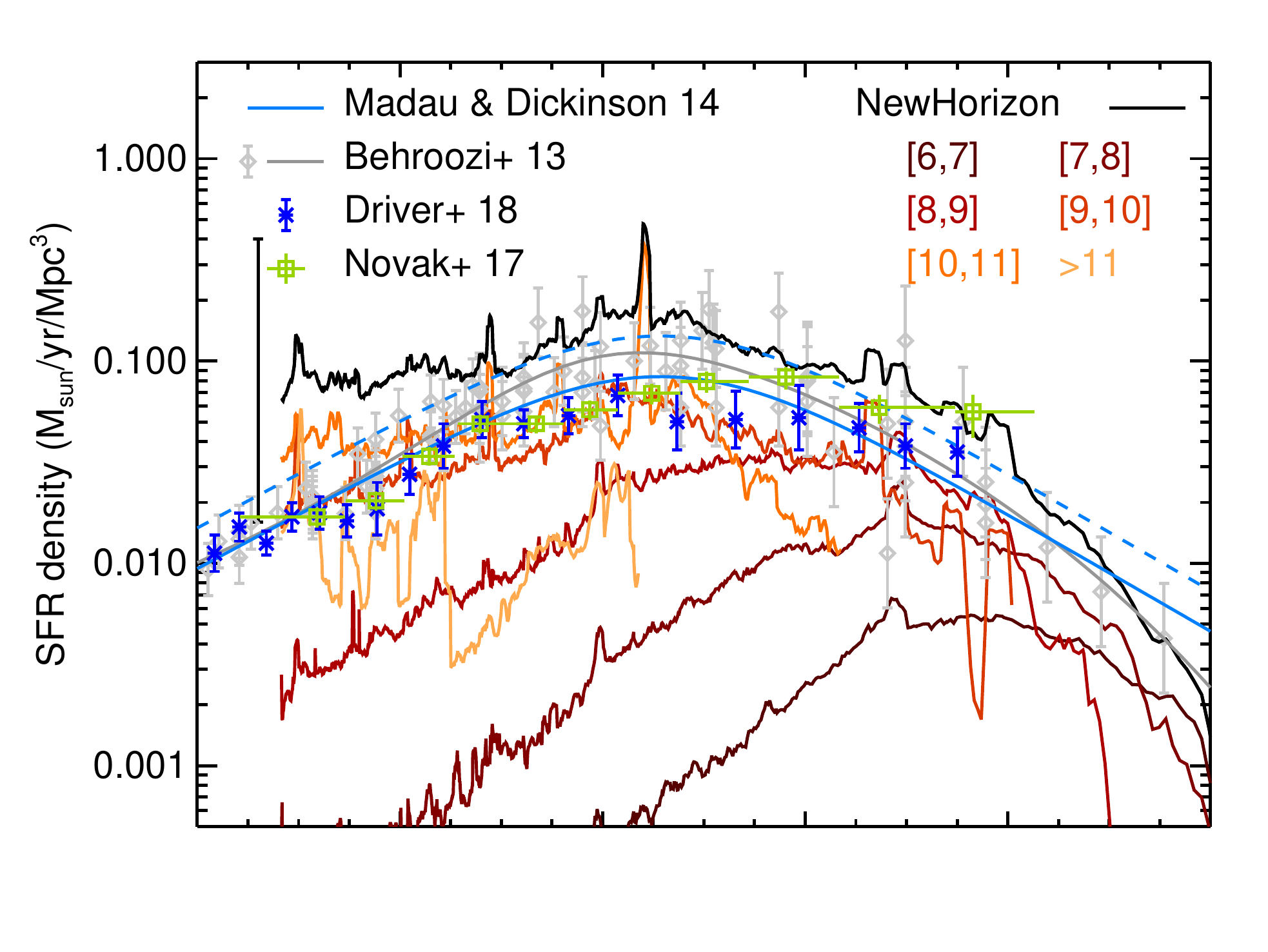}\vspace{-1.35cm}
\centering \includegraphics[width=0.49\textwidth]{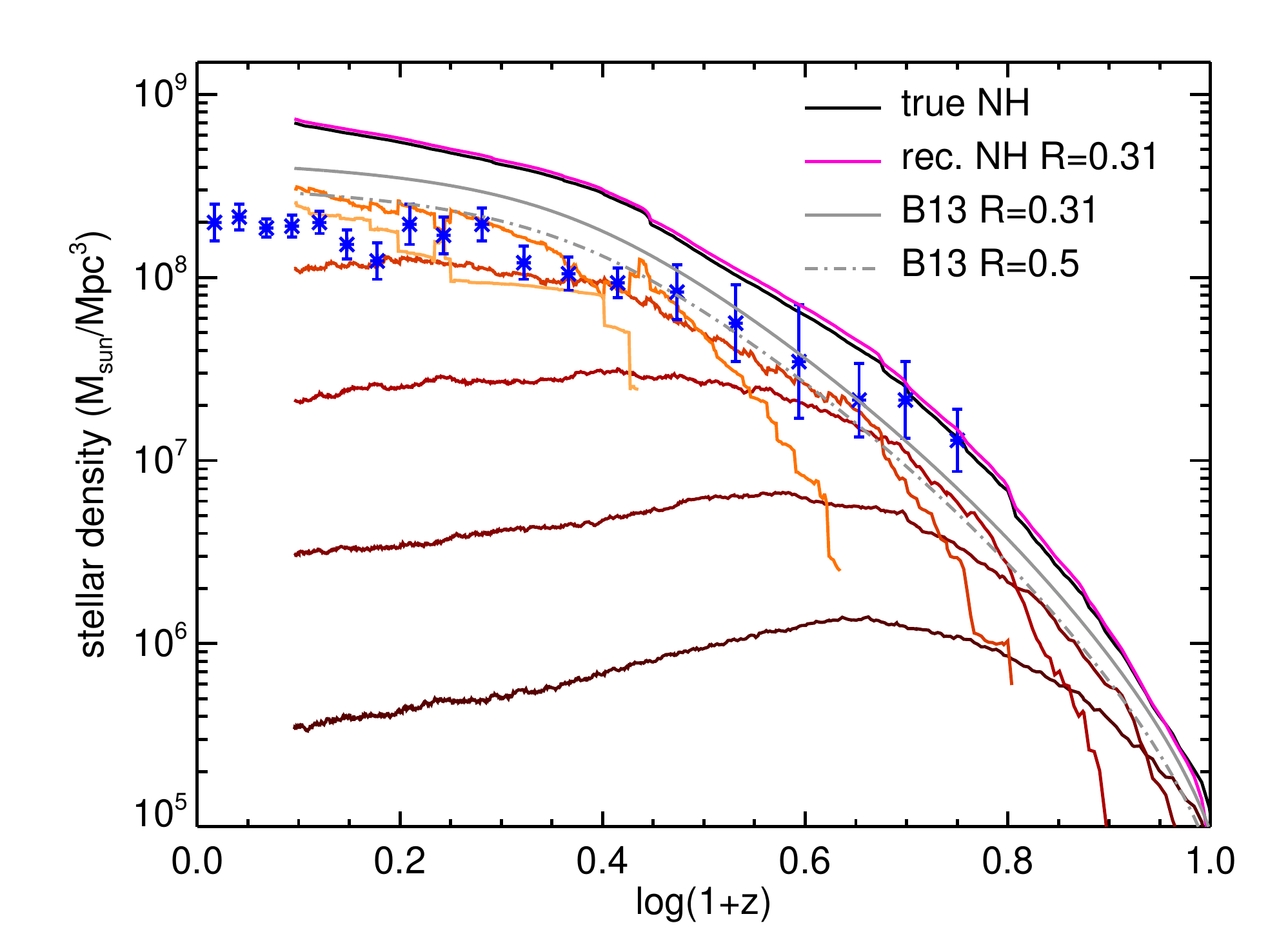}
\caption{Cosmic SFR density (top panel) and stellar density (bottom panel) as a function of redshift in \nh\ (solid black or red lines) compared to observations~(\citealp[][B13]{behroozietal13},~\citealp{madau&dickinson14},~\citealp{novaketal17},~\citealp{driveretal18}) as indicated in the panels. All observational quantities are shown for a~\cite{chabrier05} IMF except for the dashed blue line, which indicates the~\cite{madau&dickinson14} fit for a Salpeter IMF as originally assumed in their analysis. For the stellar density, the reconstructed result from the SFR density for \nh\ (magenta line) and for the fit from B13 (grey lines) using two different return fractions $R$ are also shown.
The red coloured lines indicate the cosmic SFR densities of the different galaxy stellar mass bins as indicated in the panel in $\log \rm M_\odot$ units. 
The large error bar in black in the SFR density panel corresponds to the estimate of the cosmic variance (see text for details).
The cosmic SFR in \nh\ shows the qualitative expected trend over time; however, there is a systematic offset by a factor $1.5-2$ with respect to the observational data assuming standard Chabrier-like IMF. 
The cosmic stellar density in \nh\ is broad agreement with the data with a slight overestimate at low redshift, although there is an important uncertainty on the simulated cosmic SFR (and thus stellar density) due to the large cosmic variance associated with the simulated volume.}
\label{fig:cosmicsfr}
\end{figure}

\begin{figure}
\centering \includegraphics[width=0.49\textwidth]{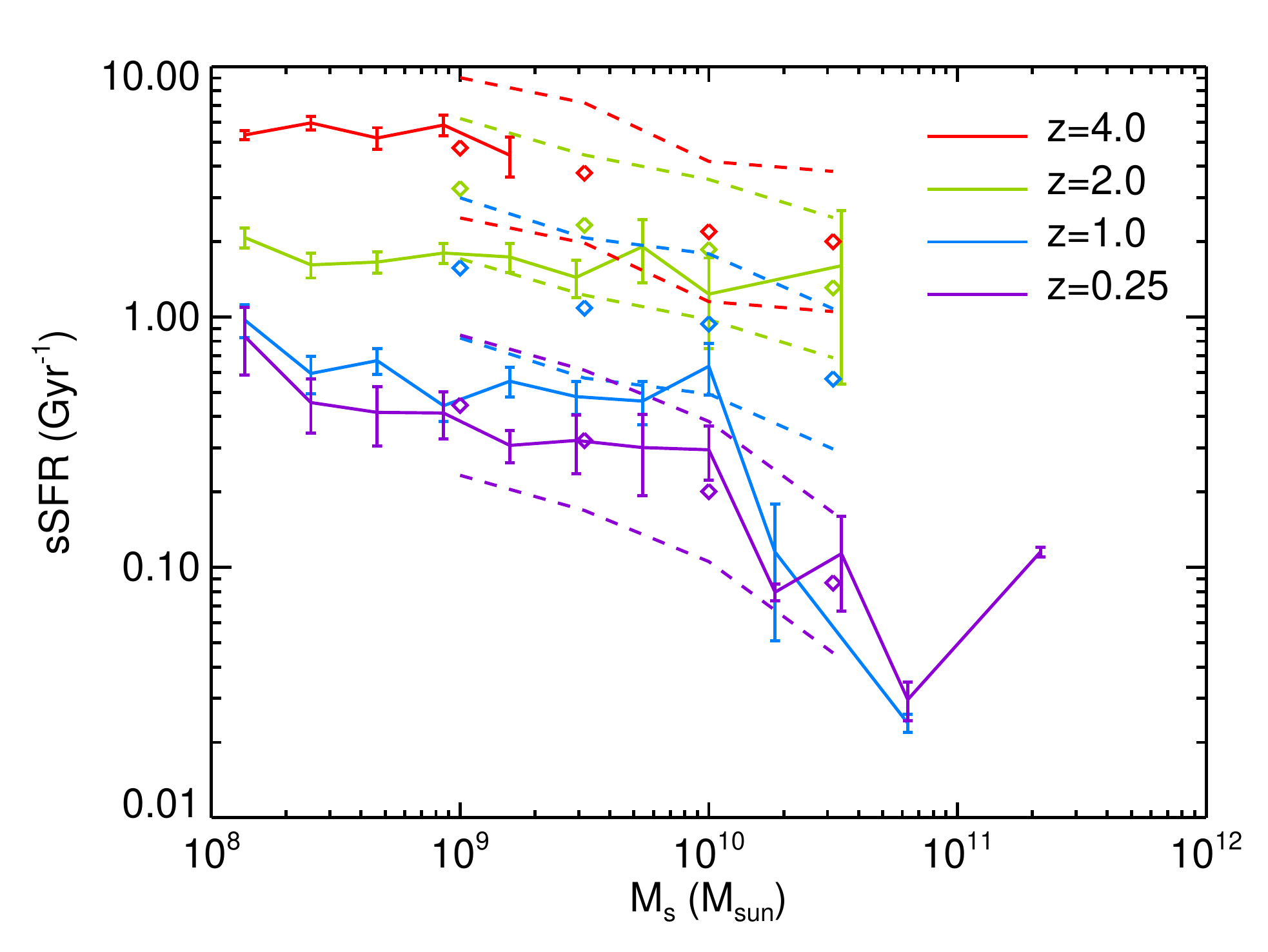}
\caption{sSFR as a function of galaxy stellar mass at different redshifts (as indicated in the panel) in \nh\ with solid lines. The error bars stand for the error around the mean. 
The symbols correspond to the best fit from~\cite{behroozietal13} from their collection of observational data; their uncertainty is shown in dashed lines. The simulated sSFR show very little evolution with stellar mass at the two highest redshifts, while there is a significant quenching at the lowest redshifts for stellar masses above $M_{\rm s}>10^{10}\,\rm M_\odot$. Simulated sSFR show a fair level of agreement with the observations.
}
\label{fig:ssfrvsmgal}
\end{figure}

Figure~\ref{fig:cosmicsfr} shows the cosmic SFR density as a function of redshift in \nh\ compared to observations.
The cosmic SFR density is obtained by summing all star particles formed at a given time over the last 100 Myr within the entire volume of the simulation, which are associated with a galaxy (sub)structure, whether pure or not\footnote{To avoid arbitrary volume correction to select for purity of the zoom-in region, we use the whole galaxy sample and the initial volume of the simulation, that is $(16\,\rm Mpc)^3$ to measure the volumetric quantities of this section.}.
Since stellar particles lose mass owing to stellar feedback, we compensate for this mass loss when reconstructing the SFR.
The observational data in Fig.~\ref{fig:cosmicsfr}~\citep{behroozietal13, madau&dickinson14,novaketal17,driveretal18} are scaled to a Chabrier IMF whenever necessary (i.e. cosmic SFR is decreased by a factor 1.6 when going from a Salpeter to Chabrier IMF).
Several datasets were selected to illustrate the typical variation from inter-publication variance and IMF assumptions~\citep[see e.g.][for a discussion]{behroozietal13,madau&dickinson14}. 
The obtained \nh\ cosmic SFR density is slightly above the observational values collected by~\cite{behroozietal13}. For the  \nh\ SFR values, the maximum offset is  at the lowest redshift, although  the numerical sampling is worse in this case and concentrated over a few rather massive $M_{\rm s}\ge 10^{10}\,\rm M_\odot$ objects.
To estimate the effect of cosmic variance, we relied on the measurement of the cosmic variance on the SFR density in Illustris~\citep[$\sim0.2\,\rm dex$ for a 35 Mpc box length;][]{geneletal14} rescaled to the corresponding smaller  volume here (the large error bar on the left-hand side of the top panel of Fig.~\ref{fig:cosmicsfr}).
This shows that, with this small simulated volume, the cosmic variance is the largest source of uncertainty on the simulated cosmic SFR density compared to observational datasets.

We also show (in Fig.~\ref{fig:cosmicsfr}) the cosmic stellar density; this value is obtained by summing over the individual mass of all the star particles in the simulation, which are again associated with a galaxy (sub)structure, whether pure or not.
Comparing the \nh\ cosmic stellar density to that directly measured in~\cite{driveretal18} produces a factor 2.5 difference at $z=0.25$, whose cosmic SFR density lies on the low side of aggregated observational values; this agrees with the mismatch that is also observed in the cosmic SFR density with the same observations.
The reconstructed cosmic stellar density is obtained from the time-integrated cosmic SFR density with an instantaneous stellar mass return of $R=0.31$, which is the value used in the simulation for SN feedback. This reconstructed stellar density is in excellent agreement with the direct measurement 
It has to be noted that the two direct measurements of~\cite{driveretal18} are self-consistent for $R=0.5$ (see their figure 16).
We also show the reconstructed cosmic stellar density from the best fit to the cosmic SFR density from the~\cite{behroozietal13} data  (Fig.~\ref{fig:cosmicsfr}), for two values of the return rate: $R=0.31$ and $R=0.5$. The
\nh\ predicted cosmic SFR density and stellar density\footnote{The cosmic matter density of the simulated zoom-in region is a factor 1.2 in excess to the average cosmic matter density, which contributes to the total excess in cosmic SFR and stellar density.} compared to~\cite{behroozietal13} are within a factor of 2 (for $R=0.31$); thus these values are in reasonable agreement with this set of data.

The contribution to the cosmic SFR and stellar density is further subdivided into separate galaxy stellar mass bins as indicated in the panels of Fig.~\ref{fig:cosmicsfr}.
Low-mass galaxies $M_{\rm s}<10^9\,\rm M_\odot$ dominate the SFR and stellar mass budget in the early Universe, while intermediate mass galaxies $10^9\le M_{\rm s}/{\rm M_\odot}< 10^{11}$ take over below the peak epoch of star formation (typically at $z\simeq1.5$); Milky Way-like galaxies dominate the cosmic SFR by the lowest redshift $z=0.25$, which is in qualitative agreement with previous theoretical predictions~\citep[e.g.][]{betherminetal13,vogelsbergeretal14,geneletal14}.
Although only a few galaxies contribute to
this range of mass, the highest mass bin $M_{\rm s}$ seems to marginally contribute to the cosmic SFR at low redshift\footnote{The sharp increase at $\log(1+z)\simeq0.25$ in the cosmic SFR density of the $M_{\rm s}/{\rm M_\odot}\ge 10^{11}$ mass bin corresponds to a galaxy moving from the $10^{10}\le M_{\rm s}/{\rm M_\odot}< 10^{11}$ mass bin to the previous mass bin; there is a dominant contribution from this single galaxy to the cosmic SFR density in that mass bin.}; however, it represents nearly half of the cosmic stellar density, thereby highlighting the role played by satellite infall (stars formed ex situ) in the assembly of massive galaxies~\citep{delucia&blaizot07,oseretal10,duboisetal13,duboisetal16}.

The specific SFR (sSFR) of individual galaxies can be computed by measuring the stars younger than $100$ Myr within their effective radius $R_{\rm eff}$ (see Section~\ref{section:reff} for the calculation of $R_{\rm eff}$) and dividing by the current stellar mass within $R_{\rm eff}$ at the given redshift.
Figure~\ref{fig:ssfrvsmgal} shows the resulting mean sSFR as a function of galaxy stellar mass for different redshifts\footnote{Changing the timescale over which SFRs are measured to 10 Myr increases the uncertainty in the mean relation of the simulated points without changing the trends. Changing the measurement radius to $2R_{\rm eff}$ or $3R_{\rm eff}$ decreases the mean sSFRs by a few tens of per cent.}. 
Galaxies are usually separated into a main sequence of star-forming (active) galaxies and quenched galaxies that are passively evolving, showing a significant evolution over mass and redshift of their their respective fraction~\citep[see e.g.][]{kavirajetal07,muzzinetal13, wetzeletal13,furlongetal15,fossatietal17}; in particular, the bulk of the quenched galaxies found in central galaxies are more massive than a few $10^{10}\,\rm M_\odot$ or in satellites hosted by massive groups or clusters.
Only active galaxies are selected based on their level of sSFR, that is with a sSFR above $0.01\,\rm Gyr^{-1}$.
Including inactive galaxies slightly changes (by up to 20 \%) the mean sSFR at $z=0$ and $z=1$ for galaxies with stellar mass $M_{\rm s}\le 10^9 \,\rm M_\odot$. The exact criterion for the separation between the two population is somewhat arbitrary, but the idea is that quenched galaxies should clearly stand out from the main population of galaxies~\citep[see e.g.][for how distinguishing between active and passive galaxies affects the quenched fraction]{donnarietal20}.
The sSFR at $z=4$ and $2$ show no trend with stellar mass, while the low redshift relations at $z=1$ and $0.25$ show a significant decrease (quenching) at $M_{\rm s}>10^{10}\,\rm M_\odot$.
These simulated values are compared to the best-fit relation from~\cite{behroozietal13} obtained from a collection of observational data (see references therein). 
The simulation agrees fairly well with the data at $z=4,2$ and $0.25$, but the values at $z=1$ are significantly below those of the data; however, this corresponds to the decrease in the cosmic SFR density right before a peak that almost doubles the overall SFR in galaxies more massive than $M_{\rm s}=10^9\,\rm M_\odot$.
Nonetheless, it has to be noted that there is a systematic offset between the \nh\ sSFR and the data. Simulated sSFR are on average systematically lower than the data. This leads to a slight inconsistency with the cosmic SFR density, which is higher than the values found in the data.
Similar tensions are noted within the data themselves~\citep[see Appendix C4 of][]{behroozietal19}.

\subsection{Kennicutt-Schmidt relation}

\begin{figure}
\centering \includegraphics[width=0.49\textwidth]{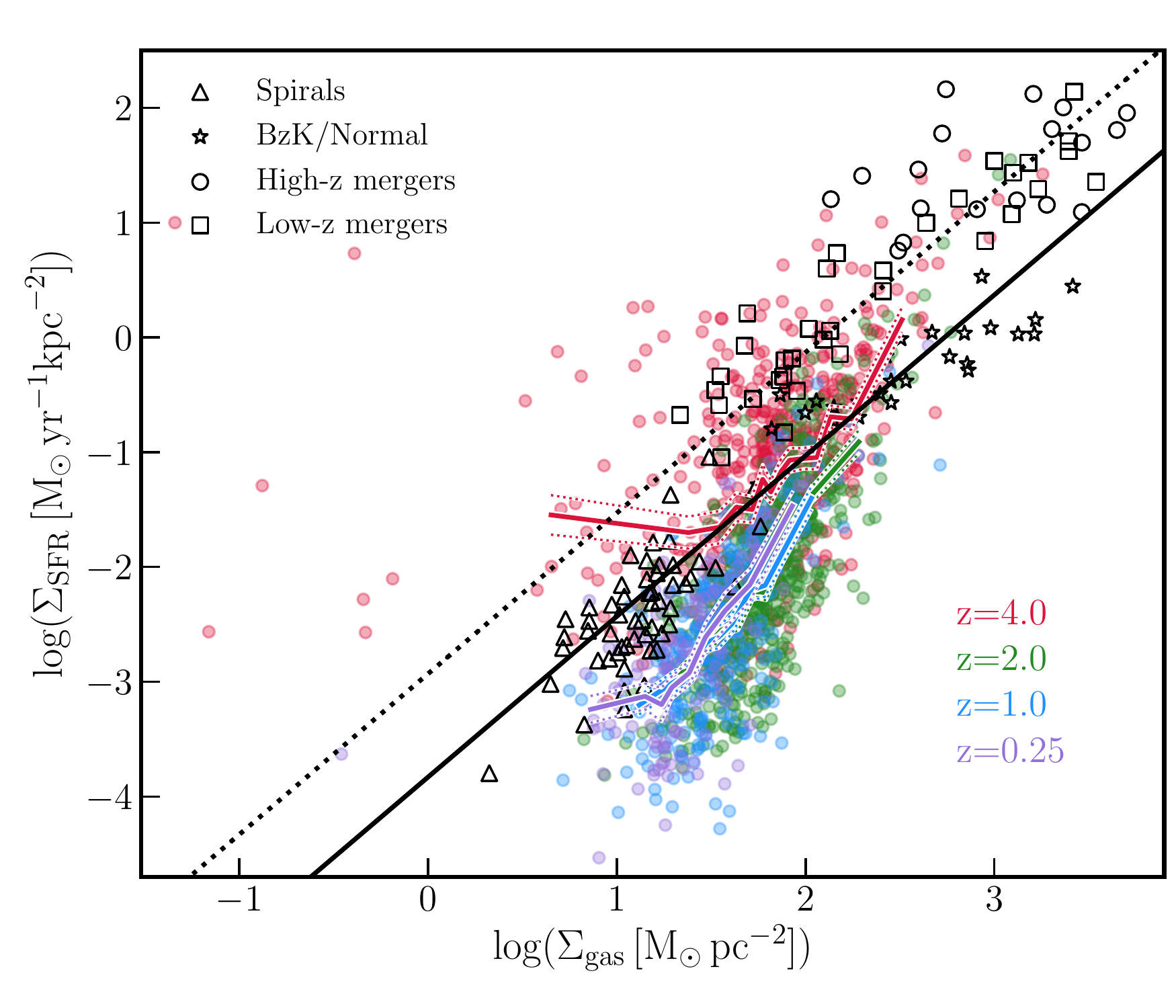}
\caption{Kennicutt-Schmidt relation compared to observations. Filled coloured circles correspond to \nh{} galaxies at various redshifts. The mean and error on the mean are shown with solid and dotted coloured lines, respectively. The empty black symbols correspond to observational datasets: triangles are local spirals \citep{kennicutt98}, stars are high-$z$ BzK/Normal galaxies \citep{daddietal10a,tacconietal10}, circles high-$z$ mergers \citep{boucheetal07,bothwelletal09}, and squares low-$z$ mergers \citep{kennicutt98}. The solid black and dotted black lines represent the sequence of discs and starbursts, respectively, from \cite{daddietal10b}. The \nh\ tool is able to capture the galaxy main sequence, although the SFR surface densities are slightly  lower than in the observational data. There is a larger fraction of high SFR galaxies at higher redshift in qualitative agreement with observations.}
\label{fig:ks}
\end{figure}

Figure~\ref{fig:ks} shows the Kennicutt-Schmidt relation of surface density of SFR (\sigmasfr) as a function of surface density of total (HI+H$_2$) gas (\sigmagas) for galaxies in the \nh{} simulation at redshifts $z=4,2,1,0.25$ and with $M_{\rm s} > 10^{7}\,\rm M_\odot$, compared to observations. The quantities \sigmasfr and \sigmagas are computed within $R_{\rm eff}$. The SFR is obtained by summing over all stellar particles with stellar age below 10 Myr and the HI+H$_2$ gas is selected as gas with density $n > 0.1 \, \rm  cm^{-3}$ and temperature T < 2$\times 10^4$ K.
The observational data shown in Fig.~\ref{fig:ks} include $z \sim 1.5$ BzK-selected normal galaxies \citep{daddietal10a} and $z=1-2.3$ normal galaxies \citep{tacconietal10}, high-$z$ submillimetre  selected  galaxies \citep[SMGs;][]{boucheetal07,bothwelletal09}, IR-luminous galaxies (ULIRGs), and spiral galaxies are taken from the sample of \cite{kennicutt98} as compiled in \cite{daddietal10b} with a consistent choice of the conversion factor used in Fig. 10 to derive molecular gas masses from CO luminosities ($\alpha_{\rm CO}$) and a \cite{chabrier03} IMF.

With decreasing redshift, the population of simulated galaxies as a whole moves roughly along the sequence of discs \citep[solid line,][]{daddietal10b} towards lower values of \sigmagas and \sigmasfr. Qualitatively, simulated galaxies occupy comparable regions of the Kennicutt-Schmidt parameter space, reproducing the observed diversity of star-forming galaxies; however there are some notable differences. At $z \sim 4$, there are galaxies at low \sigmagas$<10\,\rm M_\odot\,pc^{-2}$ and high \sigmasfr$>10^{-2}\,\rm M_\odot\,yr^{-1}\,kpc^{-2}$, which seem to be offset from the bulk of the population. 
The reason for this offset is their low gas fraction that is $\lesssim 0.1$ regardless of their stellar mass and SFR. These galaxies cover the entire stellar mass range, having similar SFR and size to galaxies with comparable \sigmasfr and higher \sigmagas.  
By redshift $z \sim 3$, there are no galaxies left in this region of the parameter space. Below $z = 3$, the number of galaxies on the canonical sequence of starbursts \citep{daddietal10b} decreases with decreasing redshift. The slope of the average \sigmagas-\sigmasfr relation does not evolve strongly between $z=1-3$; 
however, the slope is offset from the sequence of discs by $\sim 0.5$ dex. At $z=0.25$, the lowest available redshift, the slope is in qualitative agreement with observations of local spirals, albeit it still has an offset. We have checked that when considering star-forming gas only, that is gas with density $n > 10 \, \rm  cm^{-3}$ and temperature $T < 2 \times 10^4$ K, the average \sigmagas-\sigmasfr relation at $z < 3$ follows the sequence of discs (Kraljic et al., in prep.).

\subsection{Galaxy-to-halo mass relation}

\begin{figure}
\centering \includegraphics[width=0.45\textwidth]{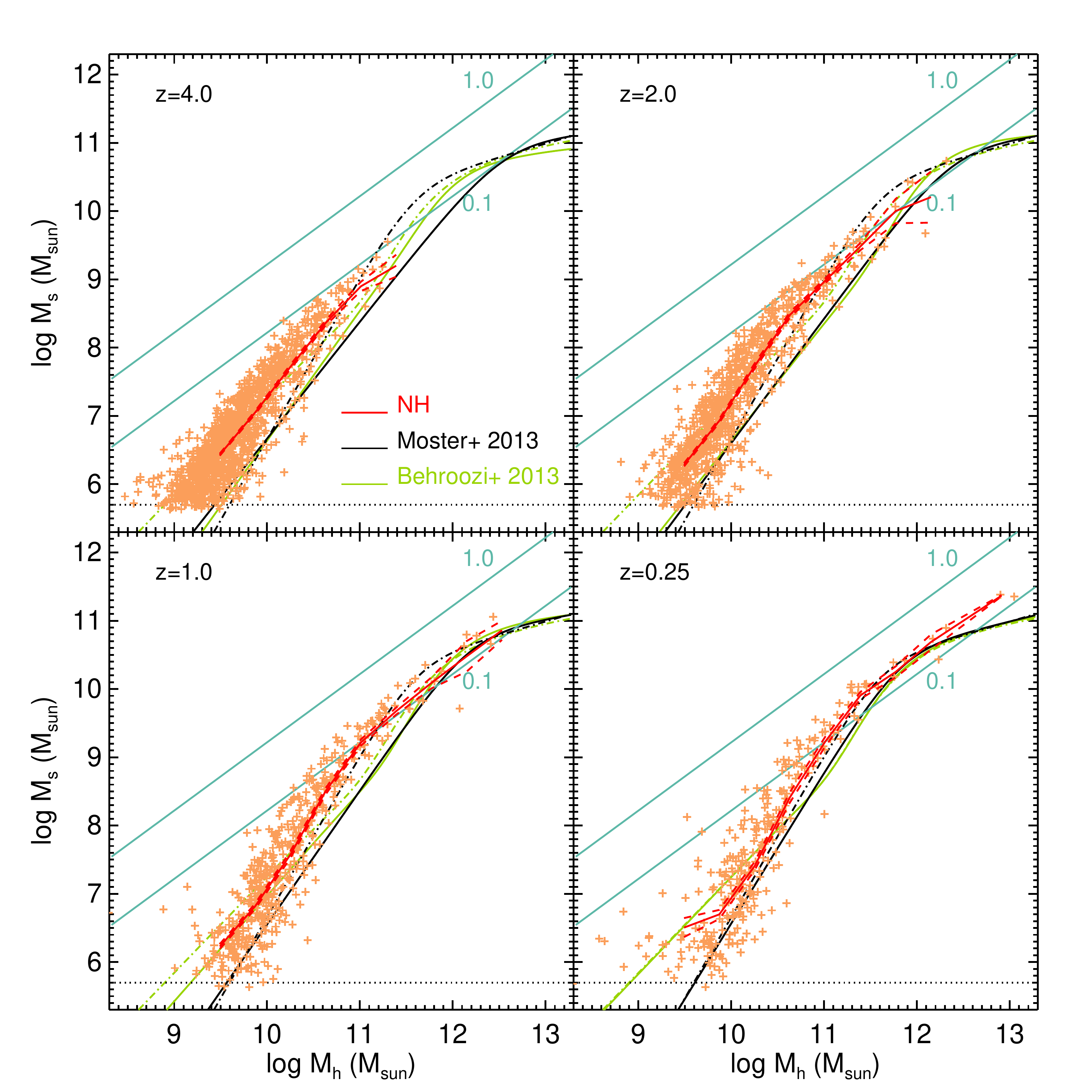}\vspace{-0.2cm}
\centering \includegraphics[width=0.45\textwidth]{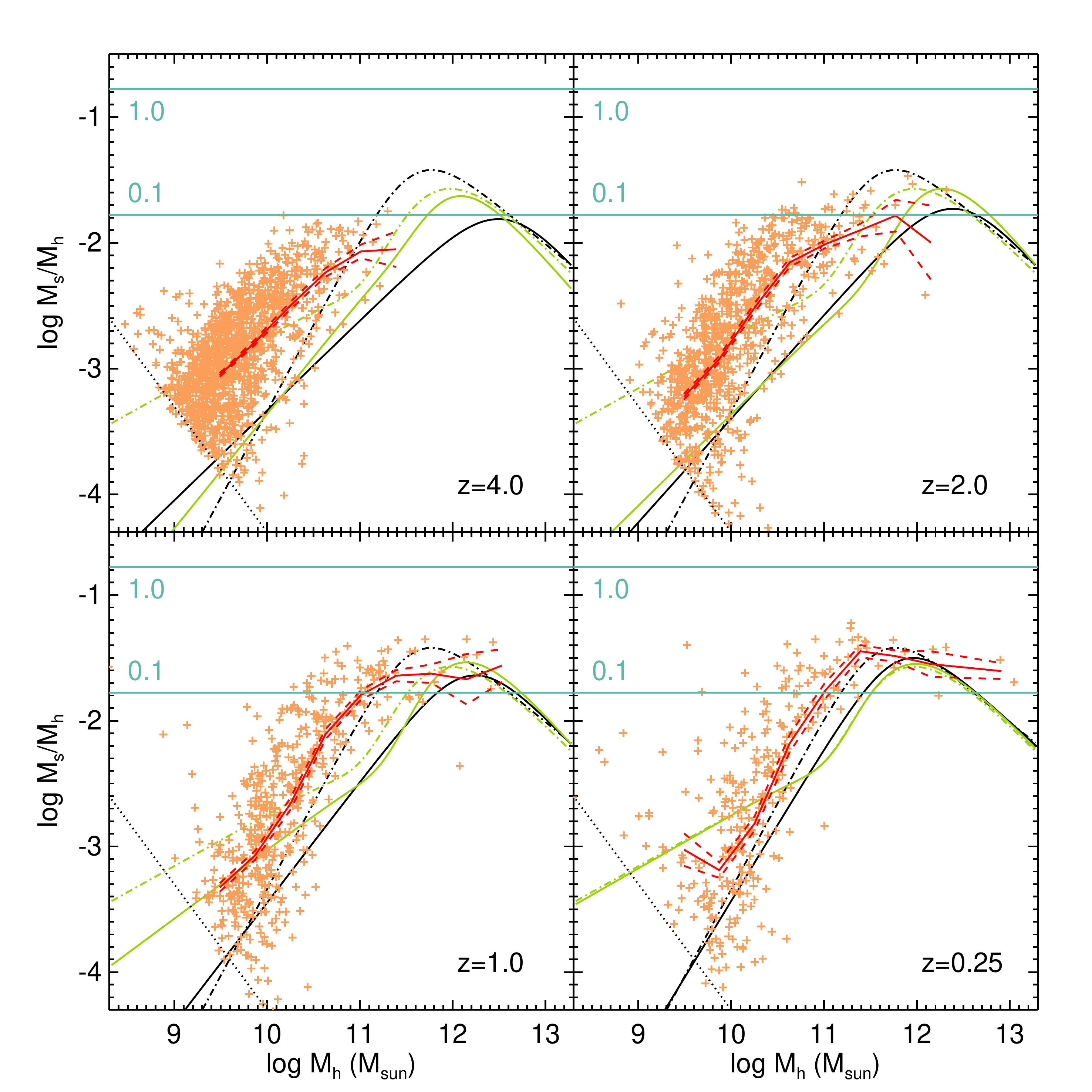}
\caption{Stellar-to-halo mass relation (top) and baryon conversion efficiency $M_{\rm s}/M_{\rm h}$ at different redshifts (as indicated in the panels). The solid red lines represent the average with their error of the mean (dashed) with individual points as plus symbols. The solid black line indicates the semi-empirical relation from~\cite{mosteretal13} and the green line from~\cite{behroozietal13} at the indicated redshift and dot-dashed lines the respective relations at $z=0$. The cyan lines stand for the constant star formation conversion efficiencies. The dotted lines correspond to the minimum stellar mass detected by the galaxy finder. The simulated relation between central stellar mass and halo mass is in fair qualitative agreement with the semi-empirical relations, where the increase in the baryon conversion efficiency with mass up to the peak at Milky Way halo mass is captured; however this efficiency is significantly overestimated below that peak.}
\label{fig:msmh_redshift}
\end{figure}

Fig.~\ref{fig:msmh_redshift} shows the stellar mass of galaxies as a function of their host DM halo mass at redshifts $z=4,2,1,0.25$ compared to the semi-empirical relation from~\cite{mosteretal13} and~\cite{behroozietal13}.
The mass of the DM halo is obtained by taking the total (DM, gas, stars, and MBHs) mass enclosed within the radius corresponding to $\Delta_{\rm c}$ times the critical density at the corresponding redshift, where the $\Delta_{\rm c}$ of the analytical form is taken from~\cite{bryan&norman98}.
This procedure is similar to that of~\cite{behroozietal13}, but differs from~\cite{mosteretal13}, where $\Delta_{\rm c}$ is fixed at 200.
We note that this is a $\sim0.1-0.2 \,\rm dex$ increase in halo mass compared to the value of the virial mass obtained by the AdaptaHOP halo mass decomposition.
AdaptaHOP galaxies are considered in this work; hence, satellites (and in situ stellar clumps) are not considered when the total stellar mass is measured, but these clumps only constitute a small fraction of the total galaxy mass as discussed in Section~\ref{section:clumps} (typically 10 per cent and 1 per cent of the total stellar mass at the most extreme redshifts, resp. $z=4$ and $z=0.25$).
Satellite galaxies are connected to their host subhalo mass at the current redshift; thus these galaxies should not be compared directly with the semi-empirical constraints, whereby they reconstruct the relation with the halo mass before it becomes a subhalo.
Satellite galaxies (not shown here) have systematically a larger value with respect to their host subhalo mass because the DM particles are first stripped by gravitational interactions with the main halo well before the more concentrated stellar mass becomes affected~\citep{penarrubiaetal08,smithetal16}.

There is a fairly good qualitative agreement of the stellar-to-halo mass relation with the general trends from~\cite{mosteretal13} and~\cite{behroozietal13} at all redshifts for the population of main halos. The baryon conversion efficiency, that is the ratio of $M_{\rm s}/M_{\rm h}$, shows a maximum at near to the Milky Way mass at a few $10^{11}\,\rm M_\odot$, although this value is slightly below the expected peak of the semi-empirical relations. This ratio steeply decreases with the decreasing halo mass and the ratio plateaus around the Milky Way scale, while it is expected to decrease above this mass.
However, see~\cite{kravtsovetal18} for the underestimated stellar light component at those group and cluster scales together with IMF variation effects.
The simulated stellar masses are still significantly above the relation; however there is a better agreement at the higher masses $M_{\rm h}>10^{11}\,\rm M_\odot$.

\subsection{Size-to-galaxy mass relation}
\label{section:reff}

\begin{figure}
\centering \includegraphics[width=0.29\textwidth]{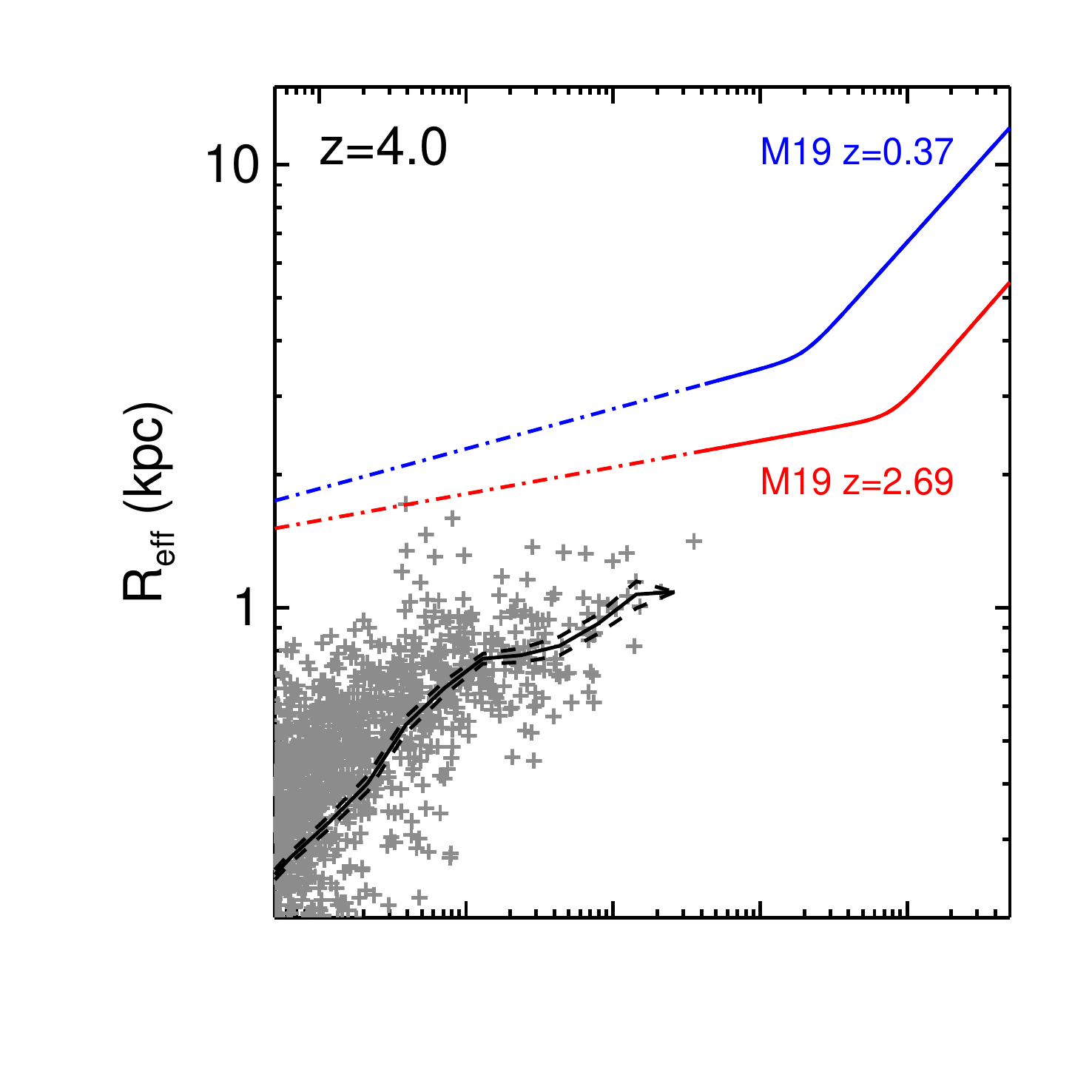}\hspace{-1.825cm}
\centering \includegraphics[width=0.29\textwidth]{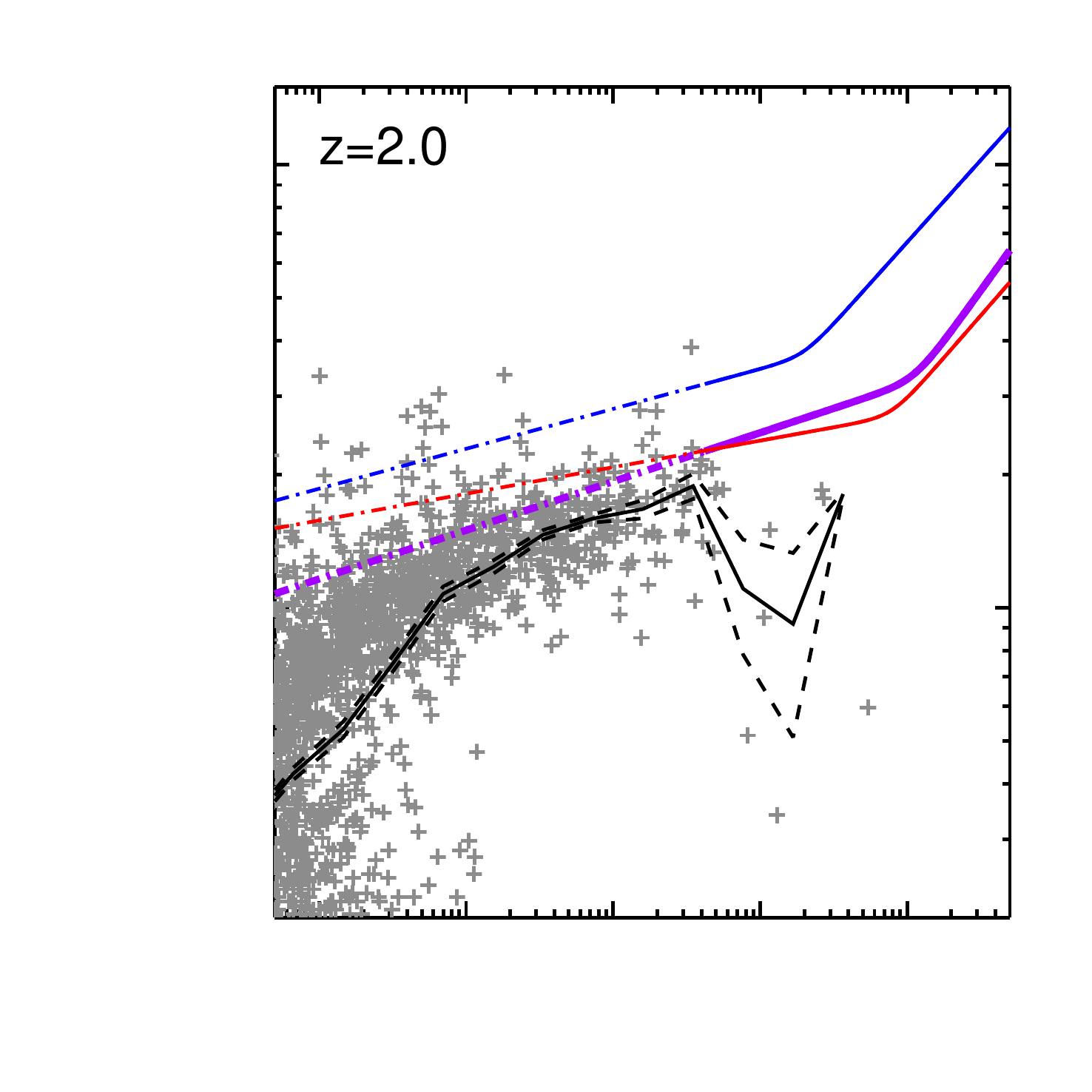}\vspace{-1.3cm}
\centering \includegraphics[width=0.29\textwidth]{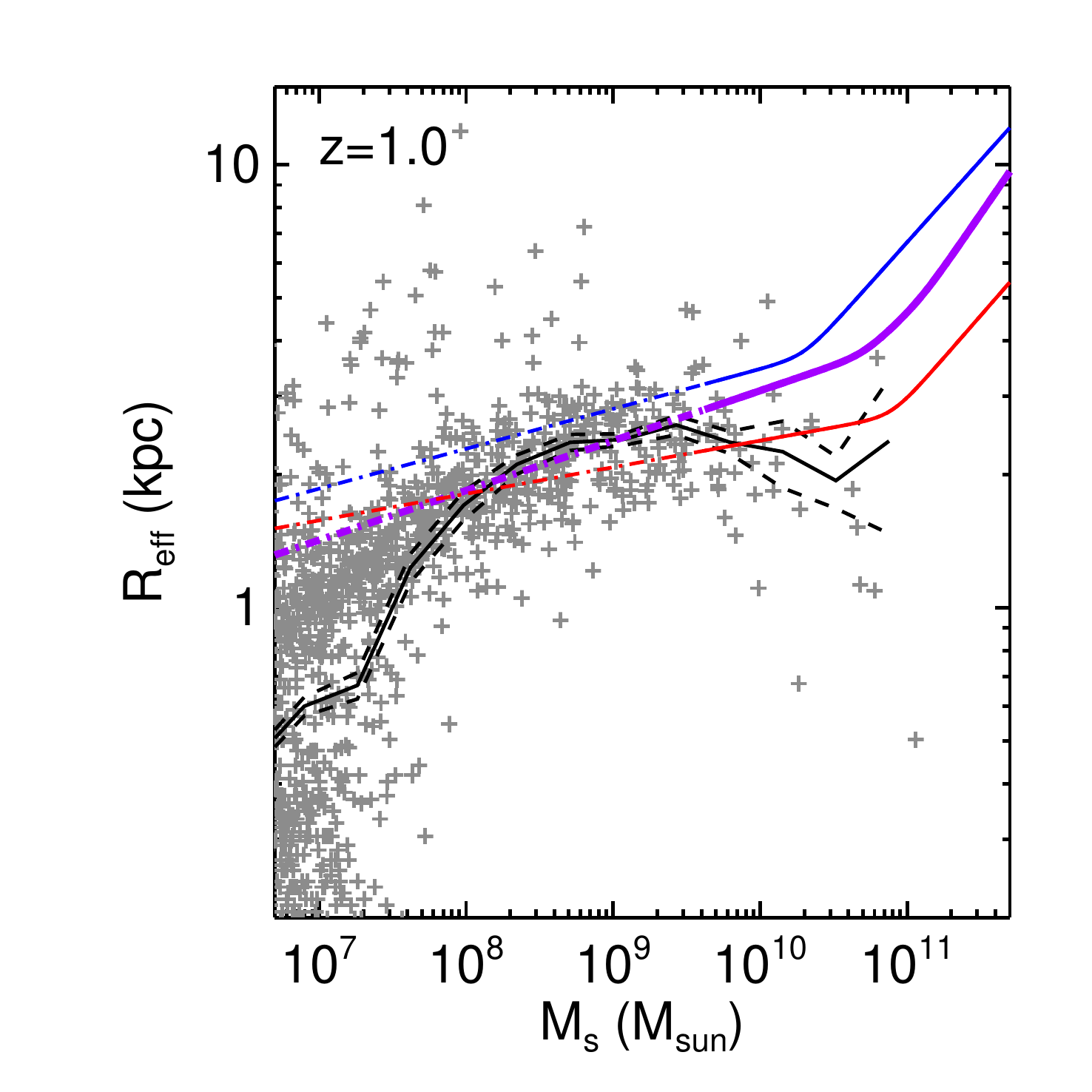}\hspace{-1.825cm}
\centering \includegraphics[width=0.29\textwidth]{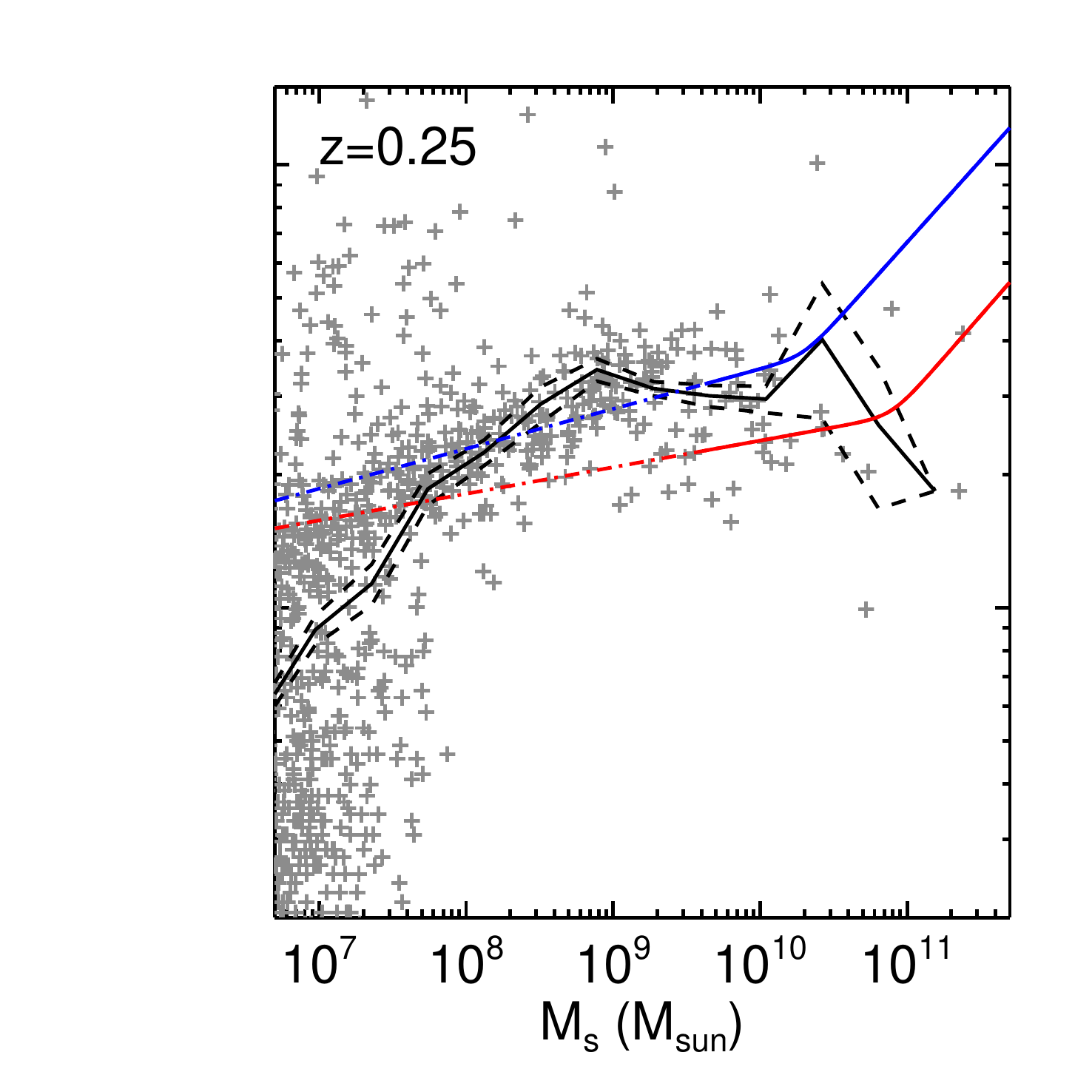}
\caption{Effective radius as a function of stellar mass in \nh\ at different redshifts for individual galaxies (grey crosses) and for the average (solid line) with standard deviation (dashed lines). The sizes are computed with the geometric mean of the x, y, z effective radius using the AdaptaHOP classification. Observations from~\cite{mowlaetal19} are shown as a purple line at the current redshift when available and with their most extreme redshift fits at $z=0.37$ and $z=2.69$ in blue and red, respectively, which are extrapolated (triple dot-dashed line) beyond their range of available data (solid line) to guide the eye.
Galaxies in \nh\ are more compact at high redshift as opposed to low redshift in good agreement with  observations.
}
\label{fig:rvsmg_redshift}
\end{figure}

Galaxy effective radii are obtained by taking the geometric mean of the half-mass radius of the projected stellar densities along each of the Cartesian axis.
For this measurement, we consider AdaptaHOP galaxies since HOP galaxies can largely overestimate the effective radius of galaxies at the high-mass end (see Appendix~\ref{appendix:size_hop_vs_ahop}) when satellites orbit around centrals and are connected by the diffuse stellar light.
At the same time, star-forming clumps are also removed but as they only represent a small fraction of the total stellar mass, they do not have a significant impact on the determination of the effective radius.
We note that this procedure tends to reduce the scatter of the relation, but in this work we are mostly interested in investigating to what extent the observed mean relation is reproduced.

Figure~\ref{fig:rvsmg_redshift} shows the effective radius of \nh\ galaxies for four different redshifts with simulated data points in grey plus signs and its average and error around the mean with black lines, compared to the observational relation obtained by~\cite{mowlaetal19} in purple at the corresponding redshift, which we have linearly interpolated from the two contiguous redshifts provided in~\cite{mowlaetal19}.
To guide the eye, and because there is no observational data available beyond $z=2.69$, we also overplotted the observational relation at the lowest ($z=0.37$ in blue) and highest redshift ($z=2.69$ in red).
There is overall good agreement between the simulated size-mass relation and the observations at all redshifts.
We note that galaxy sizes around 1 kpc to a few kiloparsec are described with several resolution elements (resolution is $34$ pc), that is more than what is typically achieved in the low-mass range for large-scale simulations such as Horizon-AGN~\citep{duboisetal14}, EAGLE~\citep{schayeetal15}, or IllustrisTNG~\citep{pillepichetal18}.
\cite{pillepichetal18} show that with the IllustrisTNG model, the mean size obtained at their low-mass end of around a few kiloparsec is systematically scaled to 3-4 times their minimum stellar spatial resolution. Thus, the simulated low-mass galaxies (though above a few $10^7\,\rm M_\odot$) are comfortably resolved in \nh\ in comparison.
The \nh~tool recovers the size growth of galaxies with mass and the size growth of galaxies with redshift (at a given mass): as they grow in mass, galaxies tend to be more extended, and the size-mass relation produces more extended galaxies with time as measured in observations~\citep[e.g.][]{trujilloetal06,vanderweletal14,mowlaetal19}.

The large spread of simulated data points below $M_{\rm s}<10^8\, \rm M_\odot$  corresponds to embedded in situ stellar clumps for the most compact ones.
Outliers with large sizes at relatively low stellar masses are due to satellite galaxies embedded in diffuse stellar light of a much more massive companion, where the galaxy finder has failed to extract them from the background correctly.
There is formation of compact massive galaxies at high-mass end ($M_{\rm s}>10^{10}\,\rm M_\odot$) and more compact galaxies at higher redshift, with sizes below $R_{\rm eff}\lesssim 1\,\rm kpc.$ This feature is reminiscent of the nugget formation of massive galaxies that have endured a gas-rich compaction event triggering high levels of SFR before quenching~\citep[see e.g.][]{dekel&burkert14,zolotovetal15,lapineretal21}.
The \nh\ galaxies however seem to fail to reproduce the rapid rise in galaxy sizes at the high-mass end. This might partially be a consequence of the volume that is simulated corresponds to a region of average density, and because we lack the formation of dense environments that are the main producers of such massive objects, where galaxies are more likely to be more passive at low redshift and, hence, built though mergers \citep{Martin2018} that lead to a large size increase~\citep[e.g.][]{naabetal09,oseretal10,duboisetal13,duboisetal16}.

\subsection{Stellar and gas metallicities}

\begin{figure}
\centering \includegraphics[width=0.49\textwidth]{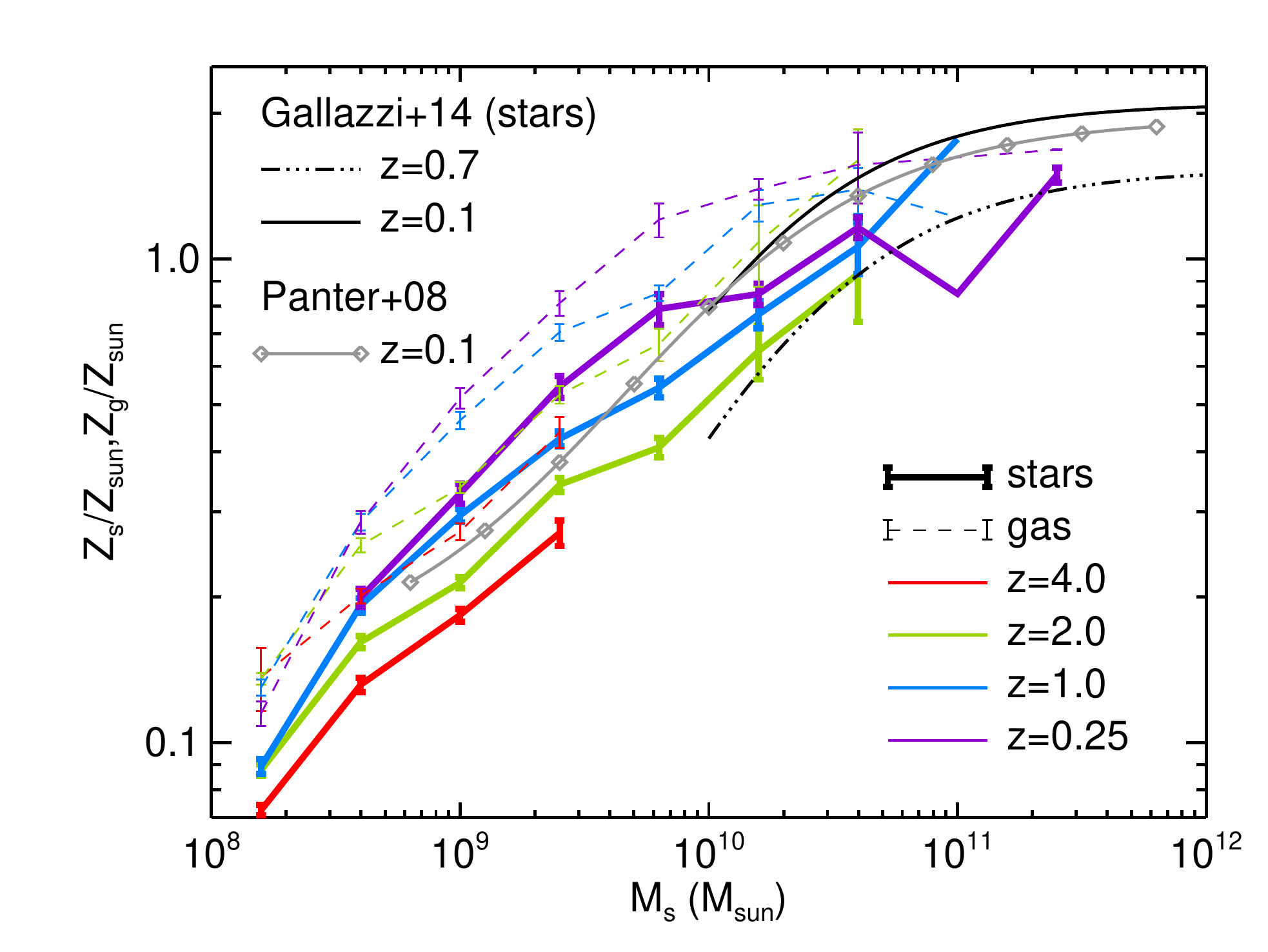}
\caption{Average stellar (coloured solid) and gas (coloured dashed) metallicity as a function of stellar mass for different redshifts in \nh\ as indicated in the panel. The error bars stand for the error around the mean. Observational fits for the stellar metallicity from~\cite{gallazzietal14} (solid black at $z=0.1$, and dashed black at $z=0.7$) and from~\cite{panteretal08} (solid grey) are also shown. The metallicity of both the gas and stellar component are decreasing with increasing redshift owing to less chemically enriched galaxies.}
\label{fig:met}
\end{figure}

The mass-weighted stellar metallicity $Z_{\rm s}$ is computed for all the stars within $R_{\rm eff}$ for each galaxies.
The value is renormalised to the solar metallicity $Z_\odot=0.01345$~\citep{asplundetal09}.
Figure~\ref{fig:met} shows the \nh\ stellar metallicities as a function of each galaxy stellar mass at different redshifts $z=4,2,1,0.25$ and is compared to the (luminosity-weighted) observations by~\cite{gallazzietal14} at $z=0.7$ and $z=0.1$ and to the (mass-weighted) observations by~\cite{panteretal08} at $z=0.1$. 
\footnote{The observational values of~\cite{gallazzietal14} and~\cite{panteretal08} are given for an assumed solar metallicity of $Z_\odot=0.02$~\citep{anders&grevesse89}, while more self-consistent calculations of the composition of the solar atmosphere give a significantly lower value of $Z_\odot=0.01345$~\citep{asplundetal09}. We have, therefore, scaled up their fitting relations accordingly.}
We note that observational estimates of metallicities barely differ whether are they mass-weighted or luminosity-weighted, as also seen in simulations~\citep[e.g. as is reported in][for the EAGLE simulation]{derossietal17}. The
\nh\ galaxies show an increase in stellar metallicity with mass and with decreasing redshift as expected from the continuous release of metals from stars and their increased level of retention in more massive galaxies~\citep[e.g.][]{tremontietal04}.
Despite the extremely crude modelling of metal release used in \nh\ (all metals are released at once after 5 Myr), the mass-metallicity relation at $z=0.25$ is consistent with observations at $z=0.1-0.7$.
However, the mass-metallicity relation in \nh\ shows a slower evolution over redshift than what is suggested by observations.
We also show in Fig.~\ref{fig:met} the metallicity of the cold gas phase, namely gas with density $n>0.1\,\rm cm^{-3}$ and temperature $T<2\times 10^4\,\rm K$.
Gas metallicity is systematically larger by $\simeq 50$ per cent than the stellar metallicity at any given redshift as an effect of the stellar metallicity being composed of very old poorly enriched stars.

\begin{figure}
\centering \includegraphics[width=0.49\textwidth]{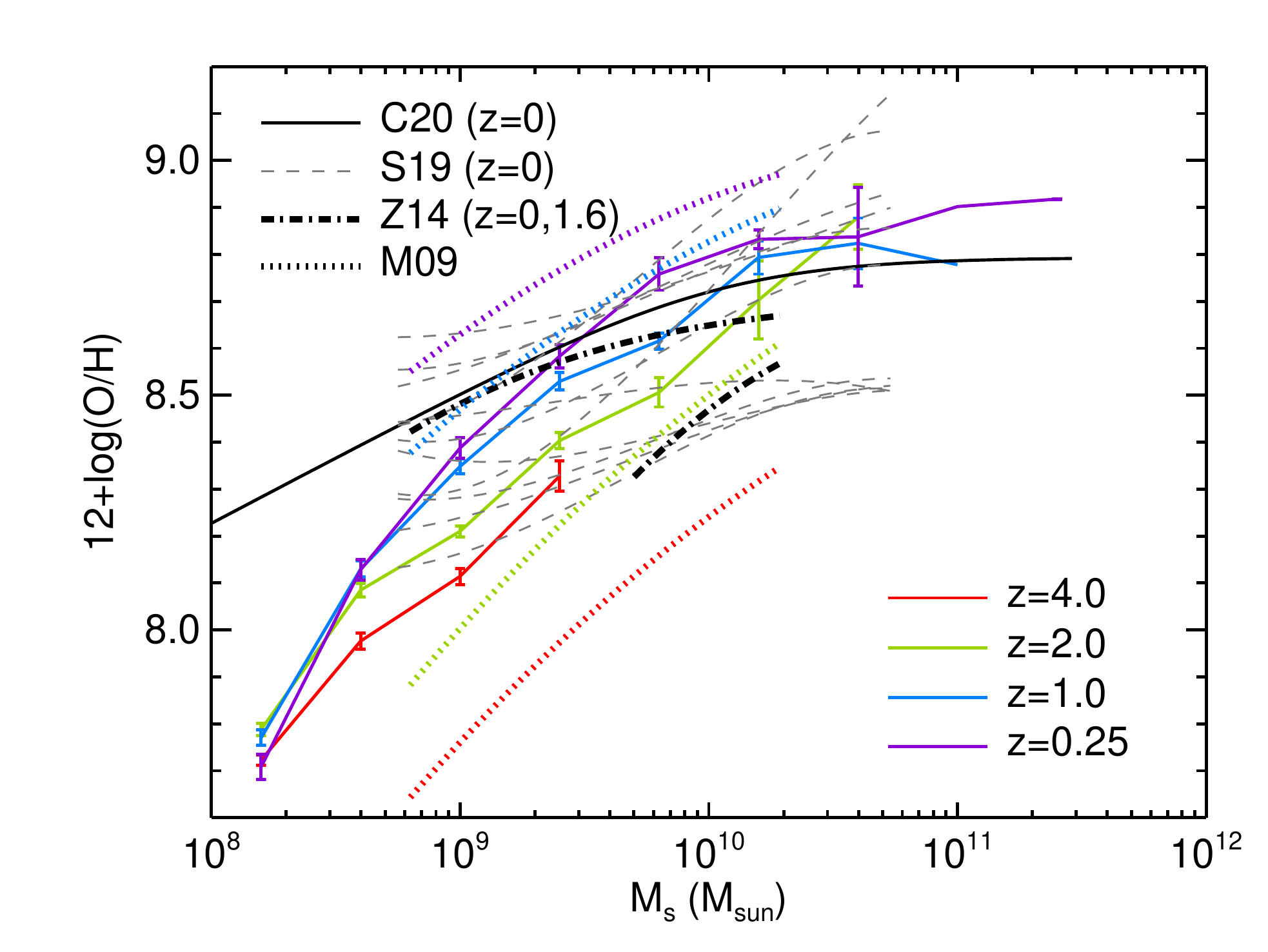}
\caption{Average abundance of oxygen (coloured solid) in the cold gas as a function of stellar mass for different redshifts in \nh\ as indicated in the panel. The error bars stand for the error around the mean. Several observational relations are also shown from~\cite{mannuccietal09} (M09; in dotted lines with four different colours corresponding to $z=0.07,0.7,2.2,$ and $3.5$),~\cite{zahidetal14} (Z14; dot-dashed line),~\cite{sanchezetal19} (S19; dashed lines with different lines corresponding to various oxygen calibrators), and \cite{curtietal20} (C20; black solid).}
\label{fig:oxabund}
\end{figure}

The gas oxygen abundance ratio and its relation to galaxy mass (or the so-called mass-metallicity relation; MZR) exhibits an evolution with redshift~\citep[e.g.][]{erbetal06,maiolinoetal08,zahidetal11,zahidetal14,yabeetal15,sandersetal18}, which is reflective of the redshift evolution of the SFR-$M_{\rm s}$ that drives the scatter of the MZR~\citep{mannuccietal10}.
To qualitatively appreciate the evolution over redshift of the MZR, we show the data from~\cite{zahidetal14} measured within fixed aperture of 10 kpc at $z=1.6$ together with the \nh\ data in Fig. 14.
We also report the various fits of the MZR obtained from the spatially resolved data of the SAMI galaxy survey within $R_{\rm eff}$ for different oxygen abundance calibrators from emission lines, which is known to be a major source of uncertainty~\citep[see details in][    ]{sanchezetal19}, and the fitting relation from~\cite{curtietal20} for the MZR of SDSS galaxies at $z=0.027$.
For \nh, even though the tool does not self-consistently following the amount of oxygen (mainly produced by massive stars), we rescale the gas metallicity values by a factor corresponding to the fractional abundance of oxygen in the solar atmosphere and further assume that the fraction of hydrogen is solar~\citep[][]{asplundetal09}: the mass fraction of hydrogen of the gas is $73.4$ per cent, and oxygen represents 43 per cent of the total mass of elements heavier than H and He.
This is, obviously, a crude estimate of the actual oxygen abundance in the simulation since the fractional amount of oxygen amongst metals varies with the age of the galaxy, and in turn with metallicity (or galaxy mass), and for instance its [O/Fe] ratio is known to increase faster with decreasing metallicity than some other significantly abundant elements such as carbon or nitrogen~\citep[see e.g.][for a compilation of observational data]{prantzosetal18}.
Figure~\ref{fig:oxabund} shows that the increase in the gas oxygen abundance with galaxy mass and time is well captured by the simulation despite the simple scaling for converting the metallicity in \nh\ into oxygen, and the choice of different apertures amongst data and our simulated galaxies. 
However, the highest redshift bin \nh\ $z=4$ seems to have a serious offset ($\sim 0.6\,\rm dex$) with respect to observations at $z=3.5$~\citep{mannuccietal09}, even though there is a large spread and large uncertainties in the observational data at this redshift.
This range of redshift where observed metallicities are a factor 10 below local values calls for a more accurate treatment of stellar yield release~\citep[see e.g.][]{mannuccietal09,prantzosetal18}.

\subsection{Stellar kinematics}

\begin{figure}
\centering \includegraphics[width=0.49\textwidth]{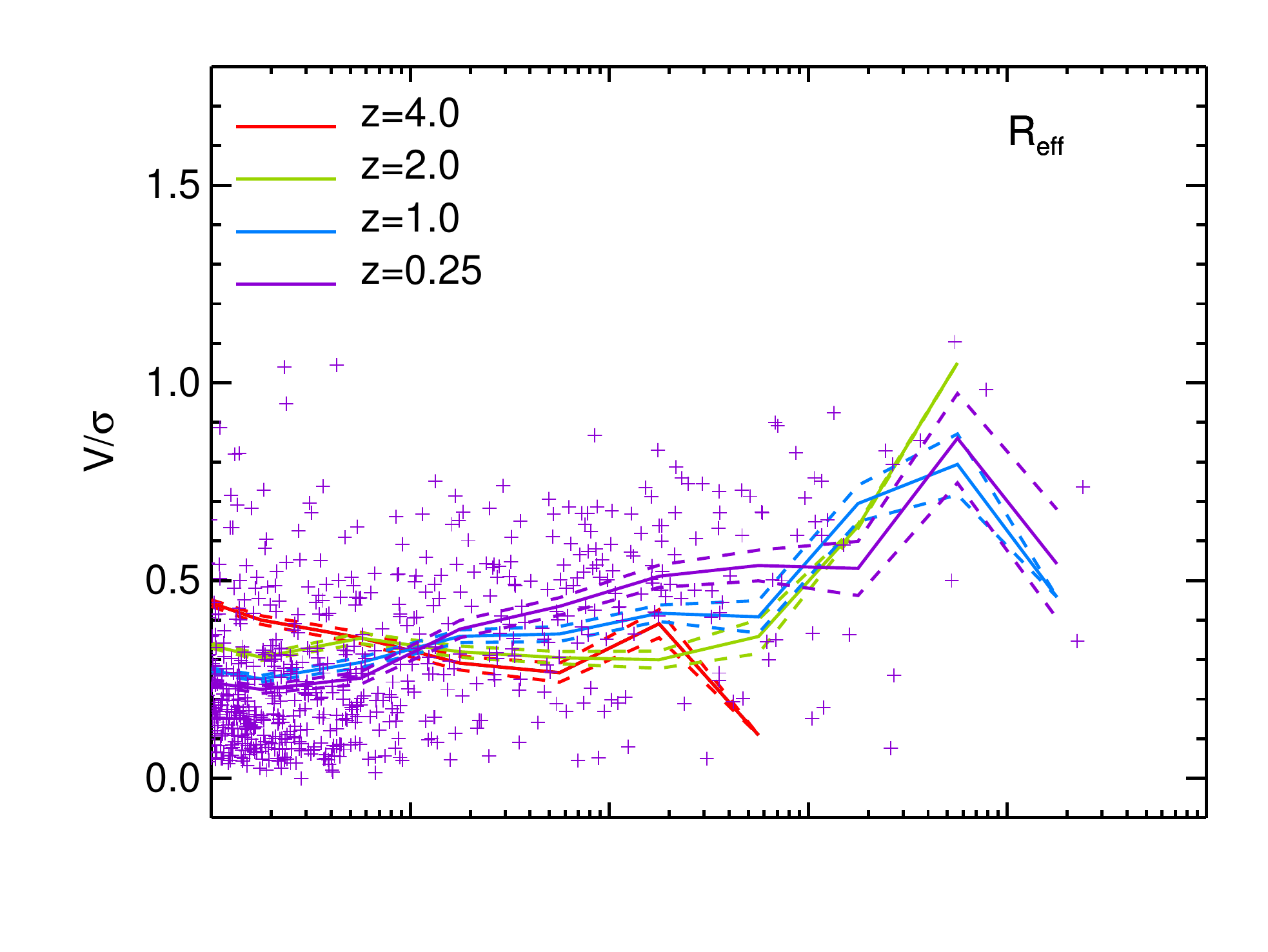}\vspace{-1.45cm}
\centering \includegraphics[width=0.49\textwidth]{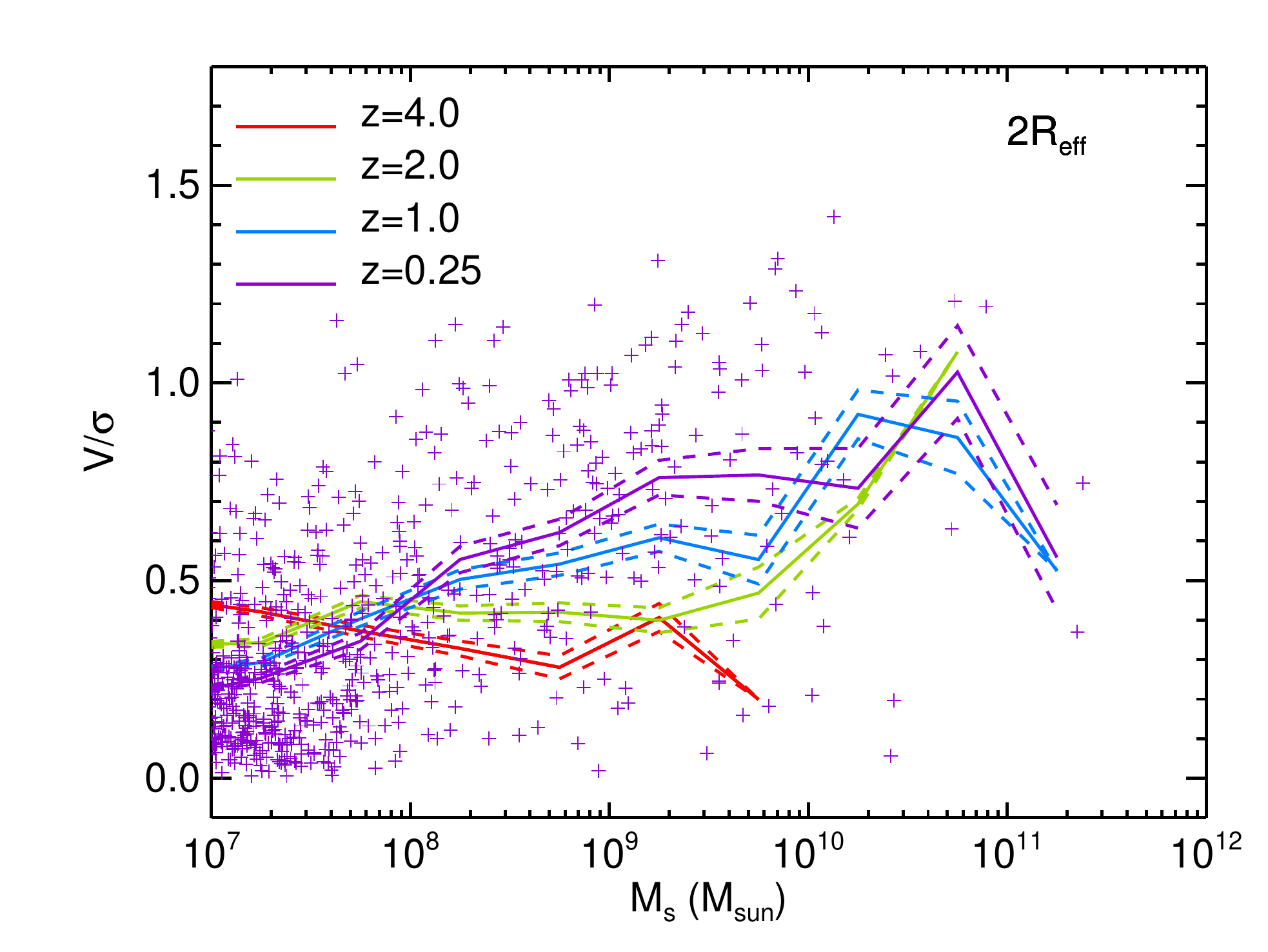}
\caption{Ratio of stellar rotation over dispersion as a function of galaxy stellar mass for different redshifts (colours) as indicated in the panels. The kinematics are measured within $R_{\rm eff}$ (top) or $2R_{\rm eff}$ (bottom). The solid lines indicate the average and the dashed lines are the error around the mean, with individual points shown at $z=0.25$ only (purple plus signs) to appreciate the scatter of the distribution. Galaxies show an increase support of stellar rotation over dispersion with time and galaxy mass (except for $z=4$) with stars outside $R_{\rm eff}$ having more rotation than inside.}
\label{fig:gakin_all}
\end{figure}

\begin{figure}
\centering \includegraphics[width=0.49\textwidth]{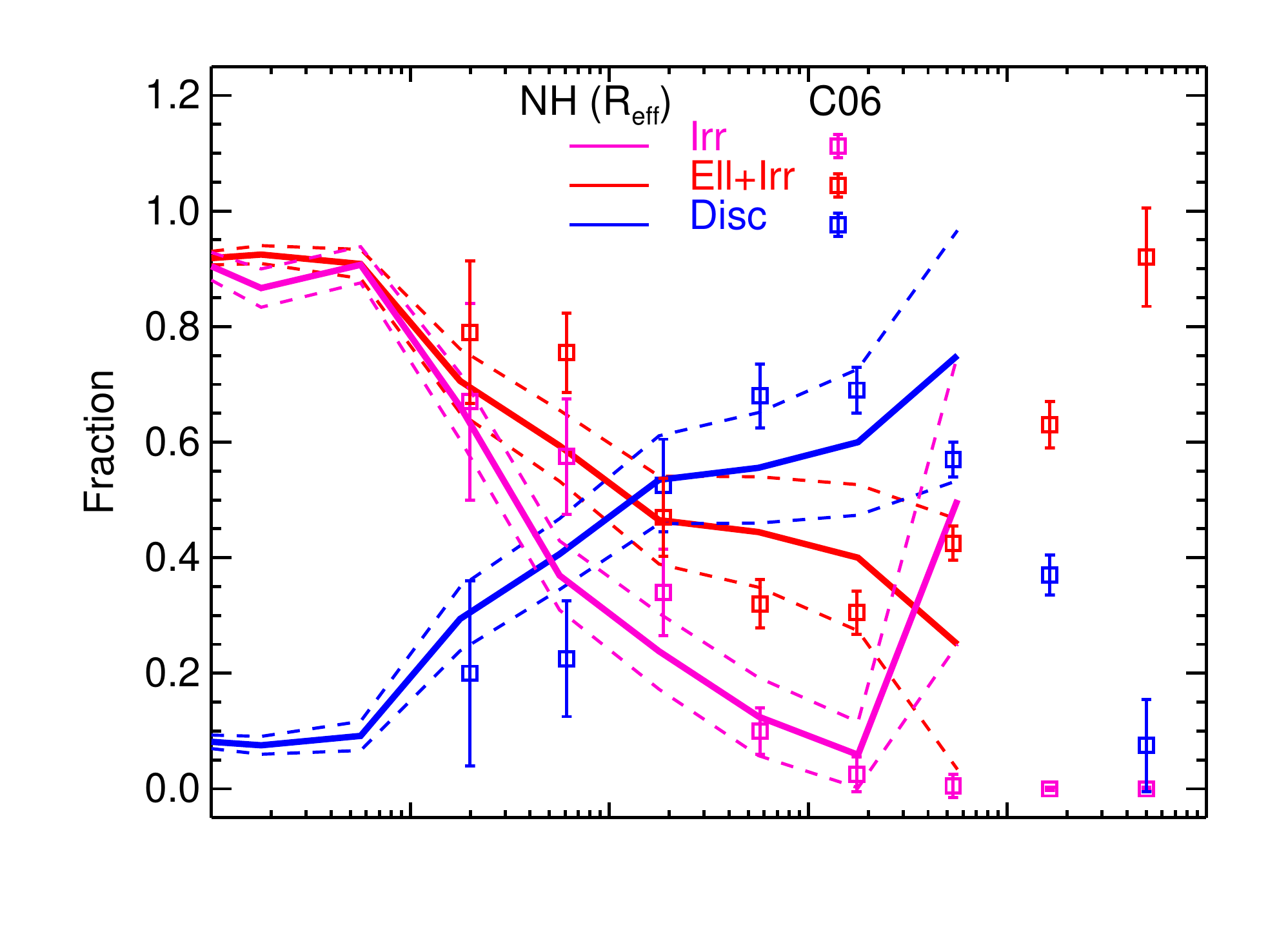}\vspace{-1.45cm}
\centering \includegraphics[width=0.49\textwidth]{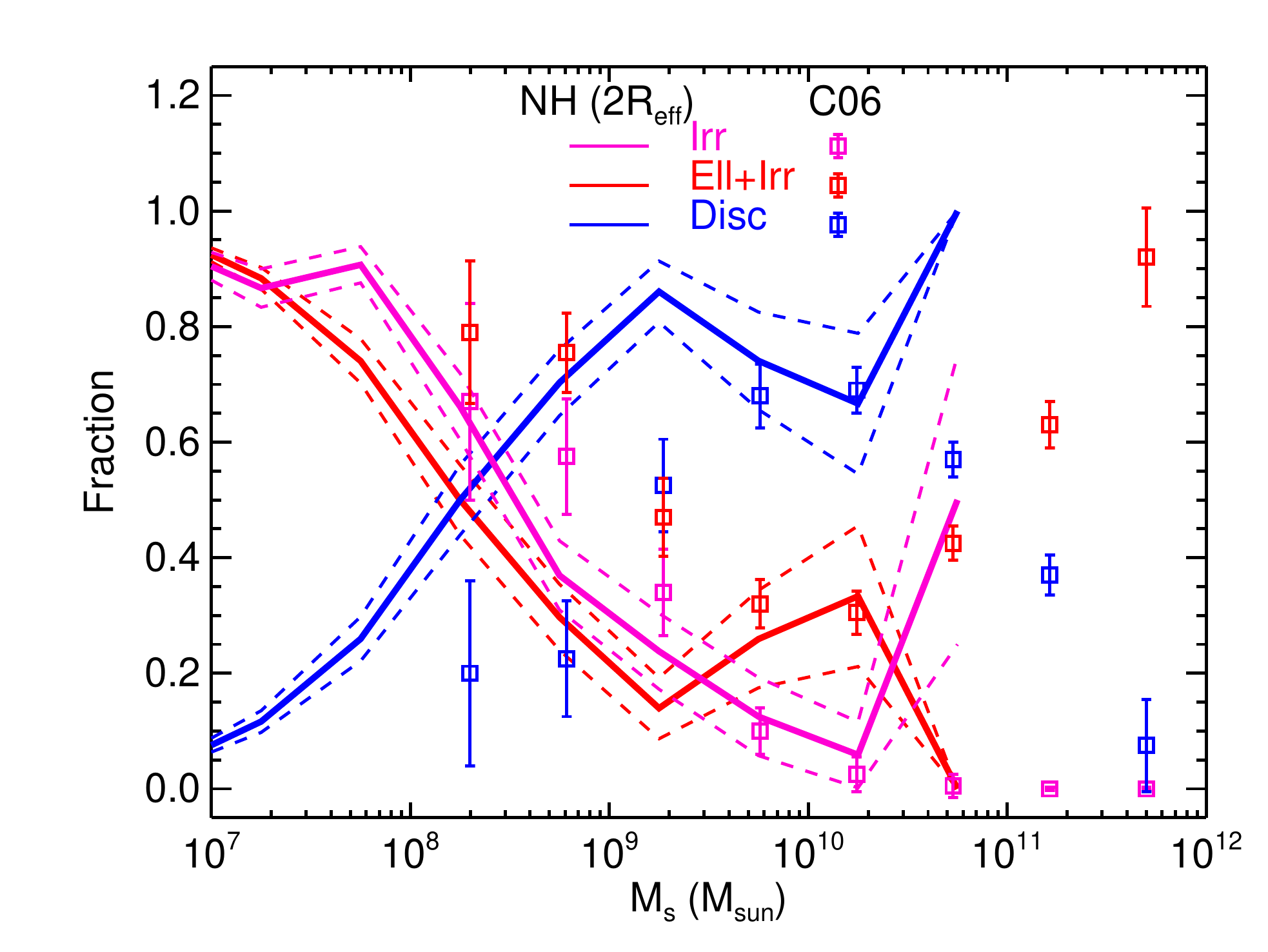}
\caption{Fraction of galaxies classified by their morphological type: $f_{\rm ell+irr}$ for elliptical$+$irregulars (red solid line; galaxies with $V/\sigma<0.5$) in \newh\ based on the stellar kinematics measured within $R_{\rm eff}$ (top) or $2R_{\rm eff}$ (bottom) and as a function of stellar mass at $z=0.25$; conversely, the fraction of discs $f_{\rm disc}=1-f_{\rm ell+irr}$ is compared to data from~\cite{conselice06}, which is at $z\simeq 0$, and is represented by symbols with error bars. The fraction of irregulars in \newh\ inferred through the asymmetry index, within the same aperture in both panels, is also shown as the magenta curve. The error bars and dashed line stand for the error on the mean. The trends of the morphological fractions with mass in \newh\ is qualitatively consistent with observational data.}
\label{fig:fellvsmg}
\end{figure}

Stellar kinematics are obtained by first computing the angular momentum vector of the stars around the centre of the galaxy. 
This vector is then used to decompose the kinematics into a cylindrical frame of reference. The stellar rotation $V$ is the average of the tangential component of velocities, while the 1D stellar dispersion $\sigma$ is the dispersion obtained from the dispersion around each mean component, that is $\sigma^2=(\sigma_{\rm r}^2+\sigma_{\rm t}^2+\sigma_{\rm z}^2)/3$. 
The kinematics are computed from the AdaptaHOP extracted stars within two different radii $R_{\rm eff}$ or $2R_{\rm eff}$ to exemplify the effect of aperture on the measured kinematics.

Figure~\ref{fig:gakin_all} shows the rotation/dispersion ratio for stars within different radii and various redshifts. 
Independently of radius, the ratio increases with decreasing redshift and with stellar mass (except at high redshift $z=4$). The stellar component is more rotationally supported over time.
We note that these relations have a significant scatter, illustrated by the distribution of individual points for $z=0.25$ in Fig~\ref{fig:gakin_all}, of around 0.3.
Galaxies also exhibit a stronger rotational support with respect to dispersion when more distant stars are taken into account. Galaxy interiors are more supported by dispersion, while the outskirts are more rotationally dominated when $V/\sigma$ is measured either in $R_{\rm eff}$ or $2R_{\rm eff}$.
This is observationally confirmed~\citep[e.g.][]{emsellemetal11,naabetal14,vandesandeetal17} and expected since central regions of galaxies are probing the bulge component mostly supported by dispersion, while the outskirts of the galaxy have a significant rotating disc components in cases where the galaxy is strongly discy.
Nonetheless, elliptical galaxies could eventually show a reverse trend, where central regions have a significantly large amount of rotation with kinematically decoupled cores~\citep{krajnovicetal13,krajnovicetal15,coccatoetal15}. For example these regions are rejuvenated by a recent episode of star formation fed by counter-rotating filaments~\citep{algorryetal14} or mergers~\citep{boisetal11,moodyetal14}, while the large stellar halo of the elliptical is more likely to be dispersion-dominated.

Alternatively it is possible to compute the fraction of dispersion-supported, elliptical, or irregular galaxies $f_{\rm ell+irr}=1-f_{\rm disc}$, by positing that a galaxy is elliptical or irregular when $V/\sigma<0.5$ (and conversely a disc when $V/\sigma\ge0.5$), where the exact threshold value is arbitrary and where in this case this value is chosen to best fit the observational data.
It has to be noted that the classification with kinematics does not allow us to distinguish between irregulars and ellipticals (both sum up to $f_{\rm ell+irr}$) but this is done using the asymmetry index (see below).
This fraction of elliptical/irregular galaxies can be compared qualitatively with observational morphological (visual) classifications from~\cite{conselice06}.
The exact value of the fraction of each morphological type with mass depends on the adopted threshold in $V/\sigma$, nonetheless the obtained trends with mass are robust against reasonable variations in the adopted thresholds as show in Appendix~\ref{Appendix:fell}.
Figure~\ref{fig:fellvsmg} shows the fraction of elliptical/irregular galaxies as a function of galaxy stellar mass at the lowest redshift of \nh,\ that is $z=0.25$ compared to the observations at $z=0$ for kinematics measured in different radii.
There is a fair agreement of the simulated data at $R_{\rm eff}$ with observations with a similar stellar mass trend. 
The agreement is weaker for $2R_{\rm eff}$, but the level of agreement also depends on the arbitrary threshold value on $V/\sigma$ adopted (here 0.5). 
There are fewer elliptical/irregular galaxies as galaxies increase in mass up to the maximum galaxy masses probed here: $M_{\rm s}\simeq10^{11}\,\rm M_\odot$.
For the most massive galaxies, owing to the limited volume of the simulation and the lack of groups of galaxies, it is impossible to conclude whether our obtained fraction of ellipticals is consistent with observations.

We further classify \newh\ galaxies as irregulars, using the asymmetry index $A_{\rm r}$ from~\cite{conseliceetal00} on the rest frame $r$-band extracted image of each individual galaxy~(see \citealt{martinetal21} for details). Since dwarf irregulars have significantly higher asymmetries than other classes of dwarf \citep[see][]{conselice14}, we define irregular galaxies using an arbitrary cut of $A_{\rm r}>0.3$; we note that we use the regular, un-smoothed, definition of asymmetry rather that the shape asymmetry as described in \cite{martinetal21}.
The exact value of $A_{\rm r}$ to be used when compared to observations might differ since $A_{\rm r}$ is sensitive to the point spread function and resolution in the observations and simulations, respectively.
However, the qualitative trend of the fraction of irregulars with stellar mass is robust against realistic variations in the threshold value of $A_{\rm r}$ (see Appendix~~\ref{Appendix:fell}).
The fraction of irregular galaxies in \newh, shown as the light blue curve in Fig.~\ref{fig:fellvsmg}, is consistent with the observational result from~\cite{conselice06} with more irregular galaxies at the low mass. 
This is the result of fewer star-forming regions, and thus it provide galaxies with more patchy and more irregular star formation and mass distribution~\citep{faucher-giguere18}.
The \nh\ data have lately been analysed in terms of morphology by~\cite{Park2019}, who use the circularity parameter to decompose disc and dispersion components of stars and pin down the origins of the discs and spheroids of spiral galaxies.

\begin{figure}
\centering \includegraphics[width=0.49\textwidth]{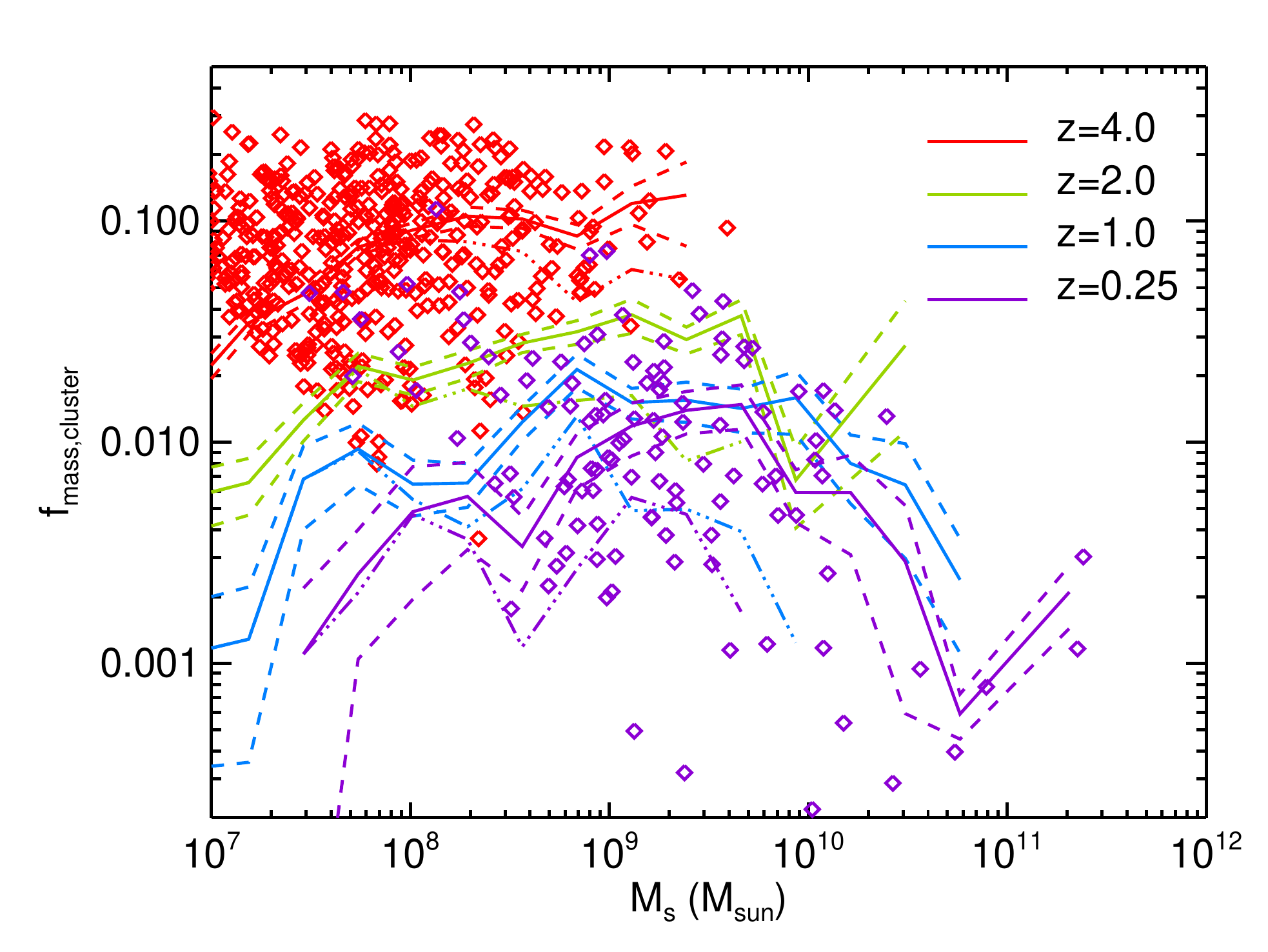}
\caption{Fraction of stellar mass in stellar clusters as a function of galaxy stellar mass at different redshifts as indicated in the panel. The solid lines indicate the mean values, and the dashed lines stand for the error on the mean. The data points are overplotted for the most extreme redshifts to appreciate the scatter in the distribution of sampled values. Galaxies at high redshift are more clumpy than at low redshift at fixed stellar mass.}
\label{fig:smass_clump}
\end{figure}

\subsection{Stellar clusters}
\label{section:clumps}

As a result of the high spatial resolution of the simulation, clumpy star formation located in large gas complexes is naturally captured (see Fig.~\ref{fig:galim_off}). 
Depending on the amount of gas in galaxies and the level of turbulence, star formation can proceed in massive clouds or in a more diffuse fashion within smaller mass clouds.
In this work, we measure the fraction of stellar mass locked into stellar clusters using a catalogue of stellar substructures, which puts a minimum detectable stellar cluster mass at $5\times 10^5\,\rm M_\odot$.
Therefore, it should be noted that a non-negligible fraction of the remaining `diffuse' stellar mass can be contained in stellar clusters with lower mass, passing below our detection threshold of the structure finder.
Also, the AdaptaHOP substructure finder considers (sub)structures solely based on the density distribution of stars; therefore stellar clusters can certainly be bound or not with some non-negligible level of contamination (stellar particles from different phase-space) from the background host.
Nonetheless, since stars form from high-gas densities, at large efficiencies for bound ($\alpha_{\rm vir}<1$) gas clumps, they form relatively bound until secular evolution~\citep[e.g.][]{gnedin&ostriker97,baumgardt&makino03} or sudden gas removal~\citep{yuetal20} eventually disrupt them.
For this analysis, we also used a lower number of star particles to detect substructures at 10 instead of 50; hence the minimum cluster mass is $10^5\, \rm M_\odot$ in that case. The main qualitative evolution with mass and redshift is not affected (see Appendix~\ref{Appendix:stellarclumpnpart10} for further comments).
It also has to be noted that, by construction, the AdaptaHOP galaxy finder cannot remove the most massive structure, which can either be a bulge, but could also be an off-centred stellar cluster in an irregular galaxy.
Finally, to minimise the contribution from satellites that are already connected to the main progenitor, only substructures within $2R_{\rm eff}$ and with a mass lower than $20\%$ that of the main galaxy are taken into account as a stellar cluster.
We note that there is a mix of accreted and in situ formed stellar clusters that contribute to the overall mass of the galaxy and we do not distinguish between these~\citep{mandelkeretal14}.

\begin{figure}
\centering \includegraphics[width=0.45\textwidth]{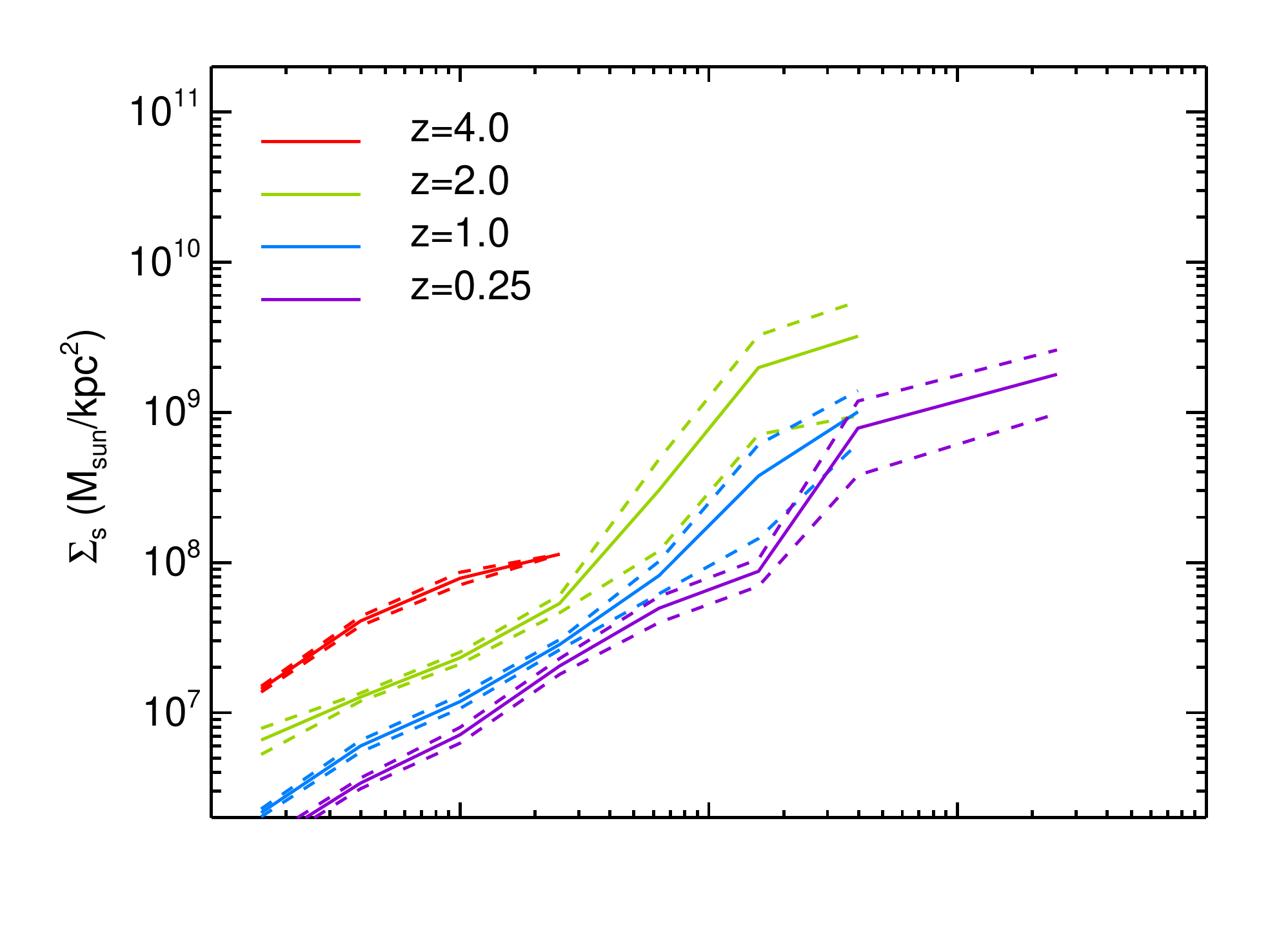}\vspace{-1.35cm}
\centering \includegraphics[width=0.45\textwidth]{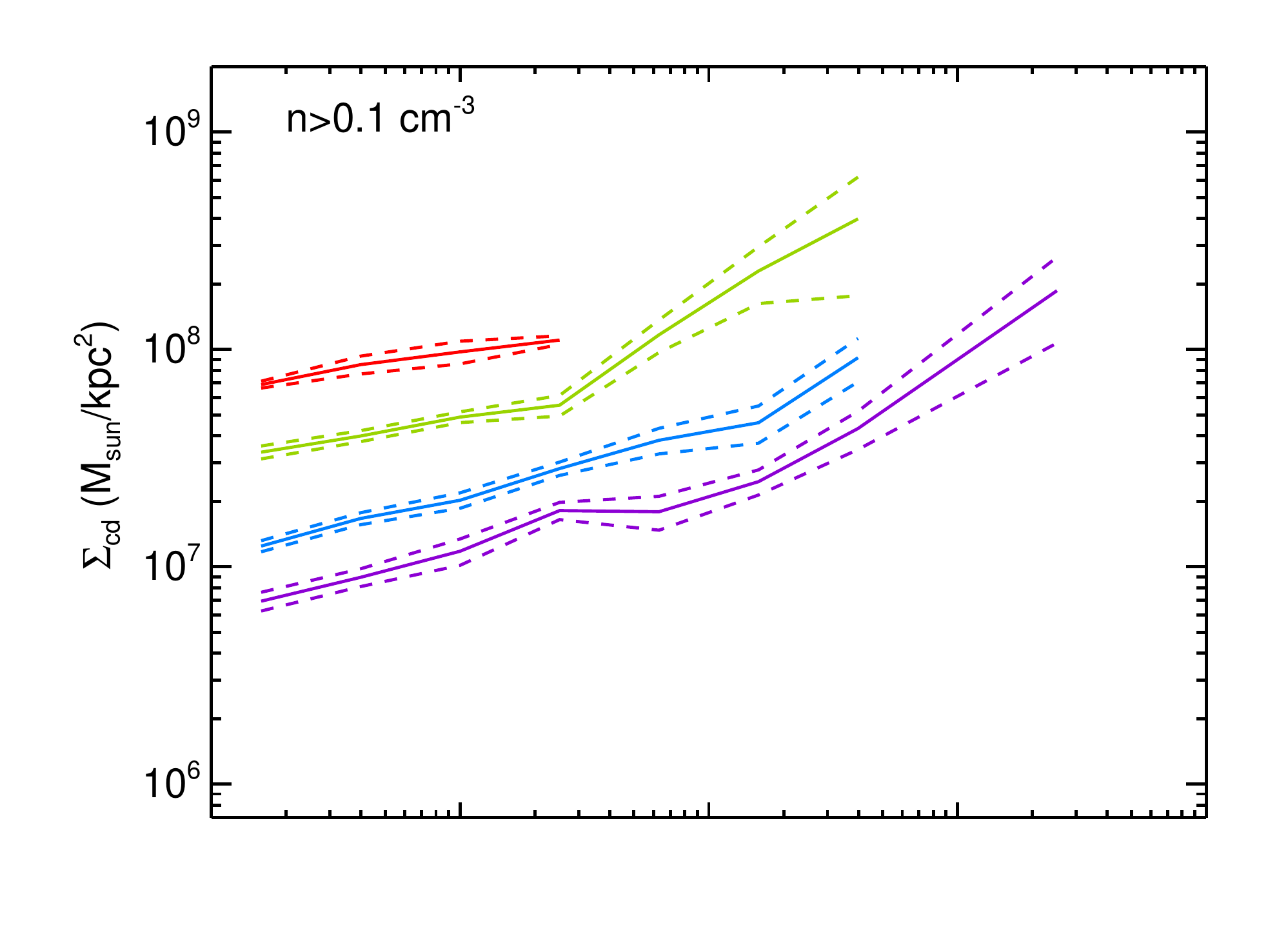}\vspace{-1.35cm}
\centering \includegraphics[width=0.45\textwidth]{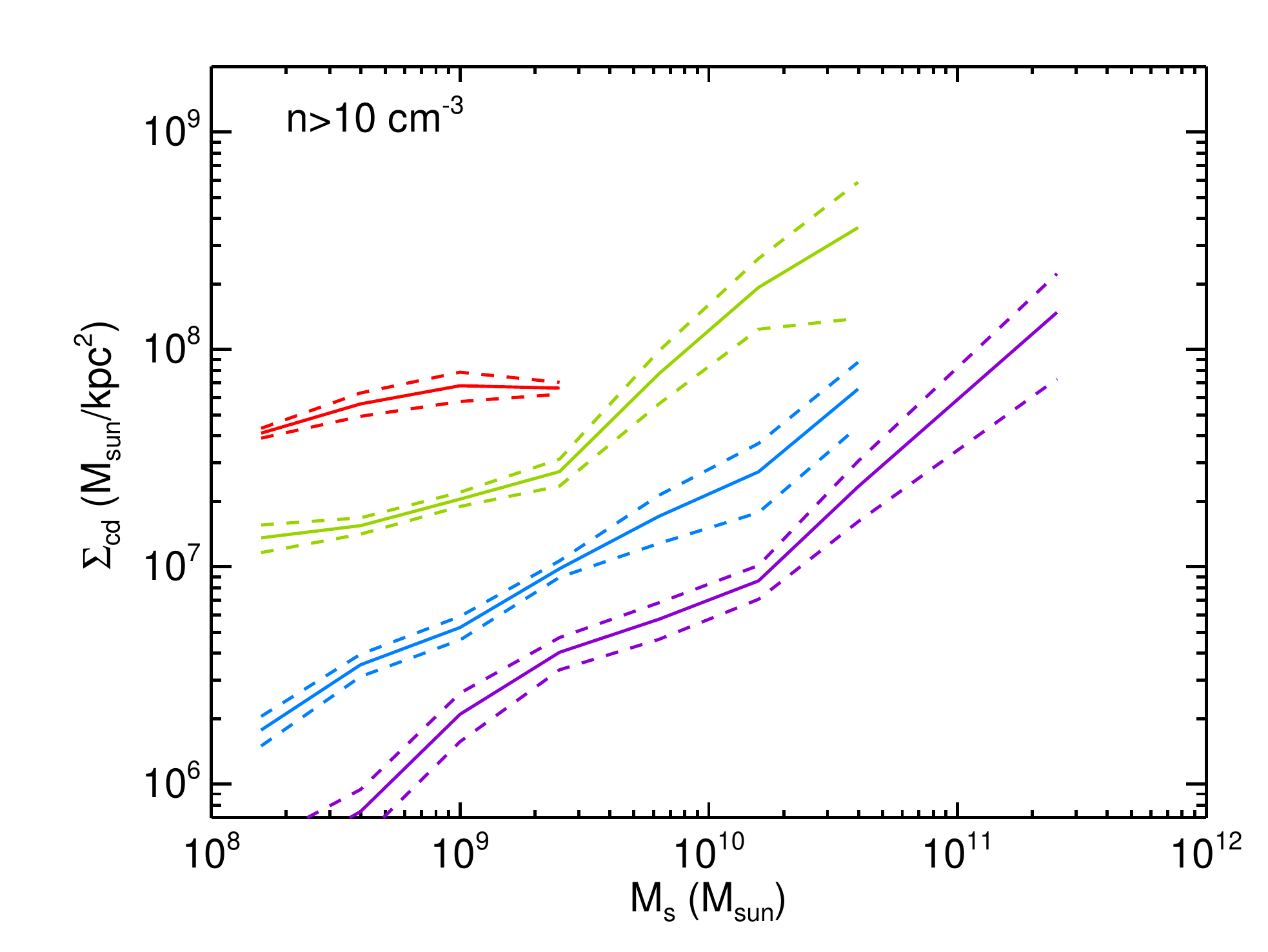}
\caption{Surface densities within $2R_{\rm eff}$ as a function of galaxy stellar mass at different redshifts for stars (top), HI+H$_2$ gas ($n>0.1 \, \rm  cm^{-3}$ and $T<2\times 10^4\, \rm K$, middle), and H$_2$ molecular gas ($n>10 \, \rm  cm^{-3}$ and $T<2\times 10^4\, \rm K$, bottom). The solid lines represent the mean values, and the dashed lines stand for the error on the mean. Galaxies are denser at high redshift at fixed stellar mass and denser in more massive galaxies at fixed redshift. }
\label{fig:surf_comp}
\end{figure}

\begin{figure*}
\centering \includegraphics[width=0.42\textwidth]{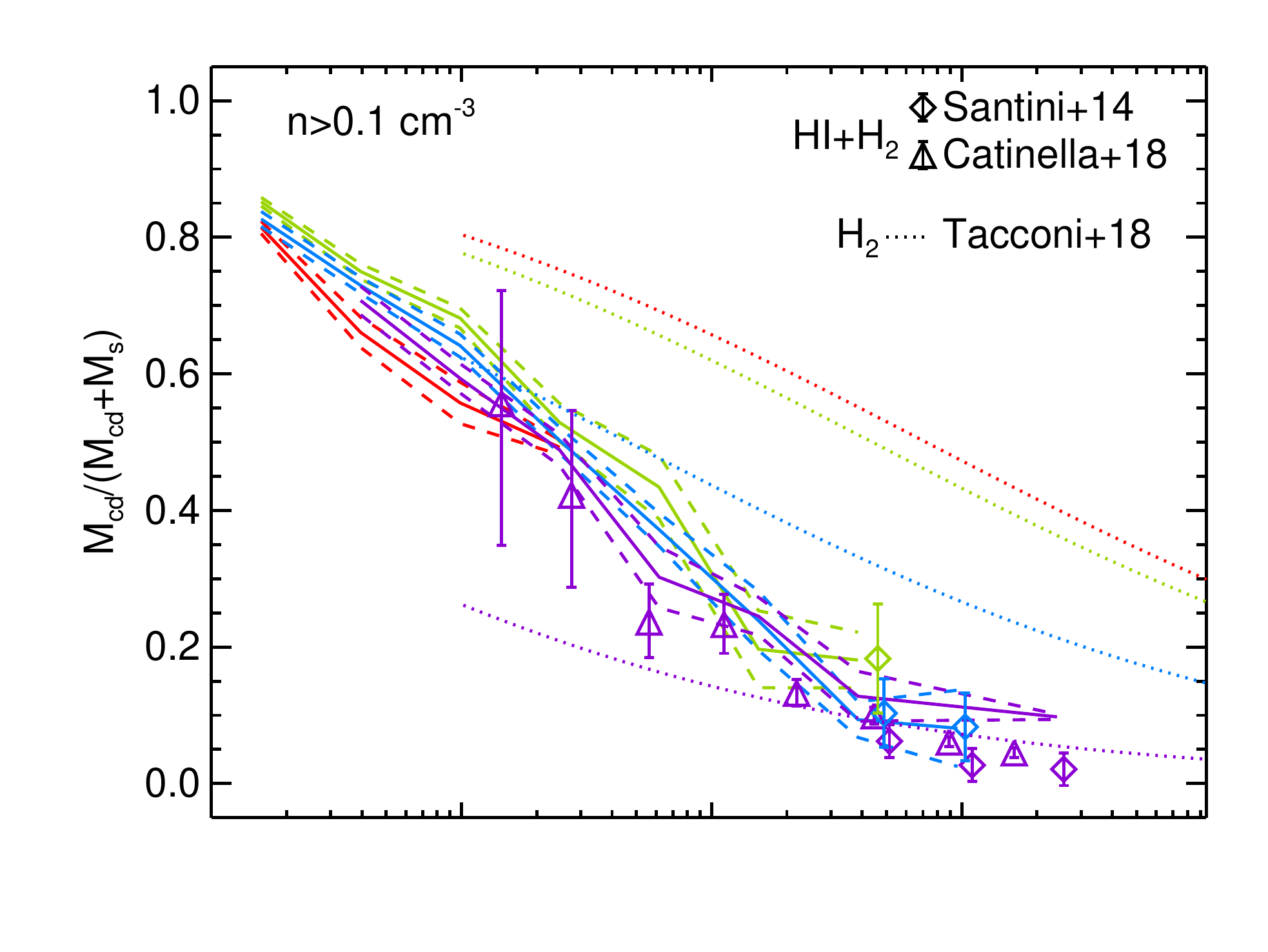}\hspace{-1.75cm}
\centering \includegraphics[width=0.42\textwidth]{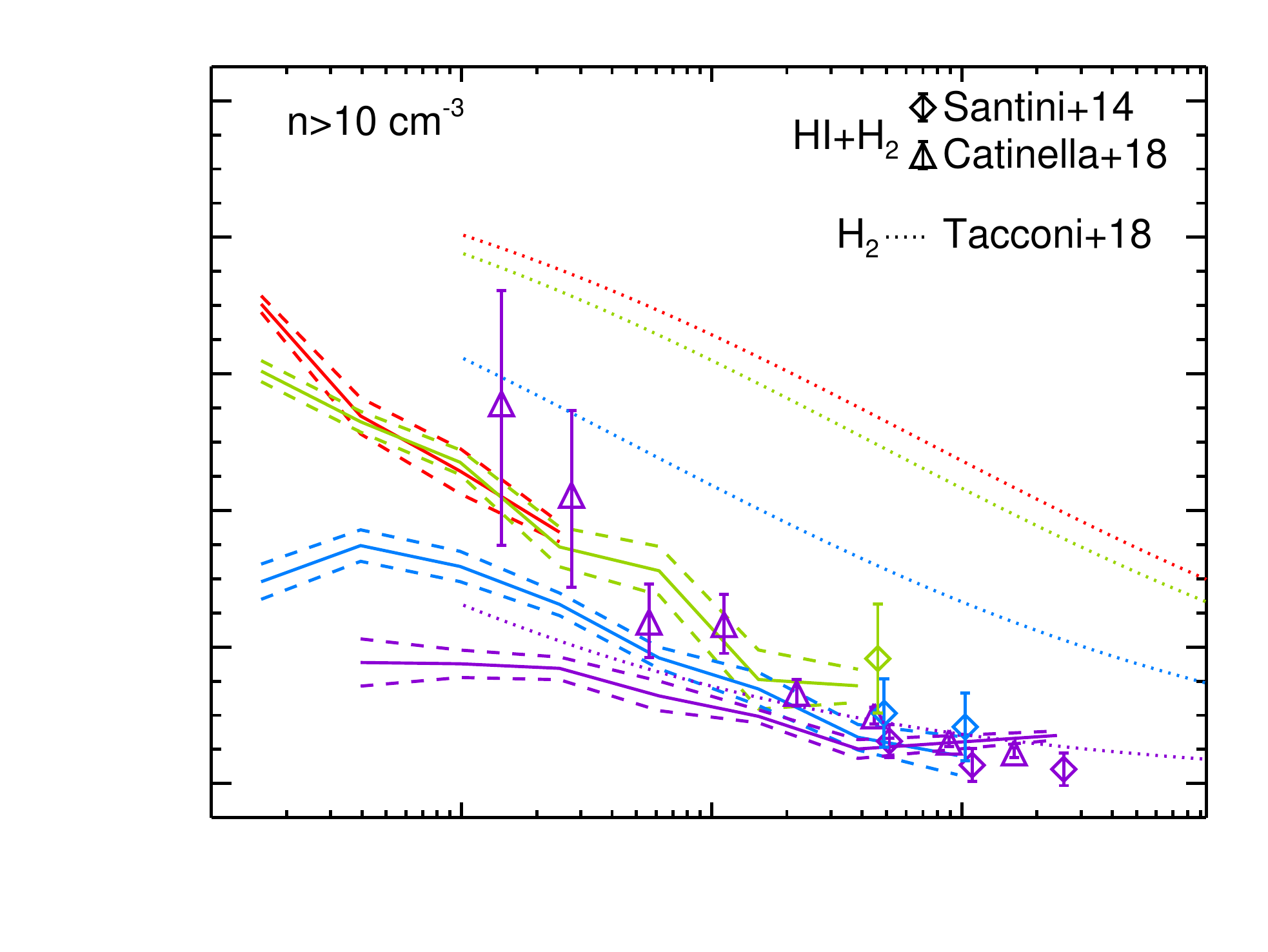}\vspace{-1.25cm}
\centering \includegraphics[width=0.42\textwidth]{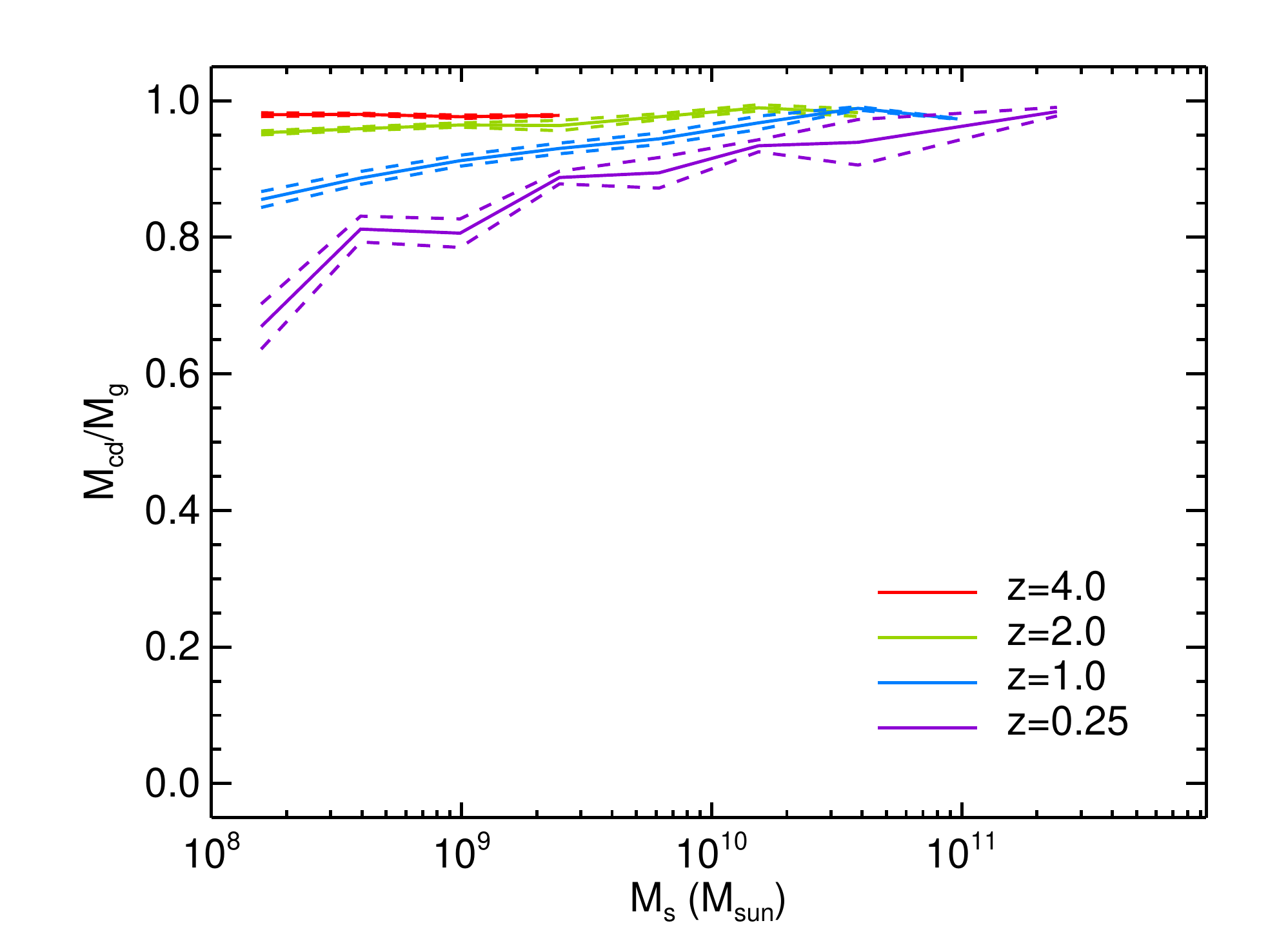}\hspace{-1.75cm}
\centering \includegraphics[width=0.42\textwidth]{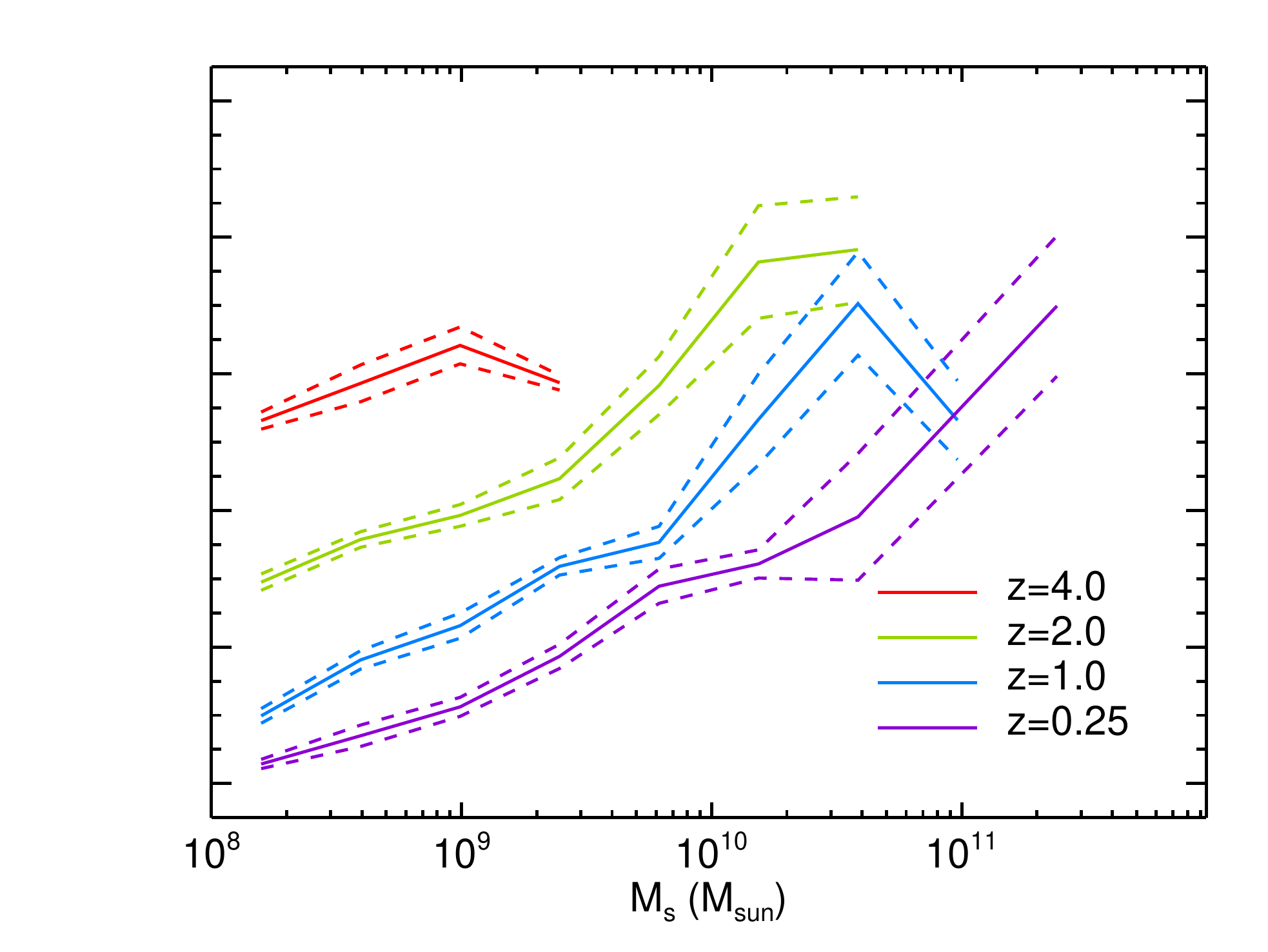}
\caption{Ratio of gas over baryon content (top panels) within $2R_{\rm eff}$ as a function of galaxy stellar mass at different redshifts for the cold dense star-forming gas with density higher than $0.1$ or $10 \, \rm  cm^{-3}$ (left and right, respectively) and temperature lower than $2\times 10^4\, \rm K$, and the ratio of this cold selected gas over total gas (bottom panels). The solid lines indicate the mean values, and the dashed lines stand for the error on the mean. The points with error bars correspond to the observations from~\cite{santinietal14} (sampled from their interpolated curves in their figure 12) at various redshifts and from~\cite{catinellaetal18} from local galaxies for the HI$+$H$_2$ gas, while the dotted line shows the data from~\cite{tacconietal18} at various redshifts for the H$_2$ molecular gas only. The \nh\ galaxies are less gas-rich at the high-mass end and also at lower redshift for a given stellar mass, in qualitative agreement with the data. But they show significantly less gas at the low-mass end compared to observations, and especially at high redshift with the data from~\cite{tacconietal18}, and in contrast they are in good agreement with the data from~\cite{santinietal14}.}
\label{fig:fbarmgal_redshift}
\end{figure*}

Figure~\ref{fig:smass_clump} shows the mass fraction of stars contained in those relatively massive clumps as a function of galaxy stellar mass and redshift.
High-redshift galaxies ($z=4$) contain a large fraction ($10\%$) of their stellar mass content within these stellar clusters, while this fraction decreases by an order of magnitude below $z<2$.
In addition, the low-mass galaxies have a lower fraction of stellar mass within stellar clusters at all redshifts, and at the two lowest redshifts ($z=1$ and $z=0.25$) there is a trend for the most massive galaxies to decrease their cluster mass fraction with respect to intermediate mass galaxies.
At the lowest redshift, many of the individual values of the cluster mass fraction at the low-mass end ($M_{\rm s}\lesssim10^9\,\rm M_\odot$) do not appear in the figure since their value is exactly zero.
It should be recalled that the exact value of the cluster mass fraction is affected by the capability to capture the formation and survival~\citep[see e.g.][and references therein]{pfefferetal18} of the lowest mass stellar clusters due to limited mass and spatial resolution of the simulation, and to detect the stellar overdensities with AdaptaHOP.
Despite such resolution effects, the qualitative trend of increasing mass fraction of stars inside cluster with redshift is expected~\citep{elmegreen&elmegreen05,genzeletal11} as a result of more gas-rich, compact, and turbulent galaxies that are increasingly more gravitationally unstable~\citep{cacciatoetal12,inoueetal16}, thereby forming stars into more numerous massive clusters and more efficiently; this is confirmed in sections~\ref{section:fgas} and~\ref{section:gaskin}.

\subsection{Baryonic content}
\label{section:fgas}

We decompose the baryonic content of galaxies by measuring the amount of stars and gas within twice the effective radius of galaxies; these values are also obtained within $R_{\rm eff}$ or $0.1R_{\rm vir}$ and show similar behaviour.
The gas content is further decomposed into a cold and dense gas component (identified by a subscript `cd' in gas quantities) by considering only the star formation gas with density $n\ge0.1,10 \, \rm  cm^{-3}$ and temperature $T<2\times 10^4\,\rm K$.
In the following, we distinguish between the neutral HI$+$H$_2$ gas component with the density cut-off at $0.1 \, \rm  cm^{-3}$, and the H$_2$ molecular dense component at $10 \, \rm  cm^{-3}$~\citep[e.g.][]{lupietal18,nickersonetal19}. An exact match of the ionisation and molecular states of the gas however would require a detailed treatment of radiative transfer and molecular chemistry.

In Fig.~\ref{fig:surf_comp}, we show the gas and stellar surface densities as a function of stellar mass and redshift. 
Surface densities are obtained by dividing the corresponding mass content by $\pi (2R_{\rm eff})^2$.
All surface densities -- gaseous $\Sigma_{\rm g}$, cold $\Sigma_{\rm cd}$, and stellar $\Sigma_{\rm s}$ -- increase with mass and decrease with redshift; the cold component of the gas surface density decreases faster with time than the total gas component.
The decrease of gas and stellar surface densities are consistent with less concentrated galaxies (large effective radius) over time and galaxies with lower sSFR as shown in the previous sections. 

Figure~\ref{fig:fbarmgal_redshift} shows the fraction of baryons locked into the cold gas as a function of galaxy stellar mass for different redshifts and for different gas density cut-offs.
The total cold gas fraction ($M_{\rm cd}/(M_{\rm cd}+M_{\rm s})$ with $n>0.1\,\rm cm^{-3}$; top left panel of Fig.~\ref{fig:fbarmgal_redshift}) strongly decreases with galaxy mass but does not evolve significantly over redshift.
The fraction of cold and denser ($n>10\,\rm cm^{-3}$; top right panel) gas shows a weaker variation with mass, in particular at low redshift, and its value is particularly low compared to the total amount of gas ($M_{\rm cd}/M_{\rm g}$, bottom right), although the fraction of gas made of dense ($n>10\,\rm cm^{-3}$) material increases with stellar mass.
Nonetheless, the fraction of cold dense gas decreases with decreasing redshift at a given stellar mass.

Observations of star-forming H$_2$ gas, via CO measurements, at $z \sim 1-4$ have shown that 
at a given stellar mass, galaxies at high redshifts tend to be  significantly more gas rich compared to their present-day counterparts \citep[e.g.][]{tacconietal06,tacconietal08,daddietal10a,tacconietal10,tacconietal13,scovilleetal17,tacconietal18,tacconietal20}. Reported baryonic gas fractions range from $\sim$ 20-80\%, a factor of $\sim$2-3 more than typically found in cosmological models of galaxy formation \citep[e.g.][]{poppingetal14,lagosetal15,daveetal17,daveetal19}. This is confirmed in this work  with the comparison of the \nh\ molecular gas to~\cite{tacconietal18} at high redshifts (top right panel of Fig~\ref{fig:fbarmgal_redshift}).
\cite{narayananetal2012} suggested that the observationally inferred gas fractions of star-forming galaxies at high redshift are overestimated because of the adoption of locally calibrated conversion factors $\alpha_{\rm CO}$. When applying a smoothly varying $\alpha_{\rm CO}$ with the physical properties of galaxies (essentially gas-phase metallicity and CO surface brightness), these authors found gas fractions of $\sim$ 10-40\% in galaxies of stellar masses in the range $10^{10}-10^{12} \, \rm M_\odot$ and a reduced scatter in stellar mass versus gas fraction by a factor of $\sim$ 2, bringing models and observations into a much better agreement. 
Similarly lower values of gas fractions have been found by \cite{santinietal14}, who study main-sequence, star-forming galaxies with 
$M_{\rm s} > 10^{10}\,\rm M_\odot$ at $z < 2$ and infer gas content from dust mass measurement, or by  
\cite{conselice13}, who derive the cold gas mass fraction by inverting the global Kennicutt-Schmidt relation for massive galaxies ($M_{\rm s} > 10^{11}\,\rm M_\odot$) at $1.5 < z < 3$. 
In the range of comparable stellar masses, gas fractions
of high-z \nh\ galaxies agree broadly with these revised measurements, hosting $\sim$ 10-40\% of star-forming gas that decreases
with increasing stellar mass and evolves weakly with redshift.
In Fig.~\ref{fig:fbarmgal_redshift}, we show a few data points obtained from the sampling of the interpolated curves from \cite{santinietal14} (see their Fig. 12). 
While their dust method used to derive the gas content of galaxies in principle traces both atomic and molecular components, the authors suggest that the bulk of the gas in studied galaxies is in the molecular phase. 
We also show in Fig.~\ref{fig:fbarmgal_redshift}, the local scaling relation from SDSS galaxies from~\cite{catinellaetal18}.\ The
\nh\ cold gas fractions agree fairly well with both local and high-z gas scaling relations for galaxy masses above a few $10^{10}\,\rm M_\odot$, but significantly differ with the local scaling relation at the low-mass end.

\begin{figure}
\centering \includegraphics[width=0.45\textwidth]{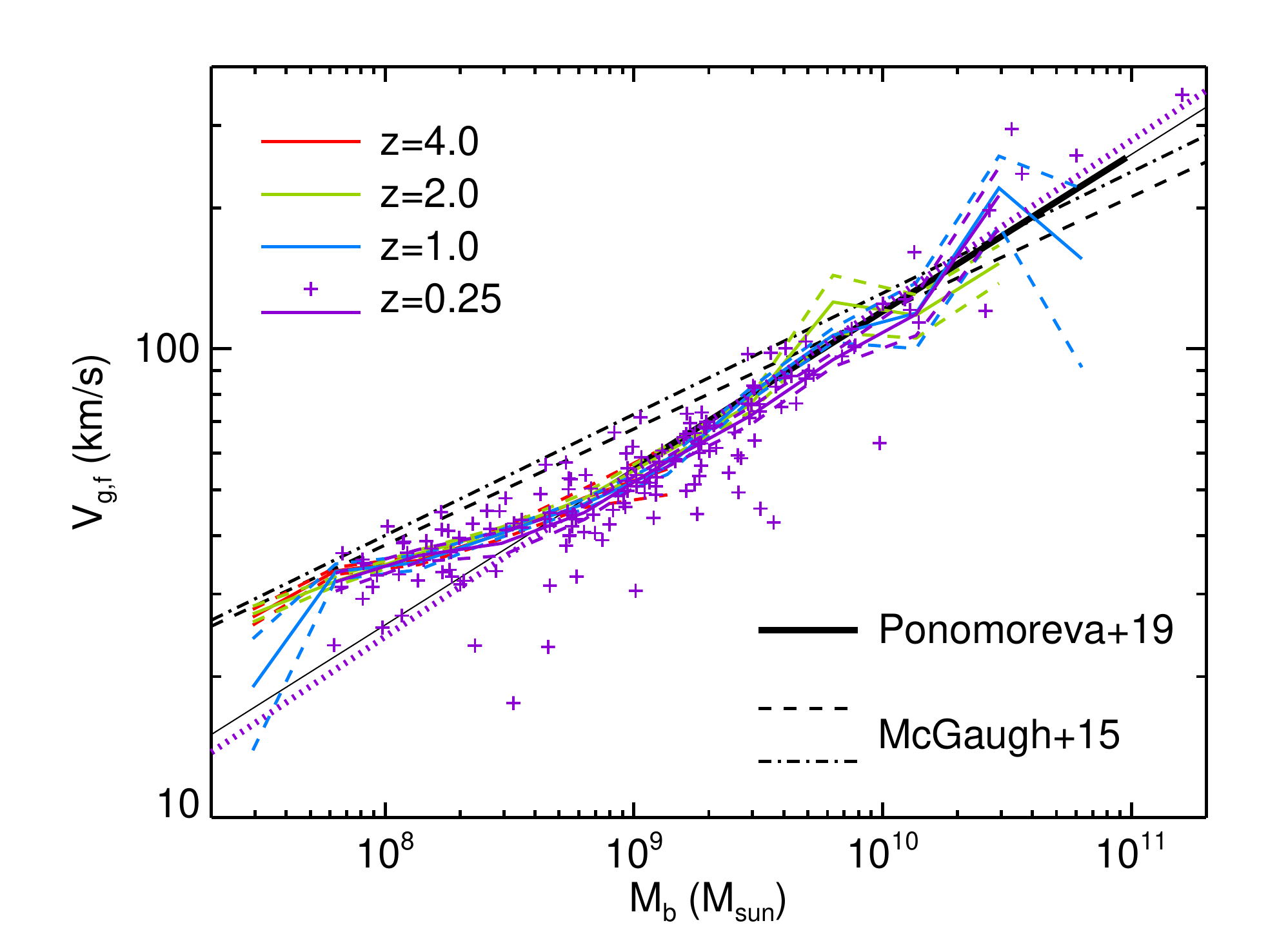}
\caption{Baryonic Tully-Fisher relation for disc galaxies in \nh\ at different redshifts as labelled in the panel with the raw data points overplotted as crosses for redshift $z=0.25$ (purple plus signs), compared to the two best fits from~\cite{mcgaugh&schombert15} (dashed black and  dot-dashed black) and the best fit from~\cite{ponomarevaetal18} (solid black; in thick their fit over the range of their data points, while the thin part is the extrapolated part). The fit to the \nh\ data at $z=0.25$ (thick dotted purple line)  shows a better level of agreement with the data from~\cite{ponomarevaetal18} (with a slope of  one-third) over those of~\cite{mcgaugh&schombert15} (slope of one-fourth).}
\label{fig:tullyfisher}
\end{figure}

\subsection{Tully-Fisher relation}

To measure the baryonic~\cite{tully&fisher77} relation~\citep[e.g.][]{bell&dejong01,mcgaugh&schombert15,ponomarevaetal18,lellietal19} in simulated galaxies, we use the decomposition of the gas kinematics in a cylindrical frame of reference. 
In this frame, rotation velocity profile is measured using only (cold) gas resolution elements with densities larger than $0.1\,\rm cm^{-3}$ and temperatures below $2\times 10^4\,\rm K$.
The flat rotational velocity involved in the baryonic Tully-Fisher relation $V_{\rm g,f}$ is obtained by measuring the average gas rotational velocity within [$1.8R_{\rm eff}$,$2.2R_{\rm eff}$].
The total baryonic mass corresponds to the total stellar and cold gas mass $M_{\rm b}=M_{\rm s}+M_{\rm g}$ within $2R_{\rm eff}$.
A small fraction of galaxies in the sample do not have cold gas that reaches the $2R_{\rm eff}$ radius. In that case,  the measured kinematics and baryonic mass are replaced by their value at $R_{\rm eff}$.
We select only disc galaxies with $V/\sigma \ge 0.5$  that are hosted within DM halos more massive than $10^9\,\rm M_\odot$; however, considering
non-disc galaxies leads to a similar relation. 

\begin{figure*}
\centering \includegraphics[width=0.45\textwidth]{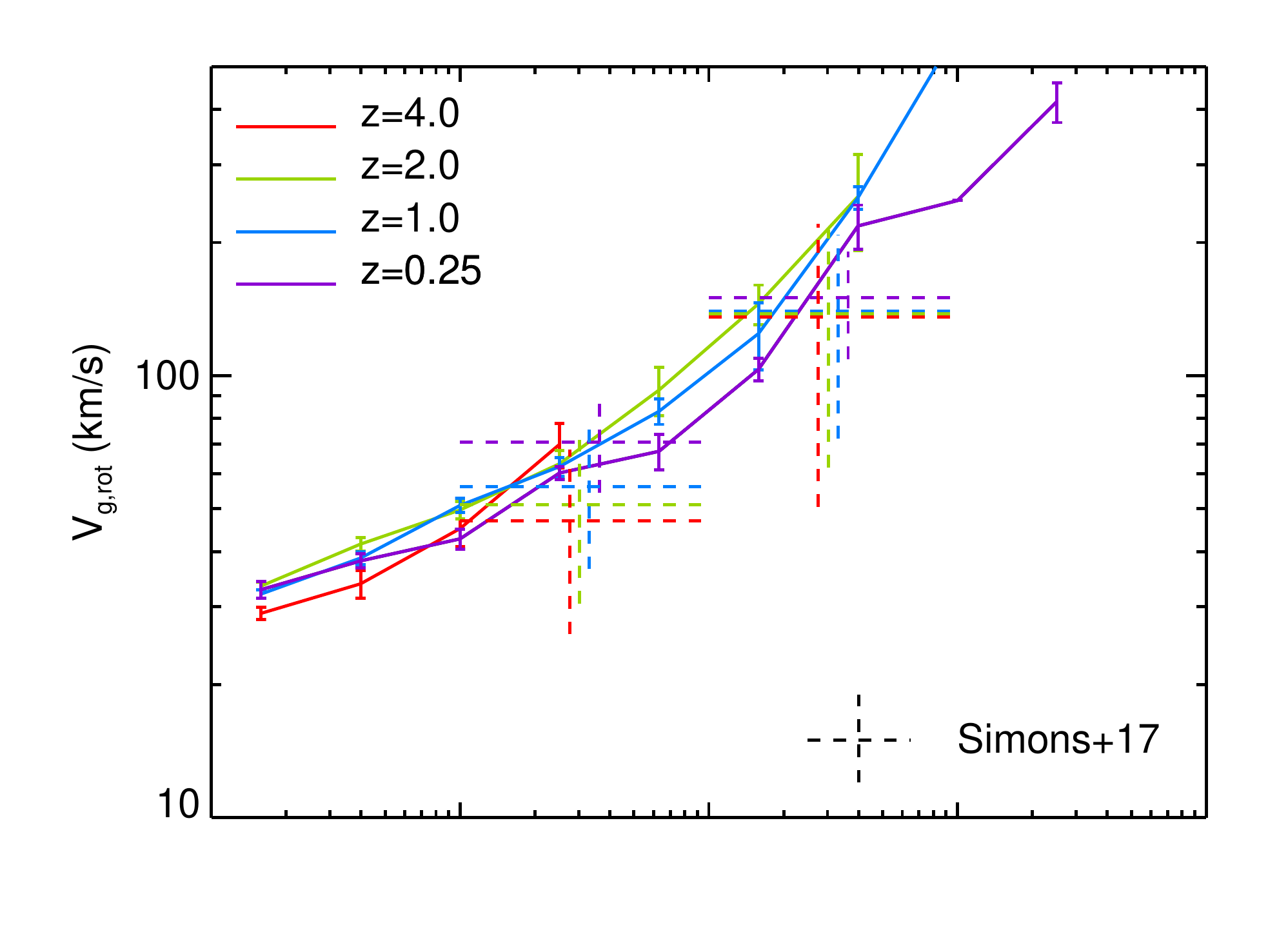}\hspace{-0.3cm}
\centering \includegraphics[width=0.45\textwidth]{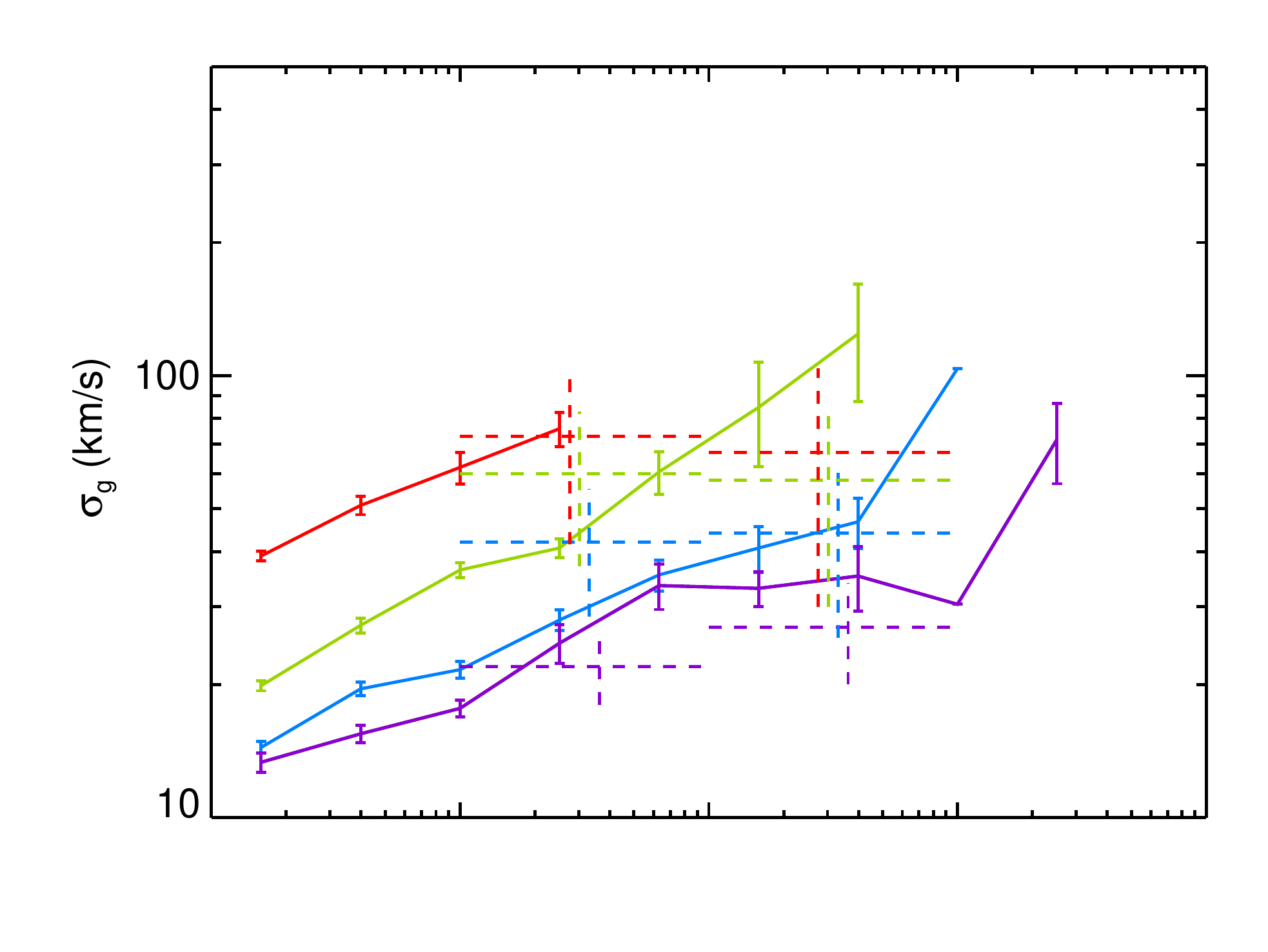}\vspace{-1.3cm}\\
\centering \includegraphics[width=0.45\textwidth]{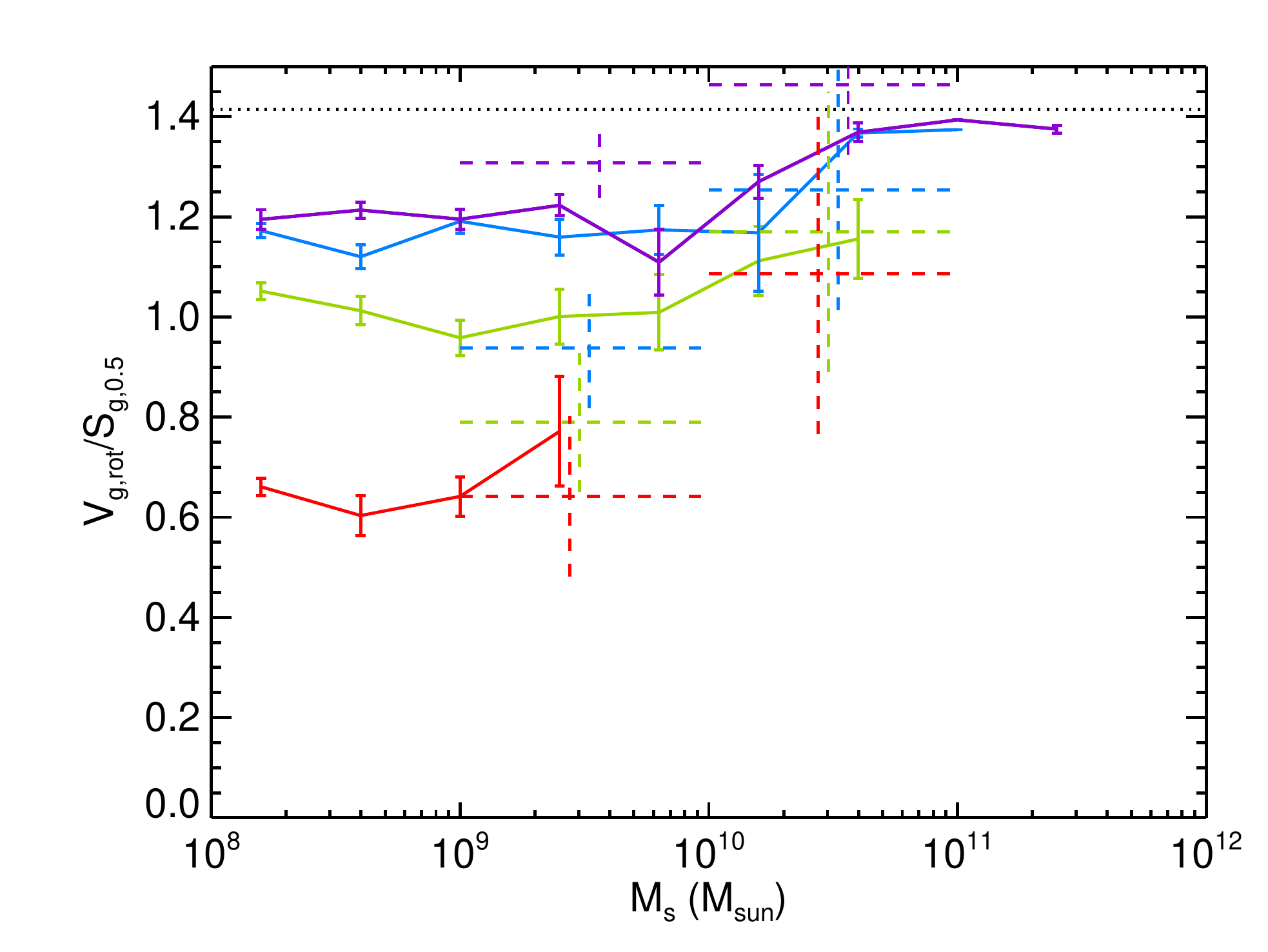}\hspace{-0.3cm}
\centering \includegraphics[width=0.45\textwidth]{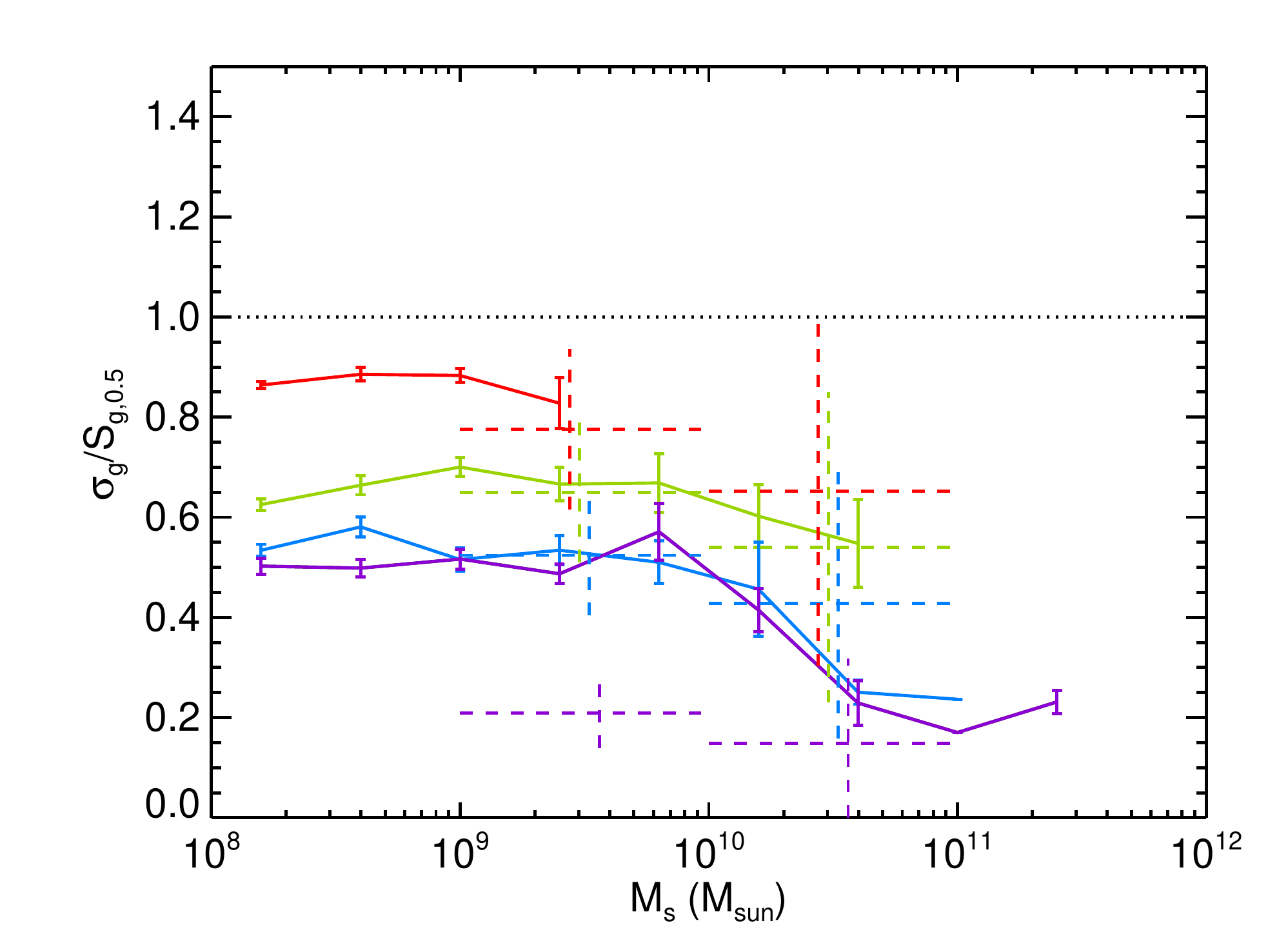}
\caption{Gas kinematics as a function of galaxy stellar mass at different redshifts for gas density  $0.01<n<10 \, \rm cm^{-3}$, temperature lower than $T<2\times10^4\, \rm K$ and within $2R_{\rm eff}$. From left to right and top to bottom: rotational velocity, velocity dispersion, velocity dispersion over $S_{\rm g,0.5}$, and rotational velocity over $S_{\rm g,0.5}$. The error bars stand for the error on the mean. The dashed plus signs in the top right panel indicate the values with uncertainties from the fit to the observational data in~\cite{simonsetal17}.
The dotted lines on the bottom panels stand for the cases of pure rotational support or pure dispersion support, respectively. The \nh\ galaxies decrease their amount of gas velocity dispersion over time for a given stellar mass, while keeping a similar amount of rotation.}
\label{fig:gaskin_comp}
\end{figure*}

Figure~\ref{fig:tullyfisher} shows the mean baryonic Tully-Fisher relation at different redshifts $z=4,2,1,0.25$ in \nh\, with the two  observational fits by~\cite{mcgaugh&schombert15} with $m\sim4$ ($V_{\rm g,f}\propto M_{\rm b}^{1/m}$) from their different combined samples of disc galaxies, and with the observational fit from~\citet{ponomarevaetal18} with $m\sim3$.
The obtained \nh\ baryonic Tully-Fisher relation shows very little evolution with redshift; the mean relations are almost indistinguishable in Fig.~\ref{fig:tullyfisher}.
Although the bulk of the relation is captured, there are a few different behaviours of the \nh\ sample with the observations, in particular with respect to the~\cite{mcgaugh&schombert15} relations: at masses between $10^8\lesssim M_{\rm b}/{\rm M_\odot}\lesssim 10^{10}\,\rm M_\odot$, \nh\ galaxies have velocities below the observed values, and the slope of the relation $V_{\rm g,f}$--$M_{\rm b}$ becomes steeper for masses larger than $10^{10}\,\rm M_\odot$.
A similar discrepancy (bending of the relation) with observations was also found in NIHAO simulations~\citep{duttonetal17} or APOSTLE/EAGLE~\citep{salesetal17}, although with different overall normalisation of the relation relative to ours.
The baryonic Tully-Fisher relation obtained from our \nh\ data at $z=0.25$ produces a slope of $m=2.8\pm 0.1$ that agrees best with the $m\sim 3$ slope from~\cite{ponomarevaetal18}.
It has to be noted that the observational slope of the relation varies with the adopted stellar mass-to-light ratio: whether a (molecular) gas component remains undetected, depending on the mass range of the sample~\citep[see e.g.][for a discussion along these lines]{ponomarevaetal18}, and how the velocity is measured. For instance in~\cite{lellietal19}, this value is $m=3.14$ if the velocity is measured at $2R_{\rm eff}$ as we adopted in this work, while it is $m=3.85$ for the velocity in the flat part of the radial velocity profile.
For comparison, the baryonic Tully-Fisher in SIMBA~\citep{glowackietal20}  differs significantly from the value in this work with a value of $m\simeq4$ rather than $m\simeq3$; the SIMBA value has a larger normalisation of the relation than in~\cite{mcgaugh&schombert15} as opposed to this work, where the normalisation is lower in most of the mass range. \cite{glowackietal20} also show that the value of the slope is sensitive to the velocity estimator.
In the EAGLE simulation,~\cite{ferreroetal17} show that the stellar Tully-Fisher relation is in excellent agreement with the observational data, even though the stellar (using $M_{\rm s}$), as opposed to the baryonic (using $M_{\rm b}$), Tully-Fisher relation, provides a less tight relation between the mass and the rotational velocity of galaxies~\citep{mcgaughetal12}.

\subsection{Gas kinematics}
\label{section:gaskin}

The kinematics of the ISM is computed by measuring gas velocities within twice the effective radius of each galaxy and for the cold non-star-forming gas only (i.e. $10^{-2}\le n/{\rm cm^{-3}} < 10$ and $T\le 2\times 10^4\,\rm K)$, which has similar properties to that observed in~\cite{weineretal06} or~\cite{simonsetal17}, who estimated the gas kinematics from nebular emission lines (amongst which H$\alpha$ and $[{\rm O \, III}]\lambda5007$).
The kinematics of the gas are decomposed along the cylindrical system of coordinates, which for each galaxy is given by the angular momentum of the selected gas elements.
Therefore, the velocity dispersion is decomposed along the three components $\sigma_{\rm g, r}$, $\sigma_{\rm g,z}$, and $\sigma_{\rm g,t}$ and is simply the mass-weighted dispersion of each velocity component around the mean value of that component; the total velocity dispersion of the gas is obtained from the three components $\sigma_{\rm g}^2=(\sigma_{\rm g,r}^2+\sigma_{\rm g,z}^2+\sigma_{\rm g,t}^2)/3$.
Similarly the rotational velocity is obtained by taking mass-weighted mean tangential velocity component $V_{\rm g, rot}=\bar{V}_{\rm g, t}$.
A proxy for the total kinematics support is built out of the dispersion and rotation using $S_{\rm g,0.5}=(0.5V_{\rm g,rot}^2+\sigma_{\rm g}^2)^{1/2}$~\citep[e.g.][]{weineretal06,kassinetal07}.
Although synthetic mocks of nebular emission lines with instrumental effects from which observed kinematics are reconstructed would certainly show differences with the direct gas kinematics measured from the simulation, 
these raw measurements on \nh\ galaxies have been tested against projection effects on $\sigma_{\rm g}$ (i.e. versus $\sigma_{\rm g,z}$), density cut-offs (i.e. versus selected gas with $n>0.1\,\rm cm^{-3}$ and $T\le 2\times 10^4\,\rm K$) or aperture (i.e. versus within $R_{\rm eff}$), which changes the estimated averaged values shown in Fig.~\ref{fig:gaskin_comp} by at most 10-20$\%$ and does not affect any of the trends obtained with mass and redshift.

Figure~\ref{fig:gaskin_comp} shows the amount of dispersion and rotational support of the cold gas as a function of galaxy mass for different redshifts.
The dispersion, rotation, and total support all increase with galaxy mass at any given redshift; however, the behaviour of these quantities over time differ significantly.
The velocity dispersion of the cold gas decreases with decreasing redshift at fixed galaxy mass, such as for galaxies with stellar mass $M_{\rm s}=10^{9}\,\rm M_\odot$, values as high as $60\,\rm km\,s^{-1}$ (and up to $100\,\rm km\,s^{-1}$ for the most massive galaxies at high redshift) at $z=4$ and $20\,\rm km\,s^{-1}$ at $z=0.25$, which agrees well with the observations~\citep{simonsetal17}.
In agreement with the observations, the rotation of cold gas in galaxies as a function of mass does not significantly evolve over time, although the small evolution with time is opposite to the small evolution in observations. 
In turn, the cold gas in galaxies is proportionally more supported by rotation ($V_{\rm g,rot}/S_{\rm g,0.5}$) over time as an effect of reduced gas velocity dispersion that is qualitatively consistent with the observations~\citep{simonsetal17}.
The ratio of $V_{\rm g,rot}/S_{\rm g,0.5}$ increases over time (while $\sigma_{\rm g}/S_{\rm g,0.5}$ decreases) and also has a sharp transition, that is an increase and a decrease, respectively, at stellar masses above $10^{10}\,\rm M_\odot$~\citep{simonsetal15}.
The obtained values of these ratios are in good quantitative agreement with~\cite{simonsetal17} for the high-mass range ($M_{\rm s}>10^{10}\,\rm M_\odot$), but $V_{\rm g,rot}/S_{\rm g,0.5}$ at $M_{\rm s}<10^{10}\,\rm M_\odot$ is two standard deviations larger the observational relation at $z=1$ and 2, and lower at $z=0.25$.
The turbulent support ($\sigma_{\rm g}/S_{\rm g,0.5}$) agrees well with the observations except for the low mass range $M_{\rm s}<10^{10}\,\rm M_\odot$ at $z=0.25$, where the amount of turbulent support is significantly larger than observed.
This decrease of the gas dispersion and hence of the ratio of dispersion to rotation over time and mass triggers the settling of galactic discs~\citep{kassinetal07,kassinetal14,ceverinoetal17,simonsetal17,pillepichetal19} as galaxies become more quiescent (see Dubois et al., in prep.).
Compared to TNG50 \citep{pillepichetal19}, \nh\ galaxy gas kinematics produce smaller rotational velocities and higher turbulent velocities at all redshifts: the rotational velocity in \nh\ is about half of that in TNG and the turbulent velocities about twice higher than in TNG.
Interestingly, TNG50 gas fractions are also a factor two smaller than in \nh, with galaxies that have a similar size-mass relation.
Therefore, the gas Toomre parameter $Q_{\rm g}$, which becomes simply proportional to $\sigma_{\rm g}/\Sigma_{\rm g}$, is of the same value, pointing towards a common saturation mechanism of the (gas) Toomre parameter despite the different models of feedback.

Turbulence can be sustained by several sources including stellar feedback~\citep{joungetal09,ostriker&shetty11,hopkinsetal14}, gravitational instabilities~\citep{agertzetal09,bournaudetal10}, or cosmological infall~\citep{elmegreen&burkert10,klessen&hennebelle10} ranging from anisotropic gas infall or mergers. 
It is not clear which process dominates the driving of the turbulence. However 
\cite{krumholz&burkhart16} and \cite{krumholzetal18} 
show that observations favour a gravity-driven source for gas-rich (or high-z) galaxies and a stellar-driven source of turbulence for gas-poor (low-z) galaxies, although the role of cosmic infall is not tested in those models.

\begin{figure}
    \centering
    \includegraphics[width=0.45\textwidth]{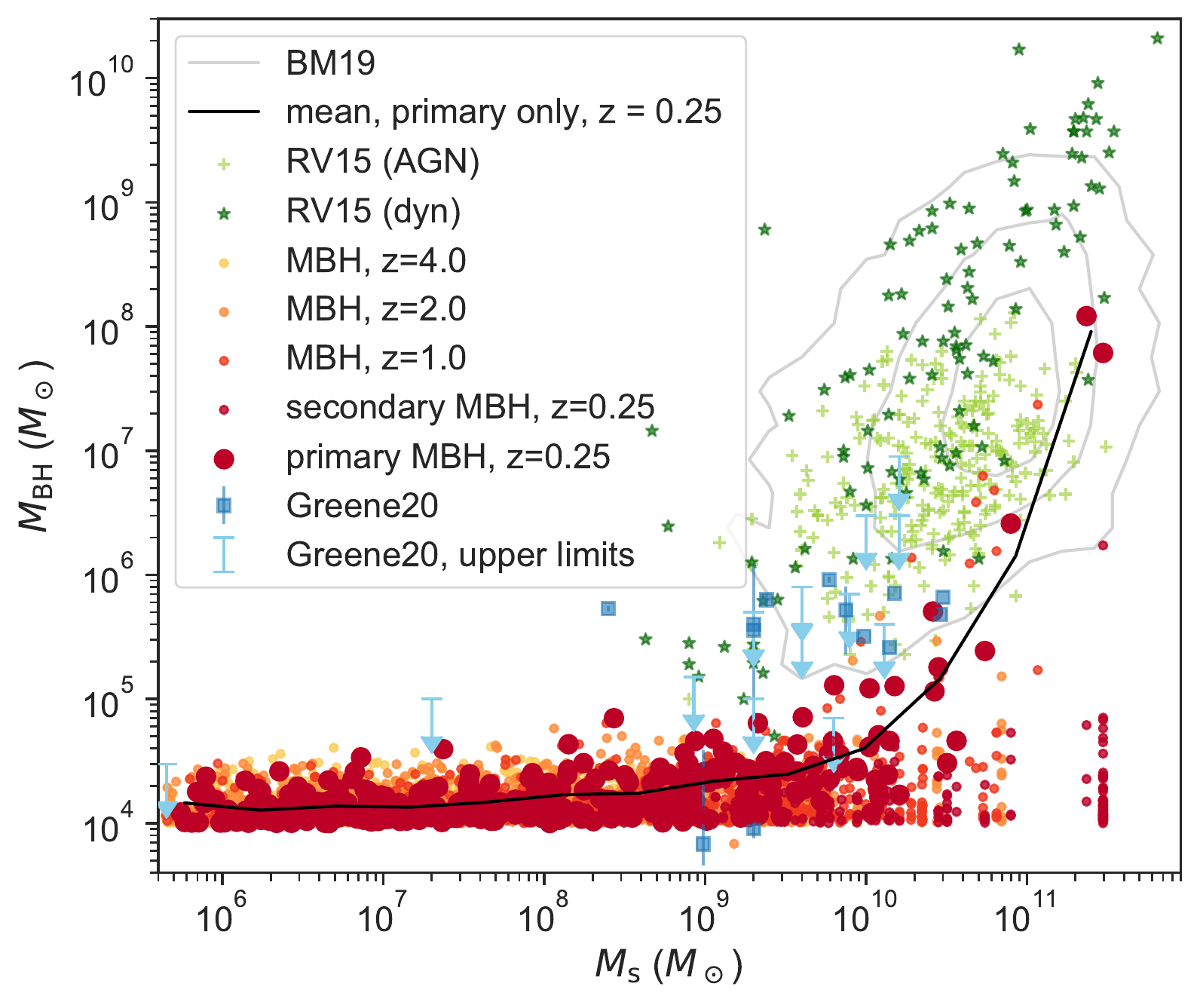}
    \caption{MBH mass as a function of galaxy mass for redshifts $z=4,2,1$ and $0.25$ as circles for all MBHs contained within 2 effective radii of their host galaxy. For the lowest redshift, $z=0.25$, big circles (dark red) highlight the most massive MBH for a given galaxy, while the small circles (dark red) show all the secondary MBHs within the galaxy. The black line denotes the mean $M_{\rm BH}$ for all primary BH within a given $M_{\rm S}$ bin at $z=0.25$ for \nh. Also shown are observations of MBH versus stellar mass for $z\sim 0-0.3$ from \citet{reines&volonteri15} (RV15; green triangles), \citet{greeneetal20} (Greene20; blue markers) and  \citet{baron&menard2019} (BM19; grey contours). The error bars in RV15 were omitted for clarity. On average, central MBHs in \nh\ grow significantly only above a stellar mass threshold of a few $10^9 \,\rm M_\odot$, with non-central MBHs generally failing to grow even in galaxies above this mass threshold.}
    \label{fig:mstar_mbh}
\end{figure}

\begin{figure*}
    \centering
    \includegraphics[width=0.80\textwidth]{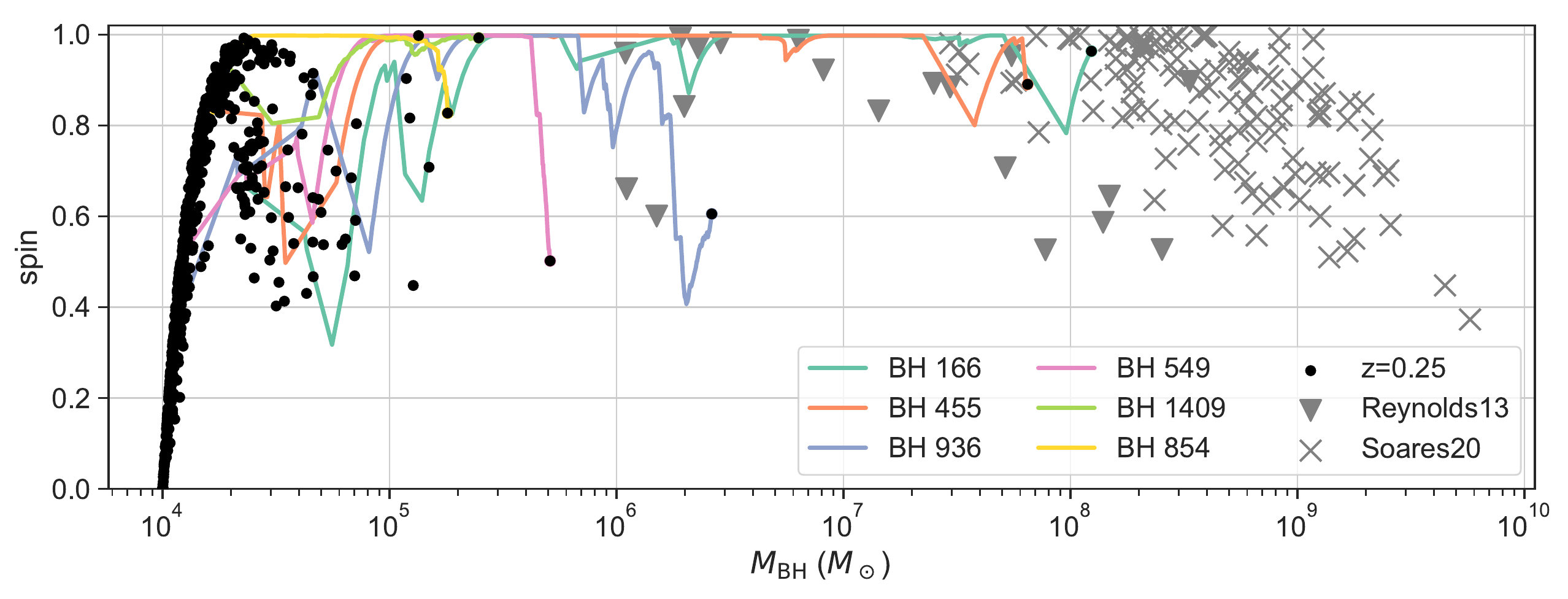}
    \caption{MBH spin evolution with MBH mass for all MBH at redshift $z=0.25$ (black markers), and the spin evolution history with mass for the six most massive MBH (coloured lines). Shown for comparison are observations from \citet{reynolds2013} (grey triangles) and \citet{soares&nemmen2020} (grey crosses). MBHs experience a fast initial spin growth phase followed by a more stochastic evolution.}
    \label{fig:mbh_spin}
\end{figure*}

\subsection{MBH-to-galaxy mass relation}

In NewHorizon, MBH formation commonly occurs even in low-mass galaxies, as can be seen by the fact that about half  of galaxies with a stellar mass $10^7-10^8 \msun$ at redshift z = 0.2 5 host at least one BH. Observationally, a statistical analyses based on X-ray observations  \citep{milleretal2015,sheetal2017} suggest that between 10 percent and 100 percent of galaxies with mass $\sim10^9 \,\rm M_\odot$ at $z=0$ host a MBH. While using dynamical masses or upper limits in nucleated local galaxies within 4 Mpc, \cite{nguyenetal2018} find that between 50 and 100 percent of galaxies with mass $\sim10^9-10^{10} \,\rm  M_\odot$ contain a MBH. By redshift $z=0.25$, \nh\ contains 583 MBHs that are located within two effective radii of a galaxy included in the catalogue. Each MBH can only be associated with a single galaxy at a given point in time, but a galaxy can contain multiple black holes and this particularly occurs for the most massive galaxies in \nh . 
This is reflected in the occupation fraction \citep{2008MNRAS.383.1079V}, which increases with galaxy stellar mass from 0.28  at $M_{\rm s}=10^6 \,\rm M_\odot$ to unity at  $M_{\rm s}= 10^{8.5}\,\rm M_\odot$. At higher galaxy masses there is on average more than one MBH per galaxy, reaching an average of two MBHs per galaxy at $M_{\rm s}=10^{10}\,\rm M_\odot$. The existence of additional `off-centre' MBHs in massive galaxies is a standard prediction of models and simulations that study the cosmic evolution of MBHs in galaxies \citep[e.g.][]{2003MNRAS.340..647I,2010ApJ...721L.148B,volonterietal16,2018ApJ...857L..22T}.  The relatively high occupation fraction in low-mass galaxies allows for MBH mergers to also occur in low-mass galaxies, following galaxy mergers. This is studied in further detail in~\cite{volonterietal20}. 

Not all MBHs are located close to the centre of their host galaxy. Overall, 38\% are located within 2.5 kpc of the centre, while the rest can be found on larger orbits. This agrees with other recent simulations of MBHs in dwarf galaxies  \citep{bellovaryetal19,pfisteretal19}. In~\cite{bellovaryetal19}, it is reported that 50 \% of MBHs in dwarf galaxies wander up to several kiloparsec from the galaxy centre. Wandering MBHs have also been found observationally, both in dwarf galaxies \citep{Reinesetal2020} and in more massive galaxies \citep{menezesetal2014,menezesetal2016,shenetal2019}. Maximum offset distances reported in observations are smaller than in simulations because they preferentially detect active MBHs, which biases their sample to centrally located MBHs that are fed by the gas supply of their host galaxy. In contrast, our sample includes all MBHs, whether active or not.

As previously shown in~\cite{duboisetal15snbh}~\citep[see also][]{habouzitetal17,boweretal17,prietoetal17,anglesalcazaretal17,mcalpineetal18,trebitschetal18,trebitschetal20}, we confirm that MBH growth is regulated by SN feedback for galaxies with stellar masses below $M_{\rm s}<5 \times 10^{9}\,\rm M_\odot$, which make up the bulk of the stellar population in \nh. As shown in Fig. \ref{fig:mstar_mbh}, as a consequence of this most of the MBH population grows little over the course of the simulation. Across all stellar mass bins, MBHs in \nh\  are under-massive in comparison to observations, lying up to two orders of magnitude below the bulk of the observed MBH masses. However,  MBHs in more massive galaxies ($>10^{11} \,\rm M_\odot$) lie closer to the locus of observational data than those in low-mass galaxies, suggesting that MBHs in \nh\ are still in the process of catching up with the observed relation.

By redshift $z=0.25$, 79 \% of MBHs have retained a mass that remains within a factor 2 of their seed mass (chosen to be $10^4 \,\rm M_\odot$ in \nh). It is only once their host galaxy reaches the transition stellar mass of $M_{\rm s}>5 \times 10^{9} \,\rm M_\odot$ that the main MBH of a galaxy (big circles, here defined to be the most massive MBH for each galaxy) grows efficiently and soon reaches the observed $M_{\rm BH} - M_{\rm s}$ relation (triangles and contours) because the deepening gravitational potential of their host galaxies allows for nuclear inflows to start efficiently feeding the central MBH. All secondary MBHs associated with a galaxy (small dark red circles for $z=0.25$) continue to struggle to grow as they are too far from the centre of their galaxy to efficiently access this increased gas supply. 

\subsection{MBH spins}

In \nh, MBH spin is followed on the fly, taking into account the effects of gas accretion and MBH-MBH mergers. As shown in Fig. \ref{fig:mbh_spin}, even though MBHs are initially formed with zero spin, they spin up quickly as their mass doubles, which is in line with earlier findings that gas accretion leads to maximally spinning MBH~\citep{duboisetal14bhspincosmo,duboisetal14bhspin,bustamante&srpingel19}. From there, spin evolution becomes more complex because mass growth is accomplished through MBH-MBH mergers and gas accretion and has the broadest scatter in the mass range $M_{\rm BH}= 2 \times 10^4 - 10^5 \, \rm M_\odot$. The MBHs are formed with zero spin and initial mass $M_{\rm BH}=10^4 \,\rm M_\odot$, and for masses up to $2 \times 10^4 \,\rm M_\odot$ the only way spin can change is by gas accretion. We note  that at low accretion rates $\chi<0.01$ jets spin MBHs down, while at higher accretion rates gas accretion can either spin MBHs up or down, depending on the relative orientation of the spin and the gas feeding the MBH. As shown in Fig. \ref{fig:mbh_spin}, for MBHs with mass up to $2 \times 10^4 \,\rm M_\odot$, spin increases with mass, implying that gas accretion with $\chi\geq 0.01$ is responsible for increasing MBH spins. Most of accretion occurs in prograde fashion, meaning that gas and MBH spin are initially aligned or align during the accretion episode, which is in line with earlier findings that gas accretion leads to maximally spinning MBH \citep{duboisetal14bhspin}. Above $M_{\rm BH}= 2 \times 10^4 \,\rm M_\odot$ spin can be affected by the combination of accretion and MBH-MBH mergers, which can modify abruptly both spin magnitude and orientation. The MBH-MBH mergers are responsible for the large scatter in the mass range $M_{\rm BH}= 2 \times 10^4 - 10^5 \,\rm M_\odot$. As shown in the coloured evolution histories, individual MBHs can repeatedly spin up and down, but as they grow more massive all MBHs in our sample spin up to spin values larger than 0.7. All MBHs with mass $>10^6 \,\rm M_\odot$ have acquired more than 80\% of the their mass through accretion, which is once again responsible for increasing MBH spins. This is in good agreement with observations, which find high spin values for the mass range $M_{\rm BH} = 10^6 - 5 \times 10^7 \,\rm M_\odot$ \citep{reynolds2013}. For higher MBH masses, spins are expected to turn over again as MBH mergers become increasingly important in MBH growth, but no MBHs in our sample grows sufficiently massive to probe this regime. The impact of mergers and gas accretion on the mass and spin evolution of the MBH population will be explored further in a companion paper (Beckmann et al. in prep.). 

\section{Conclusions}
\label{section:conclusions}

The \nh\ tool provides a compromise  to alternative strategies, namely large-volume cosmological simulations with poor spatial resolution within galaxies or selected high-resolution zoomed-in halos, by simulating an intermediate sub-volume $(16\,\rm Mpc)^3$ of average cosmic density within a larger box that is resolved at a spatial resolution good enough ($34\,\rm pc$) to capture the turbulence and multi-phase structure of the ISM.
Despite a still important effect of the cosmic variance on the various estimated properties, we believe this compromise between sampling a diversity of environments and high resolution is an important step to faithfully capture the relevant physical processes involving star formation
and to improve our understanding of the various mechanisms at work that shape galaxies over cosmic times.
We have shown that \nh\ reproduces with a good level of accuracy several of the key properties that define galaxies, which we briefly review:
\begin{itemize}
\item Galaxies simulated in \nh\ naturally exhibit a multi-phase structure with dense cold star-forming clumps embedded in warm and hot diffuse gas, where turbulence shape star-forming properties of galaxies. 
\item The galaxy mass function with the characteristic turnover at $10^{10}-10^{11}\,\rm M_\odot$ is well reproduced at low redshift when surface brightness limitations are accounted for.
\item The cosmic SFR density and stellar density compare reasonably well with observations, although they are higher than in observations (up to a factor of 2 depending on the observations), and with a significant uncertainty in the predicted value of the cosmic SFR (0.7 dex) owing to cosmic variance.
The sSFR as a function of galaxy stellar mass was measured at various redshifts and a comparison with the observations shows that the simulated sSFRs are in a fair agreement with the observations, showing that galaxies were more active in the past.
\item The simulated Kennicutt-Schmidt relation points are in fair agreement with the observed main sequence although \nh\ galaxies are a factor $\simeq 0.3\,\rm dex$ below the observations.
High-redshift galaxies show a larger amount of starburst galaxies above the main sequence in qualitative agreement with the data.
\item The stellar-to-halo mass relations show that galaxies are relatively too massive in \nh\ at the low-mass end ($M_{\rm h}$<  a few $10^{11}\,\rm M_\odot$) compared to observational data.
The relation is in good agreement at the Milky Way mass scale.
\item The size-mass relation of galaxies falls well within the range of observations with a significant positive evolution of galaxy sizes over decreasing redshift for a given stellar mass.
At the high-mass end, \nh\ produces more compact galaxies than expected and no extended ellipticals, although there is a clear lack of statistics in this mass range.
\item The metallicities of stars and gas agree reasonablly well with the observations in mass and redshift, despite a crude model for chemical evolution. However, the stellar metallicities in \nh\ show less variation with redshift than in the observations.
\item \nh\ galaxies display an increase in stellar rotation over dispersion over mass and decreasing redshift.
The fraction of ellipticals based on their stellar kinematics compares well with observations, although the exact fraction is sensitive to the radius within which the kinematics are measured along with the adopted threshold.
\item Simulated galaxies at high redshift have a larger fraction of their stellar mass enclosed in stellar clumps than at low redshift for a given stellar mass, as a result of more gas-rich, turbulent, and concentrated galaxies.
\item The \nh\ galaxies have lower stellar and gas surface densities over decreasing redshift.
The fraction of cold star-forming gas is decreasing over decreasing redshift at a given stellar mass, which makes galaxies less gravitationaly unstable.
\item The bulk of the Tully-Fisher relation is fairly well captured, where a slope of  one-third is preferred over a slope of one-fourth.
\item The gas in \nh\ galaxies settles into discs over decreasing redshift as a result of the decaying level of gas turbulence, while gas rotation remains similar (at constant mass).
\item The MBH-to-galaxy mass relation shows a sharp increase at a stellar mass of a few $10^{9}\,\rm M_\odot$. 
Observations are broadly reproduced, although MBH masses fall short.
\item Once MBHs manage to grow in mass their spin is also large ($a>0.7-0.8$) and has a sharp rise during the initial growth phase with a more chaotic evolution during self-regulation of the MBH mass, leading to a fair agreement of the MBH spin-mass relation with the scarce data in this mass range.
\end{itemize}

This paper is the first step of an introduction to \nh\ by reviewing the bulk properties of galaxies (stars and gas) and of their hosted MBHs. 
In particular, we have not investigated  the properties of the circum-galactic medium of galaxies (including hot diffuse and cold flows) and how it interacts with the galactic outflows, the shape of the DM distribution within halos and galaxies, inner substructures within galaxies (bars, bulges and pseudo-bulges, thin and thick discs, and properties of star-forming clouds of gas). In addition, the  \nh\ set-up  limits our study to a limited volume within a fairly average cosmic density, and similar studies should extend this work to various environments such as dense galaxy clusters~\citep[see e.g.][for a first attempt in a proto-cluster]{trebitschetal20}, voids, or within large-scale tens-of-Mpc wide cosmic filaments. 
Since the properties of galaxies in \nh\ are expected to depend on the adopted sub-grid models~\citep[see for instance,][]{duboisetal12,duboisetal16,kimmetal15,chabannieretal20lya,nunezetal20} and resolutions~\citep[on at least some aspects, numerical resolution even in the diffuse large-scale medium can be more important than the parametrisation of sub-grid models,][]{chabanieretal20}, it would be ideal to explore the effects of critical parameters systematically in order to assess the robustness of the results. However, it is beyond the available computing resources considering that the combination of the volume and resolution of \nh\ is already pushed to the limit.

\section*{Acknowledgements}
Y.D., S.K.Y., J.D., T.K., C.P., S.P., and M.V conceived and planned
the \nh\ simulation project. 
Y.D., J.D. and T.K. prepared the preparatory runs and
Y.D., S.K.Y., H.C., K.K. and F.B. ran the simulation.
Y.D., J.D., T.K. and M.V. developed the models for galaxy formation.
C.P. produced the high-resolution initial conditions and managed the website.
R.B. and M.V. contributed to the analysis of MBH properties.
K.K. contributed to the Kennicutt-Schmidt and gas fraction analysis.
H.C. generated mock images of galaxies.
M.-J.P. and G.M. contributed to kinematic analyses on galaxies.
G.M. and S.K. contributed to the analysis on galaxy mass functions.
R.J., G.M. and S.K. contributed to the analysis on galaxy surface brightness.
C.L. contributed to the cosmic SFR analysis.
Y.D. produced halo and galaxy catalogues and contributed to all the analysis of the paper.
The manuscript has been written by Y.D. with contributions from all co-authors.
Camille No\^us contributed to the collegial construction of the standards of science, by developing the methodological framework, the state-of-the-art, and by ensuring post-publication follow-up. 
 We thank Olivier Ilbert for assistance using \textsc{LePhare}.
 We thank the anonymous referee for his/her constructing comments that improved the clarity of this paper.
 This work was granted access to the HPC resources of CINES under the allocations  c2016047637, A0020407637 and A0070402192 by Genci, KSC-2017-G2-0003 by KISTI, and as a “Grand Challenge” project granted by GENCI on the AMD Rome extension of the Joliot Curie supercomputer at TGCC. 
This research is part of the Spin(e)  ANR-13-BS05-0005 (\href{http://cosmicorigin.org}{http://cosmicorigin.org}), Segal ANR-19-CE31-0017 (\href{http://secular-evolution.org}{http://secular-evolution.org}) and Horizon-UK projects.
This work has made use of the Horizon cluster on which the simulation was post-processed, hosted by the Institut d'Astrophysique de Paris. We warmly thank S.~Rouberol for  running it smoothly.
The large data transfer was supported by KREONET which is managed and operated by KISTI. 
S.K.Y. acknowledges support from the Korean National Research Foundation (NRF-2020R1A2C3003769). 
TK was supported in part by the National Research Foundation of Korea (NRF-2017R1A5A1070354 and NRF-2020R1C1C1007079) and in part by the Yonsei University Future-leading Research Initiative (RMS2-2019-22-0216). 
This work is partly based on tools and data products produced by GAZPAR operated by CeSAM-LAM and IAP. 
This paper is based in part on data collected at the Subaru Telescope and retrieved from the HSC data archive system, which is operated by Subaru Telescope and Astronomy Data Center (ADC) at National Astronomical Observatory of Japan. Data analysis was in part carried out with the cooperation of Center for Computational Astrophysics (CfCA), National Astronomical Observatory of Japan.
The Hyper Suprime-Cam (HSC) collaboration includes the astronomical communities of Japan and Taiwan, and Princeton University. The HSC instrumentation and software were developed by the National Astronomical Observatory of Japan (NAOJ), the Kavli Institute for the Physics and Mathematics of the Universe (Kavli IPMU), the University of Tokyo, the High Energy Accelerator Research Organization (KEK), the Academia Sinica Institute for Astronomy and Astrophysics in Taiwan (ASIAA), and Princeton University. Funding was contributed by the FIRST program from the Japanese Cabinet Office, the Ministry of Education, Culture, Sports, Science and Technology (MEXT), the Japan Society for the Promotion of Science (JSPS), Japan Science and Technology Agency (JST), the Toray Science Foundation, NAOJ, Kavli IPMU, KEK, ASIAA, and Princeton University. 
This paper makes use of software developed for the Large Synoptic Survey Telescope. We thank the LSST Project for making their code available as free software at  http://dm.lsst.org
Catalogues extracted from the simulation will be made available at this URL
\href{https://new.horizon-simulation.org/data.html}{https://new.horizon-simulation.org/data.html}.

\bibliographystyle{aa}
\bibliography{author}

\appendix

\section{Purity of halos}
\label{appendix:halopurity}

\begin{figure}
\centering \includegraphics[width=0.40\textwidth]{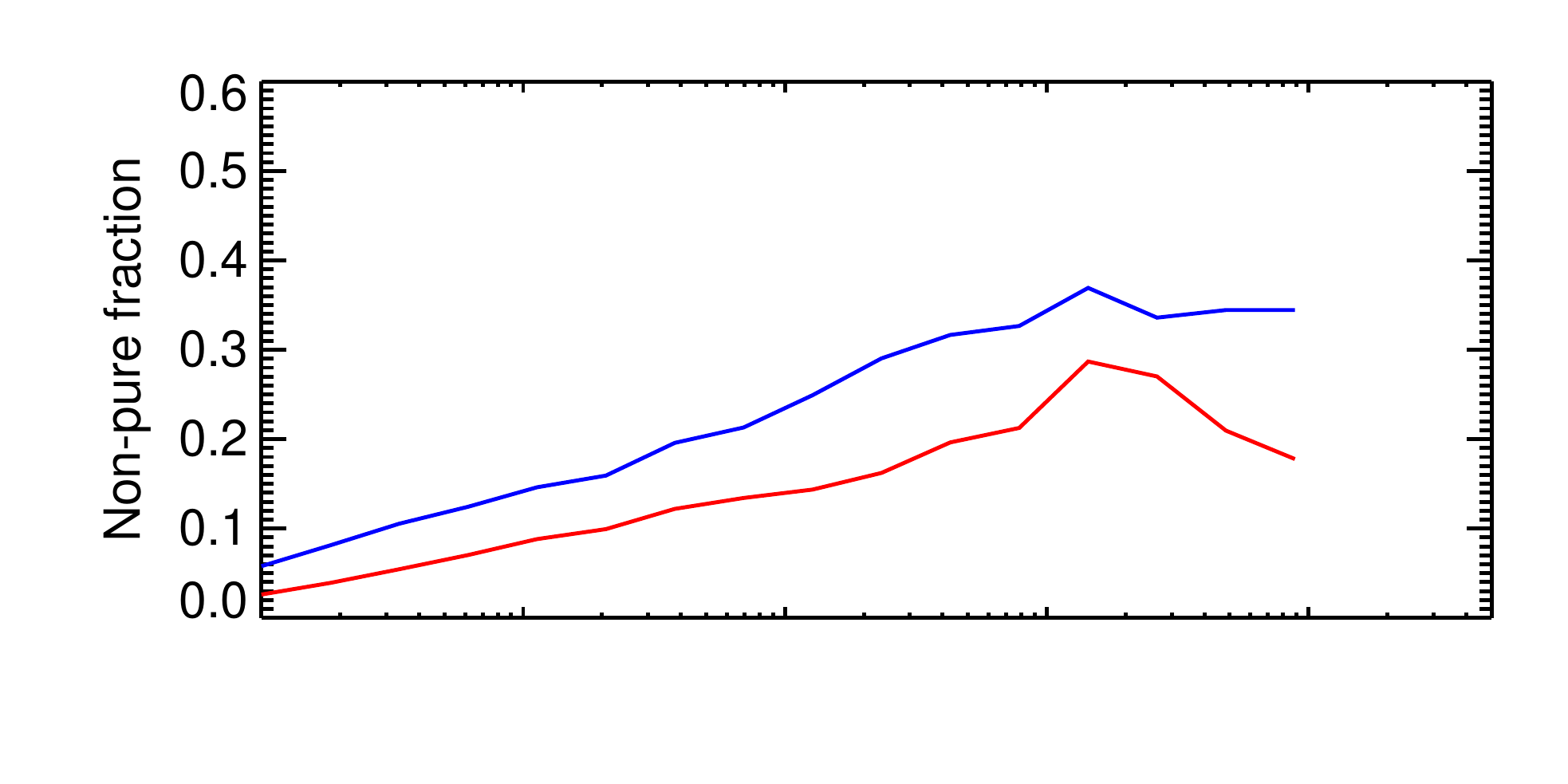}\vspace{-1.2cm}
\centering \includegraphics[width=0.40\textwidth]{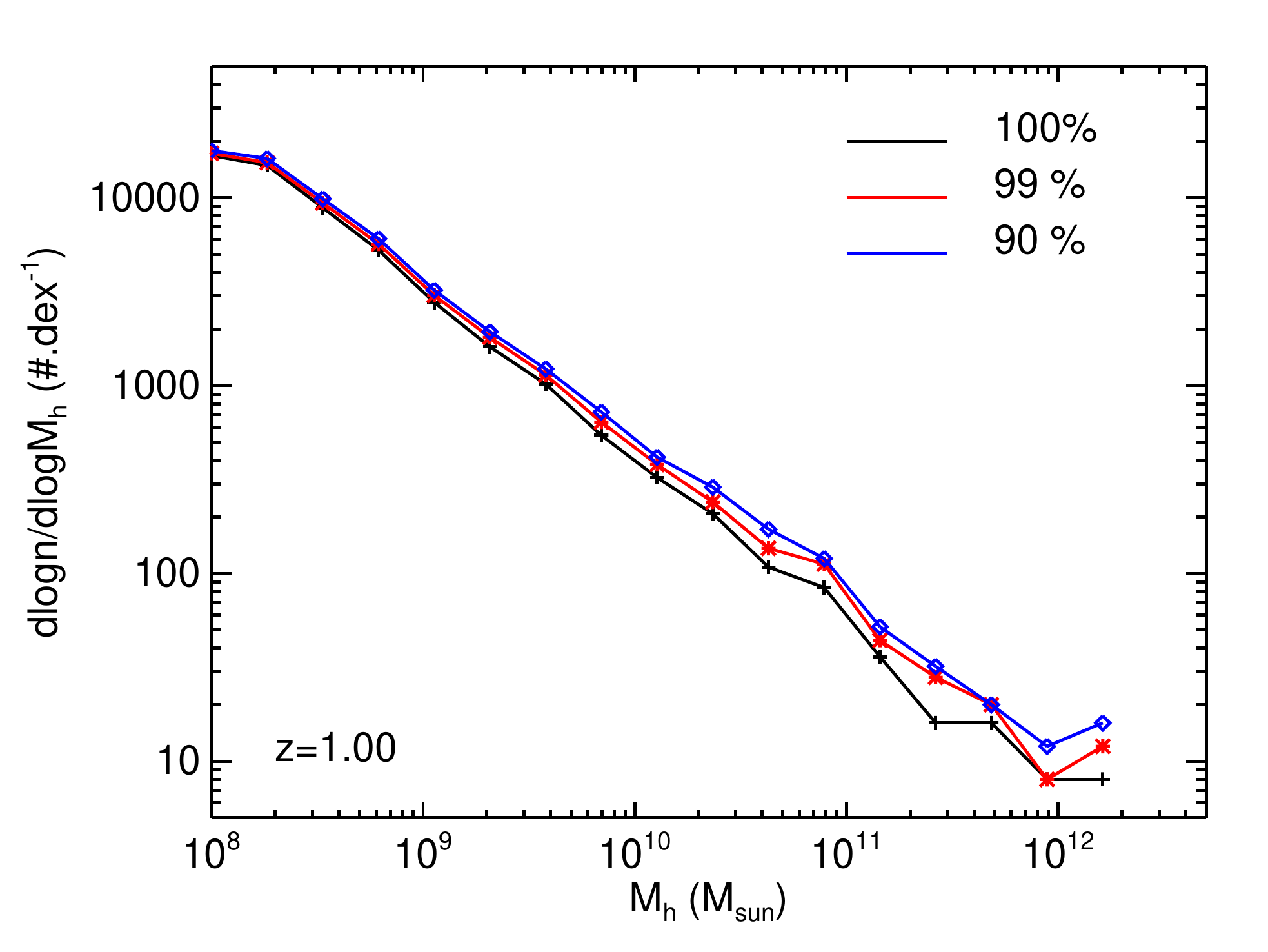}
\caption{Halo mass function (bottom panels) and fraction of non-pure halos (top panels) within the zoom simulation for halos with different levels of pollution (as indicated in the panel) for redshift $z=1$. Only  halos with a $100\%$ purity are considered in this study.}
\label{fig:halo_pollution}
\end{figure}

Figure~\ref{fig:halo_pollution} shows the halo DM mass function at redshift $z=1$ for different levels of purity of the host: for 90\%, 99\%, or 100\% of purity, where the levels of purity are computed in terms of the number fraction of high-resolution DM particles over the total number of DM particles in the halo.
The amount of massive halos can increase by up to $30\%$ if halos other than perfectly pure halos are considered.
However, in this study we only consider  halos with a $100\%$ purity.

\section{Comparison of HOP versus AdaptaHOP sizes}
\label{appendix:size_hop_vs_ahop}

\begin{figure}
\centering \includegraphics[width=0.24\textwidth]{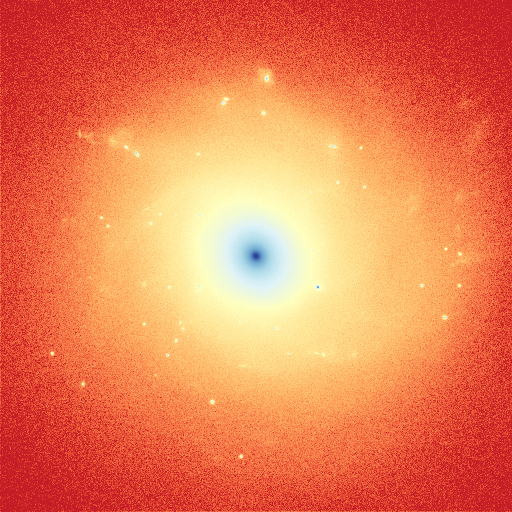}
\centering \includegraphics[width=0.24\textwidth]{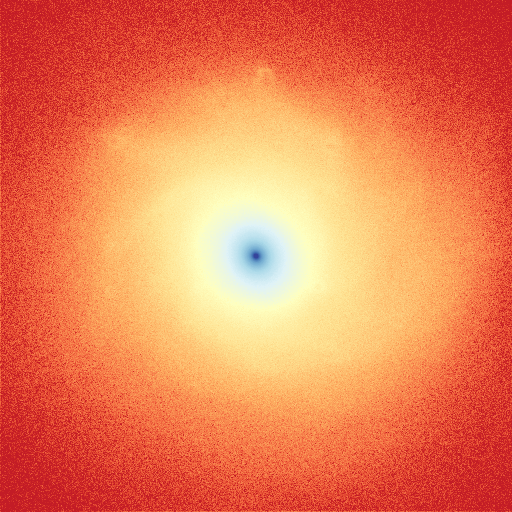}
\caption{Stellar density distribution for the most massive galaxy $M_{\rm s}=1.5\times 10^{11}\,\rm M_\odot$ at $z=1$ as identified by HOP (left panel) or AdaptaHOP (right panel). Image size is 30 kpc across.}
\label{fig:sdens_comp}
\end{figure}

\begin{figure}
\centering \includegraphics[width=0.24\textwidth]{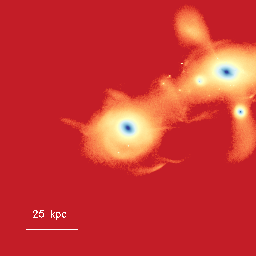}
\centering \includegraphics[width=0.24\textwidth]{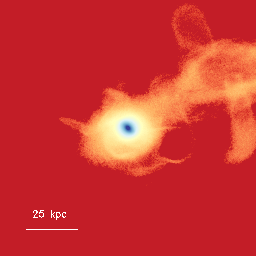}
\centering \includegraphics[width=0.25\textwidth]{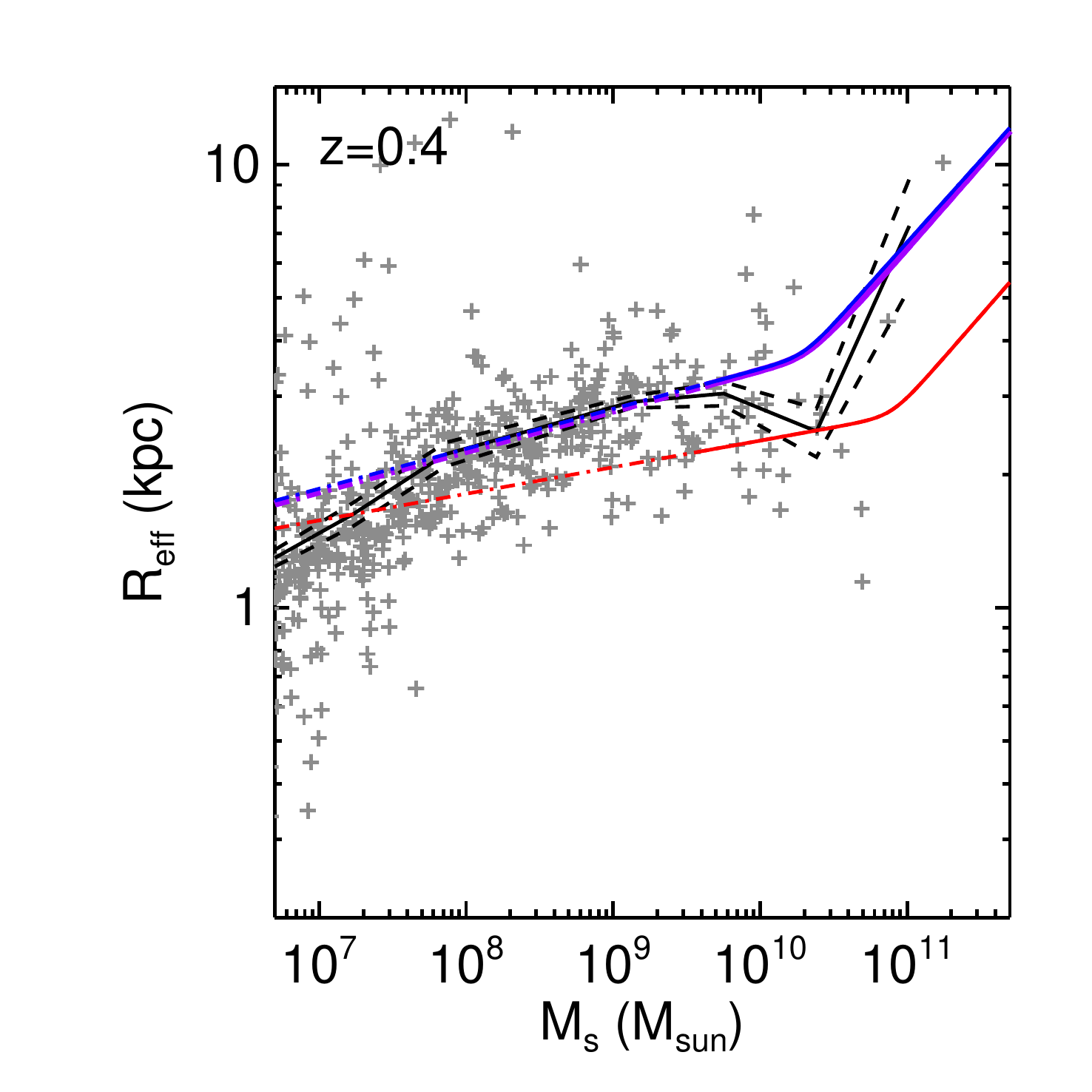}\hspace{-0.6cm}
\centering \includegraphics[width=0.25\textwidth]{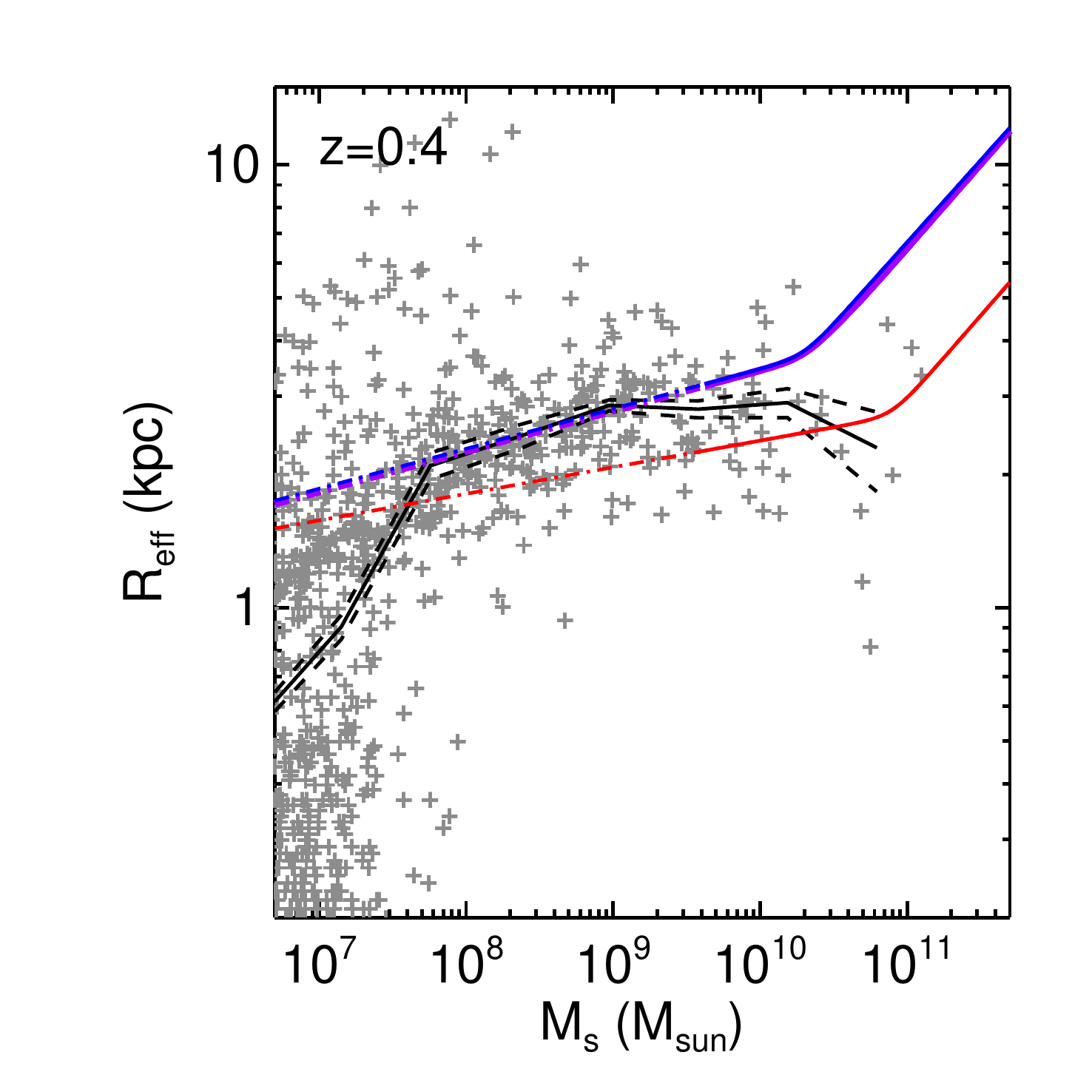}
\caption{Stellar density maps of a galaxy and companion satellite at $z=0.4$ as obtained by HOP (top left panel) and AdaptaHOP (top right panel).
Stellar mass and effective radius are $M_{\rm s}=2.14\times 10^{11}\,\rm M_\odot$ and $R_{\rm eff}=39 \,\rm kpc$, and $M_{\rm s}=1.25\times 10^{11}\,\rm M_\odot$ and $R_{\rm eff}=3.3 \,\rm kpc$ for HOP and AdaptaHOP respectively. 
The resulting size-mass relations (plotted as in Fig.~\ref{fig:rvsmg_redshift}) are affected by the galaxy finder as shown in the two bottom panels (left: HOP; right: AdaptaHOP). }
\label{fig:sizecomp}
\end{figure}

To illustrate how well the AdaptaHOP can separate galaxy substructures, in Fig.~\ref{fig:sdens_comp} we show two stellar density maps of a galaxy at $z=1$ (the most massive galaxy), extracted from the identified structures with HOP (including the stellar clumps) or with AdaptaHOP (which removes sub-structures).
With the HOP finder, multiple stellar clumps can be identified within the galaxy, while AdaptaHOP has removed all substructures.
In Fig.~\ref{fig:sizecomp}, we show two stellar density maps of a merging galaxy at $z=0.4$ as extracted from either HOP or AdaptaHOP.
Indeed, AdaptaHOP efficiently identifies stellar clumps and removes them from the main structures and also removes companion satellite galaxies\footnote{Although a haze of stars from the companion is still seen and associated with the main object. This component could be removed by using a more robust structure finder that uses information from the velocity distribution~\citep{canasetal19}.} when their stellar distribution connects to the main object.
Since HOP is not able to remove any substructure, evaluating the size of a galaxy becomes a severe issue for galaxies in a significant interaction; see the two bottom panels of Fig.~\ref{fig:sizecomp}, which leads to an incorrect evaluation of the size of the main galaxy.

\section{Fraction of morphological types: Changing the ad hoc thresholds for classification}
\label{Appendix:fell}

\begin{figure}
\centering \includegraphics[width=0.49\textwidth]{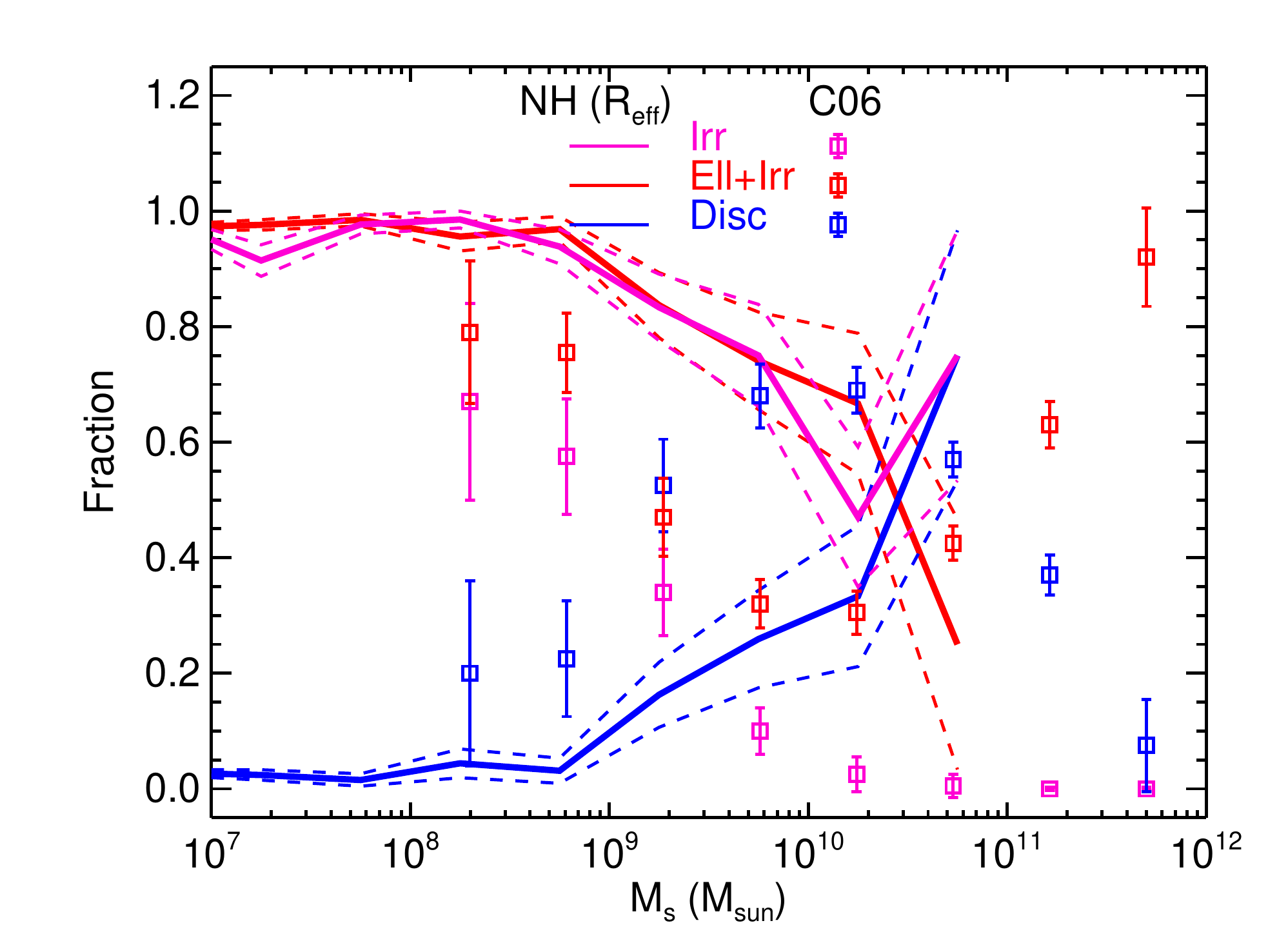}\vspace{-0.5cm}
\centering \includegraphics[width=0.49\textwidth]{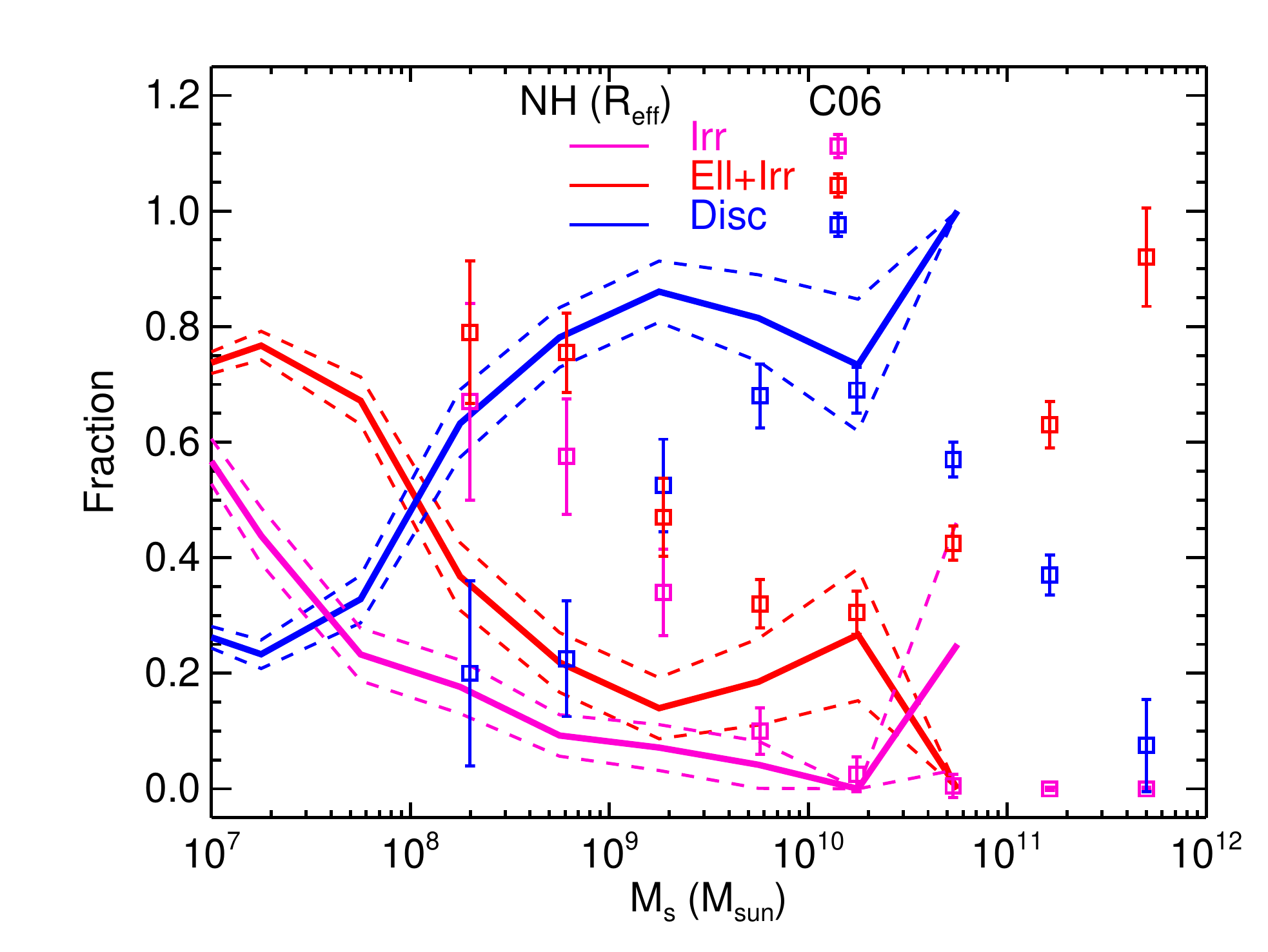}
\caption{Same as Fig.~\ref{fig:fellvsmg}: Fraction of each morphological type as a function of stellar mass with different thresholds for $(V/\sigma)_{\rm c}$ and $A_{\rm r, c}$, with 0.7 $\&$ 0.2 (top panel), and 0.3 $\&$ 0.4 (bottom panel), respectively.}
\label{fig:fell_threshold}
\end{figure}

The exact value of the fraction of galaxies per morphological type depends on the adopted thresholds for their kinematic $V/\sigma$ or their asymmetry $A_{\rm r}$ classifier. 
The adopted values of $(V/\sigma)_{\rm c}=0.5$ and $A_{\rm r,c}=0.3$ were chosen to best fit the fractions from~\cite{conselice06}. 
We show in Fig.~\ref{fig:fell_threshold} how changing those values between $(V/\sigma)_{\rm c}=0.3-0.7$ and $A_{\rm r,c}=0.2-0.4$ affect the fraction of each morphological type.
These fractions change with adopting different values for those thresholds, as expected. There are a higher (lower) fraction of ellipticals and irregulars when increasing (decreasing) $(V/\sigma)_{\rm c}$ (and conversely for the fraction of kinematically classified discs) and decreasing (increasing) $A_{\rm r,c}$, but the qualitative trends with mass are preserved.

\section{Fraction of low-mass stellar clusters}
\label{Appendix:stellarclumpnpart10}

\begin{figure}
\centering \includegraphics[width=0.49\textwidth]{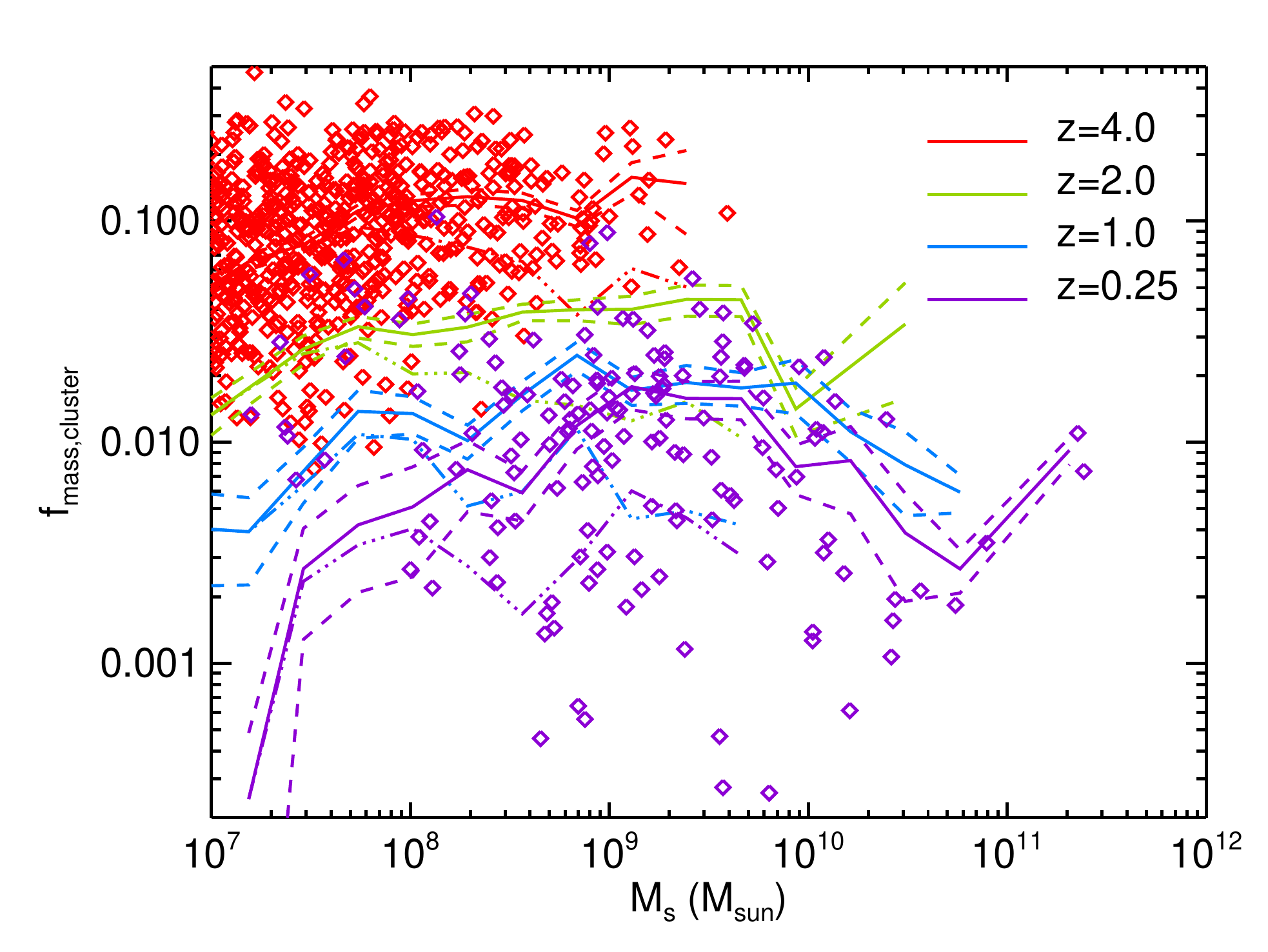}
\caption{Same as Fig.~\ref{fig:smass_clump}: fraction of stellar mass in stellar clusters as a function of galaxy stellar mass at different redshifts as indicated in the panel, except that a lower number of stellar particles are used to detect stellar clumps (10 instead of 50) in AdaptaHOP. The solid lines indicate the mean values, and dashed lines stand for the error on the mean. The data points are overplotted for the most extreme redshifts to appreciate the scatter in the distribution of values.}
\label{fig:smass_clump_npart10}
\end{figure}

The minimum number of stellar particles used to detect substructures with AdaptaHOP can affect the amount of stellar mass contained in stellar clusters. 
We evaluate the effect by lowering our fiducial number of 50 stellar particles down to 10 on the fraction of stellar mass contained in clusters in Fig.~\ref{fig:smass_clump_npart10}.
The mean relation are relatively similar at all considered redshifts, except for the most massive galaxies $M_{\rm s}\ge 10^{11}\,\rm M_\odot$ at the lowest redshift $z=0.25$ considered (to be compared with Fig~\ref{fig:smass_clump}).

\end{document}